\documentclass[11pt]{article}

\usepackage[english]{babel}
\usepackage{amsthm}
\usepackage{amsmath}
\usepackage{esint}
\usepackage{constants}
\usepackage{comment}
\usepackage{bm}
\usepackage{amsbsy}
\usepackage{amssymb}
\usepackage{enumitem}
\usepackage{citesort}
\usepackage{cite} 
\usepackage{hyperref} 
\usepackage{subfigure} 
\usepackage{fancyref}
\usepackage{graphicx}
\usepackage{wrapfig}
\usepackage{psfrag}
\usepackage{color}
\usepackage{caption}
\usepackage{float}
\usepackage{url}
\hypersetup{
   colorlinks=true,%
   citecolor=black,%
   filecolor=black,%
   linkcolor=red,%
   urlcolor=blue
}
\usepackage{epigraph}
\usepackage[active]{srcltx}
\usepackage{verbatim}
\usepackage{calc}
\usepackage{comment}
\usepackage{mathtools}
\usepackage{float}
\usepackage{import}
\usepackage{color}

\allowdisplaybreaks 

\makeatletter

\newtheorem{definition}{Definition}[section]
\newtheorem{lemma}{Lemma}[section]

\newtheorem{thm}{Theorem}[section]
\newtheorem{prop}{Proposition}[section]

\theoremstyle{plain}

\theoremstyle{remark}
\newtheorem{rem}{Remark}[section]
\theoremstyle{definition}

\providecommand{\keywords}[1]{\noindent \textbf{\textit{Keywords---}} #1}

\newcommand{\Basic}[1]{\arabic{#1}}
\newconstantfamily{Bet}{
symbol=\beta,
format=\Basic
}

\begin{document}

\title{Stabilizing Embedology: Geometry-Preserving\\ Delay-Coordinate Maps}
\date{}
\author{Armin Eftekhari, Han Lun Yap, Michael B. Wakin, and Christopher J. Rozell\footnote{The first two authors contributed equally, as did the last two authors.  AE is with the Alan Turing Institute.  HLY is with DSO National Laboratories of Singapore. MBW is with the Department of Electrical Engineering and Computer Science at the Colorado School of Mines. CJR  is  with  the  School  of  Electrical  and Computer Engineering at the Georgia Institute of Technology.
Email: crozell@gatech.edu. This work was partially supported by NSF grants CCF-0830320, CCF-0830456, CCF-1409258, and CCF-1409422; NSF CAREER grant CCF-1350954; and James S. McDonnell Foundation grant number 220020399.}
}
\maketitle

\begin{abstract}
Delay-coordinate mapping is an effective and widely used technique for reconstructing and analyzing the dynamics of a nonlinear system based on time-series outputs. The efficacy of delay-coordinate mapping has long been supported by Takens' embedding theorem, which guarantees that delay-coordinate maps use the time-series output to provide a reconstruction of the hidden state space that is a one-to-one embedding of the system's attractor. While this topological guarantee ensures that distinct points in the reconstruction correspond to distinct points in the original state space, it does not characterize the quality of this embedding or illuminate how the specific parameters affect the reconstruction.  In this paper, we extend Takens' result by establishing conditions under which delay-coordinate mapping is guaranteed to provide a stable embedding of a system's attractor.  Beyond only preserving the attractor topology, a stable embedding preserves the attractor geometry by ensuring that distances between points in the state space are approximately preserved.  In particular, we find that delay-coordinate mapping stably embeds an attractor of a dynamical system if the stable rank of the system is large enough to be proportional to the dimension of the attractor. The stable rank reflects the relation between the sampling interval and the number of delays in delay-coordinate mapping.  Our theoretical findings give guidance to choosing system parameters, echoing the trade-off between irrelevancy and redundancy that has been heuristically investigated in the literature.  Our initial result is stated for attractors that are smooth submanifolds of Euclidean space, with extensions provided for the case of strange attractors.
\end{abstract}


\section{Introduction}
\subsection{Motivation and Contribution}

Modern science is ingrained with the premise that repeated observations of a dynamic phenomenon can help us understand its underlying mechanisms and predict its future behavior.  While this idea dates back to ancient times with the observation of sunspots~\cite{wittmann.88}, today we model the behavior of a wide variety of measured phenomena from the life, physical, and social sciences~\cite{Brockwell2002,beuter.03,mosekilde.96,sulis.01,saperstein.88,grebogi.97,roe.96,wolfson.92,udwadia.06,Kantz2004} as observations arising from complex dynamical systems.  Understanding and predicting a time series is often approached by postulating a structured model for a hidden dynamical system that drives the data generation.  Linear statistical models were used in early work~\cite{yule.27} and are now reflected in standard tools such as the \emph{autoregressive-moving-average} model and the \emph{Kalman filter} (e.g., \cite{Bar-Shalom2004}). More recently, the field of \emph{nonlinear time-series analysis} models time-series data as observations of the state of a (possibly high-dimensional) deterministic nonlinear dynamical system~\cite{Kantz2004}. While the underlying dynamical system may exhibit chaotic behavior, it is often postulated as being governed by an attractor that is a low-dimensional geometric subset of the state space.

Due to the low-dimensional behavior in the underlying state space, it is reasonable to postulate that temporal dependencies in time-series observations can provide some insight into the structure of the hidden dynamical system.  This leads to a fundamental question: \emph{How much information about a hidden dynamical system is available in time-series measurements of the system state?}  The seminal \emph{Takens' embedding theorem} \cite{Takens1981,Sauer1991} asserts that (under very general conditions) it is possible to use the time-series data to reconstruct a state-space that is a topologically-equivalent image of the attractor through a simple procedure known as the delay-coordinate map.  Indeed, many algorithms for tasks such as time-series prediction and dimensionality estimation take inspiration and justification from this fundamental guarantee~\cite{Casdagli1991,Fraser1986,kazem2013support,Uzal2011,Kugiumtzis1996,PhysRevLett.106.154101,balaguer2011attracting,bello2011measuring,hamilton2016ensemble}.  While the topological guarantee of Takens' theorem provides that the delay-coordinate map is one-to-one (i.e., distinct points in the reconstruction correspond to distinct points in the original state space), it does not speak to the overall quality of the reconstruction or how this quality is affected by specific details such as the algorithm parameters, the measurement function, or the system characteristics.

Many fields of data science also rely on capturing low-dimensional structure from high-dimensional data, and recent advances have shown the value of guaranteeing \emph{geometric stability} of an embedding as a measure of quality for subsequent inference. In such a stable embedding, the embedding approximately preserves the distance between any two points in the data set of interest, which has proven to be valuable for robustness to imperfections in many forms (i.e., noise, numerical imprecision, etc.). In computer science, the Johnson-Lindenstrauss lemma constructs stable embeddings for finite point clouds using random linear projections~\cite{Dasgupta2002}. In compressive sensing~\cite{Donoho2006,Candes2006d},
the Restricted Isometry Property (RIP) condition captures the notion of a stable embedding for sparse signal families, ensuring that signal reconstruction from random linear measurements is robust to noise and stable with respect to model nonidealities~\cite{Candes2008}. For dimensionality reduction with signal families belonging to low-dimensional manifolds and more general sets, various types of stable embeddings have been constructed using adaptive nonlinear techniques such as ISOMAP~\cite{Tenenbaum2000}, adaptive linear techniques~\cite{Broomhead2001,Hegde15NuMax}, and nonadaptive linear techniques that again employ randomness~\cite{puy2017recipes,Eftekhari2015,Clarkson2008,Yap2013}.

The main contribution of this paper is to extend the notion of Takens' embedding theorem to stable embeddings, providing insight into the conditions for when time-series data can (and cannot) be used to reconstruct a geometry-preserving image of the attractor.  In addition to providing the formal foundations to justify the numerical algorithms based on delay-coordinate mapping, these results also give guidance to practitioners about how algorithm and observation design choices affect the overall quality of the representation.  In particular, examination of our main theoretical findings gives guidance to choosing these system parameters, echoing the trade-off between irrelevancy and redundancy that has been heuristically investigated in the literature.  For clarity and to gain as much insight as possible, our main result is first described for attractors that are smooth submanifolds of the Euclidean space (similar to Takens' original result) and then extended to the case of strange attractors. The remainder of the Introduction will provide a simplified version of the main result to give the flavor of the contribution from this paper, with the full technical results given in Sections~\ref{sec:main results} (smooth manifolds) and~\ref{sec:chaoticextension} (strange attractors).  To streamline readability as much as possible, the proofs and additional technical details are contained in appendices for the interested reader.

\subsection{Delay-Coordinate Maps and Takens' Embedding Theorem}
\label{sec:intro_takens}

We consider $x(\cdot)$ as the trajectory of a dynamical system in the state space $\mathbb{R}^N$ such that $x(t)\in\mathbb{R}^N$ for $t\in[0,\infty)$.  While the system has continuous underlying dynamics, we observe this system at a regular sampling interval $T>0$.  Given this sampling interval, one may define the discrete dynamics in terms of the \emph{flow} $\phi_{T}:\mathbb{R}^N \rightarrow \mathbb{R}^N$ such that $x(t+T)=\phi_{T}(x(t))$. In words, $\phi_{T}(\cdot)$ moves  the system state into the future by $T$.  We assume that during the times of interest the state trajectory is contained within a low-dimensional \emph{attractor}~\cite{Kantz2004} $\mathbb{A}$ such that $x(t)\in \mathbb{A}\subset\mathbb{R}^N$ for $t\geq0$.  The attractor $\mathbb{A}$ is assumed to be a bounded, boundary-less, and smooth submanifold of $\mathbb{R}^N$ with $\operatorname{dim}(\mathbb{A)}<N$.  The flow operator restricted to this attractor is a diffeomorphism on $\mathbb{A}$ so that there exists a smooth inverse $\phi_{T}^{-1}(x(t)) = x(t-T)$.

In applications of interest we often cannot directly observe this system state but rather receive indirect measurements via a scalar measurement function $h:\mathbb{A}\rightarrow\mathbb{R}$.  This function generates a single scalar measurement at a regular sampling interval $T>0$, producing the resulting discrete time series $\{s_i\}_{i\in\mathbb{N}}=\{h(x(i\cdot T))\}_i$, where each $s_i\in\mathbb{R}$.  The goal is to ``reconstruct'' the hidden state trajectory $x(\cdot)$ given only $\{s_i\}_i$.  To approach this task, consider the \emph{delay-coordinate map} $F_{h,T,M}: \mathbb{A} \rightarrow \mathbb{R}^M$, defined for an integer number of delays $M$ through the relation
\begin{equation}
F_{h,T,M}\left( x(i\cdot T)\right)  = \left[
\begin{array}{c}
s_i\\
s_{i-1}\\
\vdots\\
s_{i-M+1}
\end{array}
\right]  =
\left[
\begin{array}{c}
h(x(i\cdot T))\\
h\left(x((i-1)\cdot T)\right)\\
\vdots\\
h\left(x((i-M+1)\cdot T)\right)
\end{array}
\right]
=
\left[
\begin{array}{c}
h(x(i\cdot T))\\
h\left(\phi_{T}^{-1}(x(i\cdot T))\right)\\
\vdots\\
h\left(\phi_{T}^{-M+1}(x(i\cdot T))\right)
\end{array}
\right].
\label{eqn:dcm}
\end{equation}
Note that the delay-coordinate map is simply formed at a given time by stacking the last $M$ observed time-series values into a vector.  Commonly, $\mathbb{R}^M$ is referred to as the \emph{reconstruction space}.

\begin{figure}
	\centering

\subfigure[State space]
{\includegraphics[width=.3\textwidth]{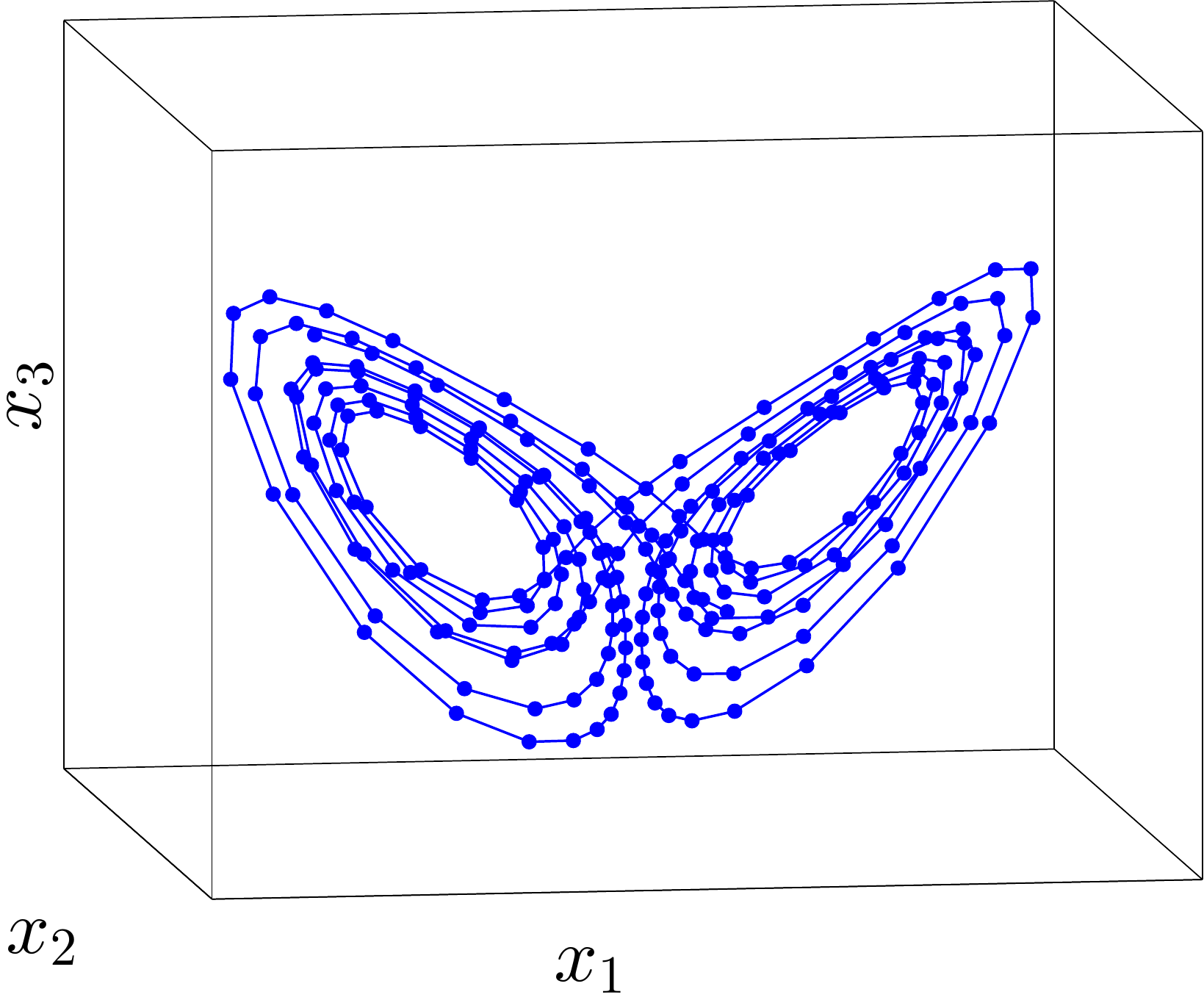}
}
\subfigure[Time series]
{\includegraphics[width=.3\textwidth]{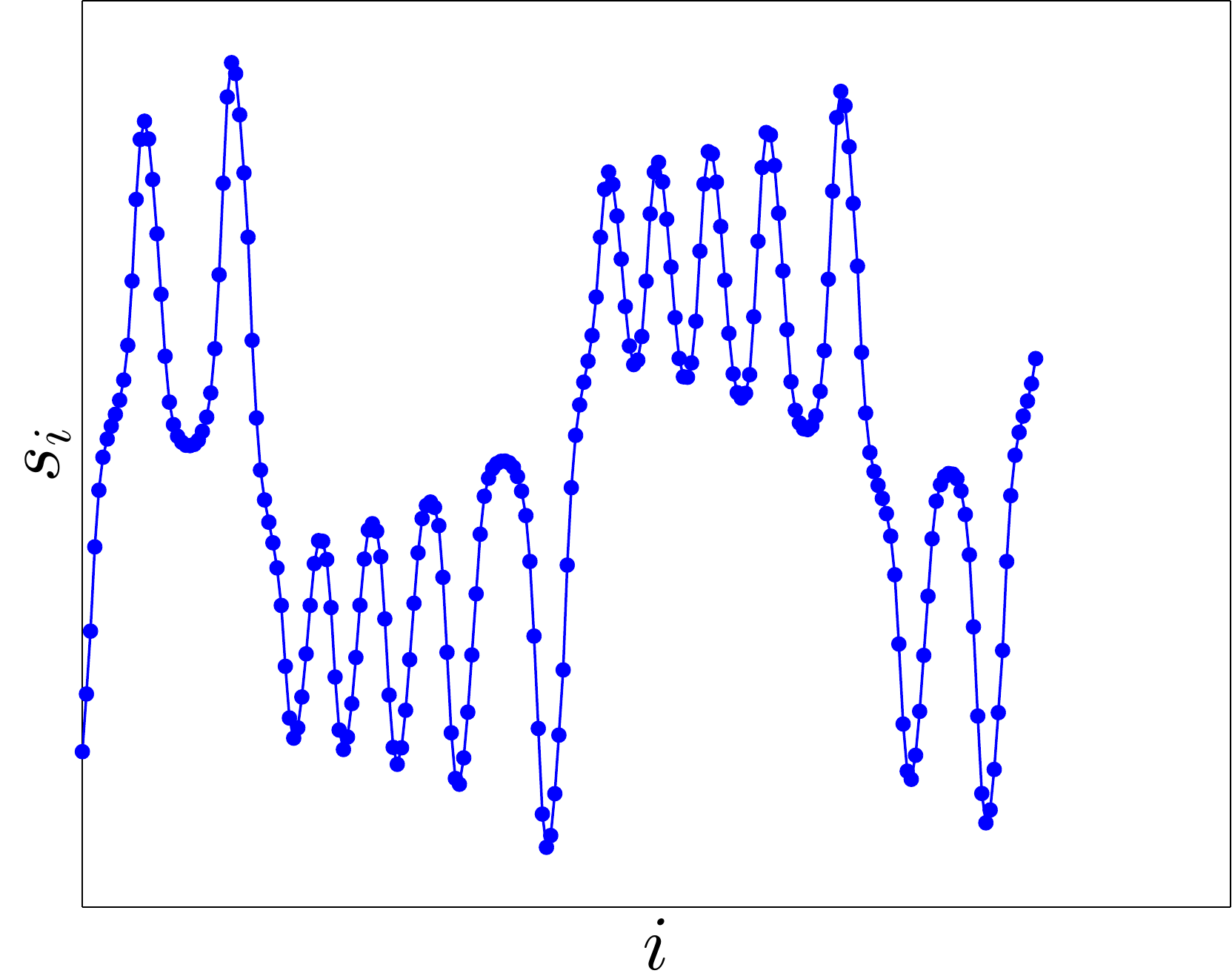}
}
\subfigure[Reconstruction space]
{\includegraphics[width=.3\textwidth]{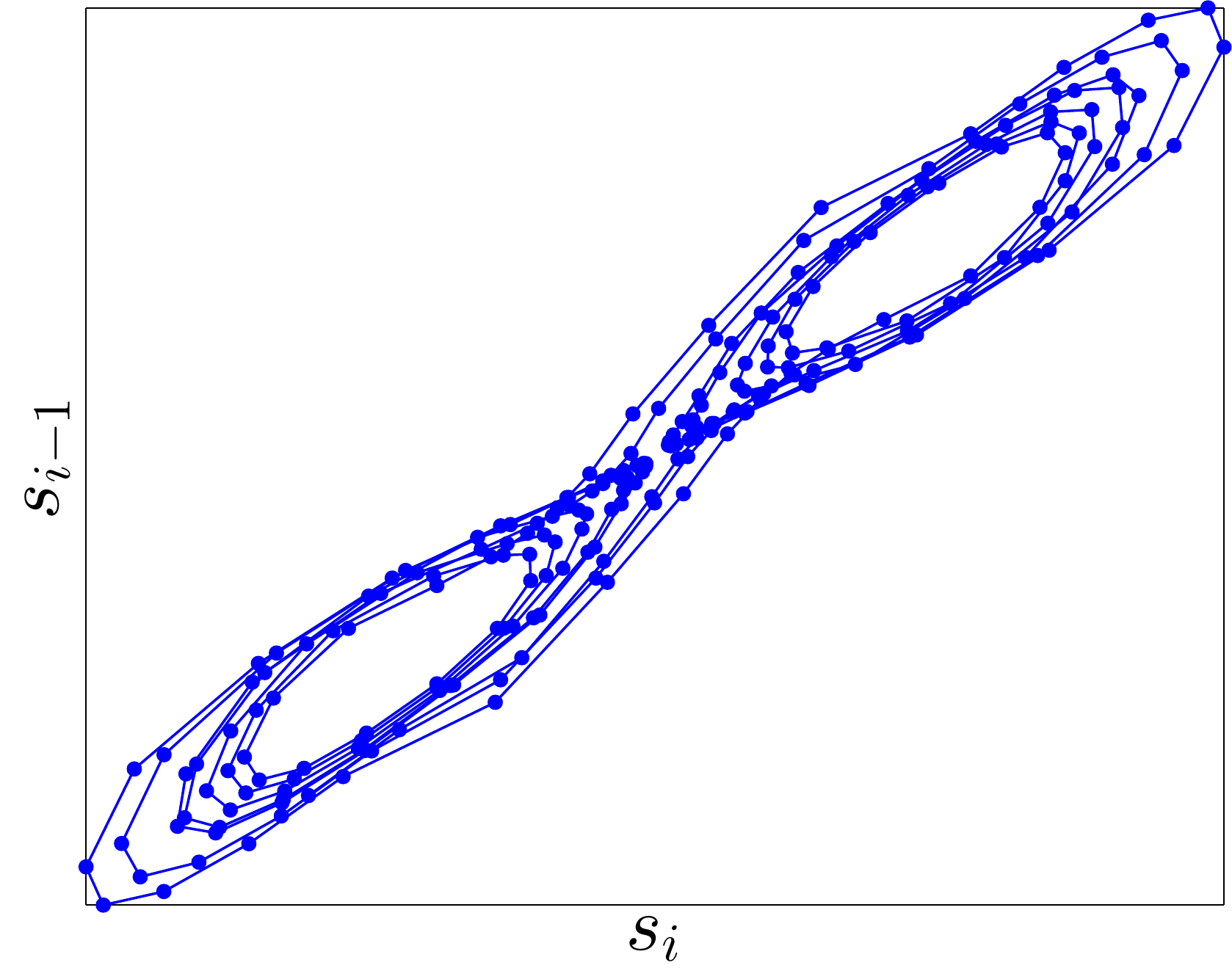}
}
	\caption{\small\sl (a) The state space trajectory of the Lorenz attractor in $\mathbb{R}^3$, demonstrating the characteristic butterfly pattern. (b) The time series obtained by a measurement function that only keeps the $x_1$-coordinate of the trajectory. (c) The delay-coordinate map points with $M=2$, recreating the butterfly pattern using only the time series.}
	\label{fig:Lorentz}

\end{figure}

Takens' embedding theorem \cite{Takens1981,Sauer1991} asserts that it is indeed possible to reconstruct the state space from the time-series data.  With this setup, Takens' result roughly states that if $M>2\cdot \operatorname{dim}(\mathbb{A)}$, then the delay-coordinate map $F_{h,T,M}(\cdot)$ resulting from almost every smooth measurement function $h(\cdot)$ embeds the attractor $\mathbb{A}$ into the reconstruction space $\mathbb{R}^M$ (i.e., the delay-coordinate map forms a diffeomorphism for $\mathbb{A}$). In other words, the topology of the attractor $\mathbb{A}$ is preserved in the reconstruction space $\mathbb{R}^M$ under the delay-coordinate map, and therefore the trajectory in the reconstruction space $F_{h,T,M}(x(\cdot))$ is (in principle) equivalent to the trajectory in the state space $x(\cdot)$. Figure \ref{fig:Lorentz} illustrates the concept of a delay-coordinate map in the case of the widely-known Lorenz attractor.  Despite this embedding guarantee ensuring that no two points from the attractor map onto each other in the reconstruction, the mapping could be unstable in the sense that close points may map to points that are far away (and vice versa).  While the simplified main result of this paper is presented for the case of attractors that are smooth submanifolds, the extensions presented in Section~\ref{sec:chaoticextension} include strange attractors such as the Lorenz attractor.

\subsection{Simplified Main Result}
\label{sec:simple}

To quantify the quality of the embedding in the reconstruction space, we seek the stronger guarantee that the delay-coordinate map $F_{h,T,M}(\cdot)$ is a \emph{stable embedding} of the attractor $\mathbb{A}$. By stable embedding, we mean that $F_{h,T,M}(\cdot)$ must act as a near-isometry on $\mathbb{A}$, in the sense that
\begin{equation}\label{eq:RIP ideal}
\epsilon_l  \le \frac{\left\|F_{h,T,M}(x)-F_{h,T,M}(y)\right\|_2^2}{M\cdot \|x-y\|^2_2} \le \epsilon_u,\qquad \forall x,y\in\mathbb{A}, ~ x \neq y
\end{equation}
for some \emph{isometry constants} $0<\epsilon_l\le \epsilon_u<\infty$. Said another way, if $\epsilon_l\approx \epsilon_u$, the stable embedding condition of \eqref{eq:RIP ideal} guarantees that the delay-coordinate map preserves the \emph{geometry} of the attractor (rather than merely its topology) by ensuring that pairwise distances between points on the attractor are approximately preserved in the reconstruction space. Since $x(\cdot)\subset\mathbb{A}$, the same would hold for the trajectory and its image, thereby guaranteeing the quality of the trajectory embedding in the reconstruction space $\mathbb{R}^M$.

The main result of this paper is to determine the conditions on the attractor $\mathbb{A}$, measurement function $h(\cdot)$, number of delays $M$, and sampling interval $T$ such that $F_{h,T,M}(\cdot)$ is a stable embedding of $\mathbb{A}$. This is a more ambitious objective than Takens' embedding theorem (leading naturally to more restrictive conditions), but with the benefit of quantifying the quality of the embedding and relating that quality to the problem-specific parameters. Roughly speaking, our main result shows that $F_{h,T,M}(\cdot)$ stably embeds $\mathbb{A}$ (in the sense of \eqref{eq:RIP ideal}) for most measurement functions $h$, provided that the following condition is satisfied:
\begin{equation}\label{eq:simple cond}
\operatorname{R}_{H,T,M}(\mathbb{A}) \gtrsim  \operatorname{dim}(\mathbb{A})\cdot  \log \left(
\frac{\operatorname{vol}(\mathbb{A})^{\frac{1}{\operatorname{dim}(\mathbb{A})}}}{\operatorname{rch}(\mathbb{A})} \right).
\end{equation}
Here, $\operatorname{dim}(\mathbb{A})$ and $\operatorname{vol}(\mathbb{A})$ are the dimension and volume of the attractor $\mathbb{A}\subset\mathbb{R}^N$, and $\operatorname{rch}(\mathbb{A})$ is an attribute of $\mathbb{A}$ that captures its geometric regularity.  To quantify the notion of ``most'' measurement functions, our result is probabilistic and holds with high probability over measurement functions drawn from a rich probability model $H$.  The \emph{stable rank} $\operatorname{R}_{H,T,M}(\mathbb{A})$ of $\mathbb{A}$ quantifies the ability of the random measurement functions to observe the system attractor.  Both reach and stable rank are well-studied concepts, and will be discussed (along with the detailed probability model $H$) in full technical detail later.

Typically, if a dynamical system is fairly ``predictable'', then $\operatorname{R}_{H,T,M}(\mathbb{A})$ grows proportionally with $M$ as the number of delays grows.  In this case, the delay-coordinate map stably embeds $\mathbb{A}$ when the number of delays scales linearly with the dimension of the attractor as in Takens' original theorem.  On the other hand, if the dynamical system is highly unpredictable, then it is likely that $\operatorname{R}_{H,T,M}(\mathbb{A})$ plateaus rapidly with increasing $M$ and it will be more difficult to stably embed this system through delay-coordinate mapping even with very long delay vectors.   In Section~\ref{sec:mainresultsection}, the main contribution of this paper precisely quantifies these conditions governing the quality of the embedding from the delay-coordinate map. As we also discuss, these conditions have a natural interpretation in the context of classical empirical methods for choosing $T$ and $M$.

\section{Background and Related Work}
\label{sec:Background}

\subsection{Takens' Embedding Theorem}

To expound on the overview in Section~\ref{sec:intro_takens}, we turn our attention to a detailed technical statement of Takens' theorem~\cite{Takens1981} showing that the delay-coordinate map $F_{h,T,M}(\cdot)$ embeds the attractor $\mathbb{A} $ (and, of course, the trajectory $x(\cdot)\subset\mathbb{A}$).
\begin{thm}
	\label{thm:Takens}
	\emph{\textbf{(Takens' embedding theorem~\cite{Takens1981})}}
    Let $\mathbb{A}\subset\mathbb{R}^N$ be a smooth, bounded, and boundary-less submanifold of $\mathbb{R}^N$, and let $M>2\cdot\operatorname{dim}(\mathbb{A})$ be an integer. For pairs $(\phi_T,h)$ where the flow $\phi_T:\mathbb{A}\rightarrow\mathbb{A}$ is a diffeomorphism on $\mathbb{A}$ and where $h:\mathbb{R}^N\rightarrow\mathbb{R}$ is a smooth measurement function, it is a generic property that the delay-coordinate map $F_{h,T,M}(\cdot)$ is an embedding (i.e., diffeomorphism) of $\mathbb{A}\subset\mathbb{R}^N$ into the reconstruction space $\mathbb{R}^M$.
\end{thm}

In this theorem, ``generic'' means that the set of pairs $(\phi_T,h)$ for which $F_{h,T,M}(\cdot)$ yields an embedding is open and dense in the set of all mappings. This topological notion of genericity was later extended to an ``almost every'' probabilistic argument by Sauer et al.~\cite{Sauer1991}. In fact, the probe space framework developed in~\cite{Sauer1991} was the inspiration for our analysis which involves drawing $h$ randomly from a subspace of measurement functions (see Section~\ref{sec:measurement}). We also note that by relaxing the manifold assumption, Takens' theorem has also been generalized to cover embedding of \emph{fractal sets} such as \emph{strange attractors}~\cite{Sauer1991} (formed by chaotic dynamical systems~\cite{Kantz2004}) and embeddings of forced systems~\cite{Stark1999}.

Note that, under Takens' theorem, $F_{h,T,M}(\mathbb{A})\subset\mathbb{R}^M$ is diffeomorphic to $\mathbb{A}\subset\mathbb{R}^N$, so that the topology of  the attractor $\mathbb{A}$ and the flow on this attractor are preserved under delay-coordinate mapping. In particular, we may trace $F_{h,T,M}(x(\cdot))$ with its samples and ``reconstruct'' the trajectory in the (often inaccessible) state space using only the time-series data.   In fact, several important characterizations of dynamical systems are preserved  under delay-coordinate mapping and can be computed directly in the reconstruction space, including the number/types of fixed points/orbits, the dimension of attractor (i.e., $\operatorname{dim}(F_{h,T,M}(\mathbb{A}))=\operatorname{dim}(\mathbb{A})$), and the Lyapunov exponents~\cite{Dechert1996,Kantz2004}.  Justified by these properties, the reconstruction space representation formed by the delay-coordinate map has been used for many practical time-series processing algorithms~\cite{abarbanel1993analysis}, including tasks such as prediction~\cite{Kantz2004,asefa2005support}, noise reduction~\cite{Schreiber1996}, chaos synchronization and control~\cite{Ott1990,Pecora1991,silva2000introduction}, system identification~\cite{Dedieu1997}, and detection of causality in complex networks~\cite{Sugihara496}.

While Takens' original theorem proves that the delay-coordinate map is theoretically equivalent to the attractor in the hidden state space, it may map close points far apart and far points close together.  This warping, though topologically equivalent, means that even small changes in the reconstruction space representation (e.g., due to noise, etc.) can amount to arbitrarily large changes in the corresponding points in the state-space. One might ask: Are there any conditions where the delay-coordinate map is guaranteed to bound the errors due to noise?  Will changes to the delay-coordinate map parameters (e.g., increasing the number of delays, decreasing the sampling interval), the system, or the measurement function affect the quality of the reconstruction or its tolerance to noise?   Takens' original theorem does not address these issues, motivating our study of geometrically stable embeddings.

\subsection{Related Work}\label{sec:prior art STE}

Prior work by Casdagli et al.~\cite{Casdagli1991} begins to methodically address the issue of noise in delay embeddings by studying the effects of the sampling interval $T$ and number of delays $M$ on asymptotic quantities defined to capture the effects of noise on the delay-coordinate map. We note that, when the image of an attractor is warped or folded (and thus, not a stable embedding), noise sensitivity can be a problem as the conditional probability  of the state given a noisy observation of the delay vector may be poorly localized.  In addition, the conditional variance of $h(x((i+1)T))$ (the next value in the time series) may increase, which reduces the ability to predict the time series.

While some approaches have been developed to perform noise reduction in the reconstruction space~\cite{schreiber1991simple}, more generally, one finds a rich literature on methods of choosing the optimal $T$ and/or $M$ that account for noise by examining quantities that can typically be interpreted as having some dependence on the distortion of the attractor. To illustrate the concept (without claiming to be an exhaustive review), methods have been proposed to choose parameters by examining how they change the neighborhood relationships between points (e.g., the introduction of ``false nearest neighbors'')~\cite{liebert1991optimal,kennel1992determining,cao1997practical}, geometric quantities (e.g., space filling) intended to separate trajectories~\cite{buzug1992optimal,rosenstein1994reconstruction}, test statistics proposed for determining whether the result is a valid embedding of the input~\cite{pecora1995statistics}, and statistics related to the predictive power of the time series (e.g., mutual information)~\cite{Fraser1986,Uzal2011}.  The work in~\cite{Kugiumtzis1996} recommends the \emph{mean orbital period} (approximated from the oscillations of the time series) as a reliable choice for the \emph{window length} (i.e., $TM$), noting that most methods for choosing embedding parameters are based on empirical arguments, use arbitrary criteria, and ultimately do not guarantee good reconstructions.  While not primarily introducing a method to choose parameters per se, our results are some of the first to provide a theoretical basis for proposed methods by explicitly illustrating the impact of parameter choices (and other problem-specific details) on a natural measure for assessing reconstruction quality:  geometric stability.

Our approach to guaranteeing the stability of delay-coordinate mapping relies heavily on recent advances in the fields of compressive sensing and geometric functional analysis. As mentioned previously, a central condition in compressive sensing is the RIP, which requires a linear measurement operator to provide a stable embedding of the sparse signal family. Of particular interest in compressive sensing are randomized linear measurement operators. In particular, when the measurement operators are constructed randomly (e.g., as a random matrix populated with independent and identically distributed Gaussian entries), the RIP can be satisfied with high probability~\cite{Baraniuk2008}. The basic compressed sensing results have been extended to various classes of structured randomized measurement functions~\cite{Rauhut2012,Krahmer2014,eftekhari2015restricted,puy2017recipes} as well as other low-dimensional models~\cite{blumensath2009sampling} such as smooth manifolds~\cite{Eftekhari2015,Clarkson2008,Yap2013}.  The present work is especially indebted to recent developments in geometric functional analysis which appeared first in~\cite{Krahmer2014} to establish the RIP for a class of structured random matrices. It is also worth noting the recent work \cite{tran2016exact}, in which sparse recovery tools are used to help identify a dynamical system in spite of large erasures in the available data.

A study of stable delay-coordinate mapping for \emph{linear} dynamical systems and with measurement functions that are \emph{deterministic} and \emph{linear}  has previously appeared in~\cite{Yap2011a}. The current result is a significant extension of this previous work by allowing general nonlinear systems and measurement functions that are both randomized and nonlinear.  However, the main result in the present work has a similar flavor to~\cite{Yap2011a}, as both papers highlight cases where the embedding quality plateaus and cannot be improved by increasing the number of delays $M$.

\subsection{Differential Geometry}
\label{sec:Differential Geometry}

Because we will consider attractors $\mathbb{A}$ that are submanifolds of $\mathbb{R}^N$, it is helpful to review the differential geometry concepts that characterize $\mathbb{A}$ and play a major role in the present results. The reader may also refer to~\cite{Spivak1975} for a more comprehensive introduction.

To any point $x\in\mathbb{A}$ we can assign a \emph{tangent subspace }$\mathbb{T}_{x}\mathbb{A}\subset\mathbb{R}^{N}$ comprised
of the directions of all curves on $\mathbb{A}$ that are tangent to $x$. The linear subspace $\mathbb{T}_{x}\mathbb{A}$ has dimension $\operatorname{dim}(\mathbb{A})$ in $\mathbb{R}^{N}$, and the union of all tangent subspaces is called the \emph{tangent bundle} of $\mathbb{A}$:
\[
\mathbb{T}\mathbb{A}:=\bigcup_{x\in\mathbb{A}}\{x\}\times\mathbb{T}_{x}\mathbb{A}.
\]
Consider a smooth map $\psi:\mathbb{A}\rightarrow\mathbb{A}$. The derivative of this map at $x\in\mathbb{A}$ is the linear operator $D\psi(x):\mathbb{T}_x\rightarrow\mathbb{T}_{\psi(x)}$ that satisfies
\begin{equation}\label{eq:def of der}
\lim_{\tau\rightarrow 0}\left\| \psi\left(x+\gamma(\tau) \right)-\psi\left(x \right) - \left[D\psi(x)\right](\gamma(\tau)) \right\|_2  =0,
\end{equation}
for every smooth curve $\gamma:[-1,1]\rightarrow\mathbb{A}$ that passes through $x$ with $\gamma(0)=x$. The \emph{normal subspace }$\mathbb{N}_{x}\mathbb{A}$ is the $(N-\operatorname{dim}(\mathbb{A}))$-dimensional orthogonal complement of $\mathbb{T}_{x}\mathbb{A}$ with respect to $\mathbb{R}^{N}$. The \emph{normal bundle} of $\mathbb{A}$ is the union of all normal subspaces:
\[
\mathbb{N}\mathbb{A}:=\bigcup_{x\in\mathbb{A}}\{x\}\times\mathbb{N}_{x}\mathbb{A}.
\]
For $r>0$, we also let $\mathbb{N}^r \mathbb{A}$ denote the \emph{open normal bundle} of $\mathbb{A}$ of radius $r$ comprised of all normal vectors of length less than $r$. For example, when $\mathbb{A}$ is the unit circle in $\mathbb{R}^{2}$ and $r \in (0, 1)$, $\mathbb{N}^r \mathbb{A}$ may be identified with an annulus of width $2r$ (around the circle).

A {\em geodesic curve} on $\mathbb{A}$ is a smooth curve that minimizes the $\ell_2$ distance between every pair of nearby points that it connects~\cite{Spivak1975}. The {\em geodesic distance} between a pair of points on $\mathbb{A}$ is the length of the shortest geodesic curve that connects them. The $\ell_2$ distance between points never exceeds their geodesic distance. Throughout, we assume that $\mathbb{A}$ is regular in the sense that, for some {\em geodesic regularity} $\operatorname{geo}(\mathbb{A})\in [1,\infty) $, we have
\begin{equation}\label{eq:assump on geod diff thm}
\|x-y\|_2\le d_{\mathbb{A}}(x,y) \le \operatorname{geo}(\mathbb{A})\cdot   \|x-y\|_2,\qquad \forall x,y\in\mathbb{A},
\end{equation}
where $d_{\mathbb{A}}(x,y)$ stands for the geodesic distance between $x,y\in\mathbb{A}$. For a circle, $\operatorname{geo}(\mathbb{A})=\frac{\pi}{2}$.

The \emph{reach} measures the regularity of a manifold and is closely related to its \emph{condition number}~\cite{Niyogi2008,Baraniuk2008,Eftekhari2015}.
\begin{definition}\emph{\textbf{(Reach of a manifold~\cite{Federer1959})}} Let $\mathbb{A}$ be a bounded and smooth submanifold of $\mathbb{R}^N$.
 The reach of $\mathbb{A}$ (denoted with $\mbox{rch}(\mathbb{A})$) is the largest number $r\ge 0$ having the following property:
The open normal bundle about $\mathbb{A}$ of radius $r$ is embedded
in $\mathbb{R}^N$ for all $r < \mbox{rch}(\mathbb{A})$. \label{def:cn SUPER}
\end{definition}
In perhaps the simplest example, the  reach of a circle with radius $\rho$ is simply $\rho$. Reach controls both local  and global properties of a manifold. Its role is summarized in two key relationships. First, the  curvature of any
unit-speed geodesic curve on $\mathbb{A}$ is bounded by
$1/\mbox{rch}(\mathbb{A})$. Second, at long {geodesic distances}, reach controls how close the manifold may curve back upon itself.
For example, supposing $x,y \in \mathbb{A}$ with $d_{\mathbb{A}}(x,y)
> \mbox{rch}(\mathbb{A})$, it must hold that $\|x-y\|_2 > \mbox{rch}(\mathbb{A})/2$.  See \cite{Eftekhari2015} for more details.

\section{Main result}
\label{sec:mainresultsection}

We are now prepared to give a precise setup and statement of the result that was summarized in Section~\ref{sec:simple}, along with additional interpretation and discussion.

\subsection{Measurement Apparatus}
\label{sec:measurement}

We first set up our framework for choosing a measurement function $h(\cdot)$ that is used to observe the trajectory $x(\cdot)$. In general we seek a result in which the choice of measurement function is not specific and arbitrary measurement functions chosen according to some probability measure will work with overwhelming probability. To do this, inspired by an approach developed in~\cite{Sauer1991}, we limit the measurement function to some subset $\mathcal{H}$ of the space of all smooth functions. To establish this subset, we fix a finite collection of $P$ \emph{measurement basis functions} $h_p:\mathbb{A}\rightarrow\mathbb{R}$, $p\in \{1,2,\dots,P\}$. For any coefficient vector $\alpha\in\mathbb{R}^P$, we define a measurement function $h_\alpha:\mathbb{A}\rightarrow \mathbb{R}$ that is a corresponding linear combination of the measurement basis functions:
\begin{equation}
h_{\alpha}(\cdot) = \sum_{p=1}^P \alpha[p]\cdot h_p(\cdot).\label{eq:rnd lin comb}
\end{equation}
We limit our attention to the class $\mathcal{H}$ of measurement functions formed by arbitrary linear combinations of this set of basis functions:
\begin{equation}\label{eq:def of H}
\mathcal{H}
:=\left\{ h_{\alpha}(\cdot)\,:\,\alpha\in\mathbb{R}^P \right\}=\mbox{span}\left[\left\{ h_p(\cdot)\right\}_{p=1}^P\right].
\end{equation}
Note that while the sum in~\eqref{eq:rnd lin comb} is linear, each basis function can be nonlinear, resulting in a rich and flexible set of potential measurement functions. For two concrete examples,
\begin{itemize}
\item $\mathcal{H}$ is the class of all linear functions on $\mathbb{R}^N$ when $h_p(\cdot)=\langle \cdot,e_p\rangle$ for $ p\in \{1,2,\dots,P\}$ with $P=N$. Here, $e_p\in\mathbb{R}^N$ is the $p$-th canonical vector in $\mathbb{R}^N$ (i.e., $e_p[p]=1$ and $e_p[n]=0$ when $n\ne p$).
\item $\mathcal{H}$ is the set of all $N$-variate polynomials of degree $K$ if $\{h_p(\cdot)\}_{p=1}^P$ is the set of all monomials of degree $K$ with $P= {K+N\choose N}$.
\end{itemize}

Our main theorem will depend on certain properties of the measurement basis functions that are revealed by defining the map $H: \mathbb{A} \rightarrow \mathbb{R}^P$, where
\begin{equation}\label{eq:def of h}
H(x) :=
\left[
\begin{array}{cccc}
h_1(x) & h_2(x) & \cdots & h_P(x)
\end{array}
\right]^*\in\mathbb{R}^P,\qquad \forall x\in\mathbb{A}.
\end{equation}
The superscript $^*$ indicates the transpose of a matrix or vector. We will require that the measurement basis functions are sufficiently well behaved in that the following three assumptions on $H$ are met:

\begin{enumerate}[label=A\arabic*.]
\item $H(\cdot)$ is a bi-Lipschitz map on $\mathbb{A}$, in the sense that $ l_H\cdot \|x-y\|_2 \le \|H(x)-H(y)\|_2 \le u_H \cdot \|x-y\|_2$ for every pair $x,y\in\mathbb{A}$, and for some $l_H,u_H\in (0,\infty)$;
\item $H(\cdot)$ is a diffeomorphism between $\mathbb{A}$ and $H(\mathbb{A})$, resulting in $H(\mathbb{A})\subset\mathbb{R}^P$ being a bounded, boundary-less, and smooth submanifold of $\mathbb{R}^P$ with $\operatorname{dim}(H(\mathbb{A}))=\operatorname{dim}(\mathbb{A})$; and
\item the nonzero singular values of $DH(x)$  belong to some interval $[\eta_{\min},\eta_{\max}]\subset(0,\infty)$, where $DH(\cdot)$ is the derivative of $H$ (see Section \ref{sec:Differential Geometry}).
\end{enumerate}

Under the above assumptions on the basis functions, the flow $\phi_T:\mathbb{A}\rightarrow\mathbb{A}$ in the state space naturally induces a flow $\phi_{H,T}:H(\mathbb{A})\rightarrow H(\mathbb{A})$ (in $\mathbb{R}^P$) specified as
\begin{equation}\label{eq:flow in H}
H(x)\overset{\phi_{H,T}(\cdot)}{\rightarrow} \phi_{H,T} ( H(x) ) := H ( \phi_T(x)),\qquad \forall x\in\mathbb{A}.
\end{equation}
As with the flow $\phi_T(\cdot)$, the induced flow $\phi_{H,T}(\cdot)$ is a diffeomorphism (but on $H(\mathbb{A})$ rather than $\mathbb{A}$).

Let $F_{h_\alpha,T,M}(\cdot)$ denote the delay-coordinate map formed with a measurement function $h_{\alpha}(\cdot)\in \mathcal{H}$, and let $F_{h_p,T,M}(\cdot)$ denote the delay-coordinate map associated with the $p$-th basis function $h_p(\cdot)$. For $x\in\mathbb{A}$, we will find it useful to collect the components of the delay-coordinate map due to the different measurement basis functions and write them as columns of a matrix such that:
\begin{align}\label{eq:def of XM}
X_{H,T,M} & :=
\left[
\begin{array}{cccc}
F_{h_1,T,M}(x) &
F_{h_2,T,M}(x) &
\cdots &
F_{h_P,T,M}(x)
\end{array}
\right]\in \mathbb{R}^{M\times P}.
\end{align}
Using \eqref{eq:rnd lin comb}, we can confirm the following useful identity:
\begin{equation}\label{eq:conn btw G and F}
F_{h_\alpha,T,M}(x)= X_{H,T,M}\cdot\alpha,\qquad \forall x\in \mathbb{A},\,\alpha\in\mathbb{R}^P.
\end{equation}
The introduction of $\phi_{H,T}(\cdot)$ above allows us to rewrite $X_{H,T,M}\in\mathbb{R}^{M\times P}$ (see \eqref{eq:def of XM}) as
\begin{align}\label{eq:other def of XM}
X_{H,T,M}
& =
\left[
\begin{array}{cccc}
H(x) &
H\left(\phi_T^{-1}(x)\right) &
\cdots &
H\left(\phi_T^{-M+1}(x)\right)
\end{array}
\right]^*\qquad \mbox{(see \eqref{eq:def of h})}\nonumber\\
& =
\left[
\begin{array}{cccc}
H(x) &
\phi_{H,T}^{-1}(H(x)) &
\cdots &
\phi_{H,T}^{-M+1}(H(x))
\end{array}
\right]^*.\qquad \mbox{(see \eqref{eq:flow in H})}
\end{align}
We also define the ``trajectory attractor'' as
\begin{equation}
\mathbb{A}_{H,T,M} :=
\left\{
\left[
\begin{array}{c}
H(x)\\
\phi_{H,T}^{-1}\left(H(x)\right)\\
\vdots\\
\phi_{H,T}^{-M+1}\left( H(x) \right)
\end{array}
\right]
\,: \,
x\in \mathbb{A}
\right\}
\subset
\mathbb{R}^{MP}.
\label{eq:traj attr}
\end{equation}
Taken together,~\eqref{eq:conn btw G and F} and~\eqref{eq:other def of XM} show that the reconstruction vectors produced by the delay-coordinate map $F_{h_\alpha,T,M}(\cdot)$ can be viewed as linear operator (which depends on $\alpha$) acting on points in the trajectory attractor. The reach of the trajectory attractor will play a role in our main result.

\subsection{Stable Rank}\label{sec:stable rank sec}

Lastly, our main result depends on a certain quantity that summarizes the quality of the measurement apparatus for a given dynamical system.  To that end, drawing from the scientific computing literature~\cite{Tropp2009a}, we first define the \emph{stable rank} of a matrix $A\in\mathbb{R}^{M\times P}$ as
\begin{equation}\label{eq:def of stable rank}
\operatorname{R}(A) := \frac{\|A\|_F^2}{\|A\|^2},
\end{equation}
where $\|A\|_F$ and $\|A\|$ are the Frobenius and spectral norms of $A$, respectively. It is straightforward to confirm that
\begin{equation}\label{eq:conn btw stbl rnk and rnk}
1\le \operatorname{R}(A) = \frac{\sum_i \sigma_i(A)^2}{\sigma_1(A)^2} \le \mbox{rank}(A),
\end{equation}
where $\sigma_1(A)\ge\sigma_2(A)\ge \cdots \ge 0$ are the singular values of $A$.
In a sense,  stable rank $\operatorname{R}(\cdot)$ is a more robust alternative to the standard rank in that it is less sensitive to small changes in the spectrum.  Supposing that $P\ge M$, two extreme cases are worth noting here.  First, if the rows of $A$ are orthonormal, then $\operatorname{R}(A)=M$.  In this case, the rows of $A$ are equal in length and have ``diverse'' directions.  Second, if the rows of $A$ are identical, then $\operatorname{R}(A)=1$.

The star of this show will in fact be the stable rank of the attractor $\mathbb{A}$, which we define to be
\begin{equation}\label{eq:s rank of h(A)}
\operatorname{R}_{H,T,M}(\mathbb{A})  := \inf_{x,y\in\mathbb{A},\; x \neq y} \operatorname{R}\left( X_{H,T,M}-Y_{H,T,M}\right),
\end{equation}
where $X_{H,T,M} \in \mathbb{R}^{M\times P}$ is defined in~\eqref{eq:conn btw G and F} (see also~\eqref{eq:other def of XM}), and $Y_{H,T,M}$ is defined analogously with $y$ in place of $x$. If $P\ge M$, then \eqref{eq:conn btw stbl rnk and rnk} dictates that
\begin{equation}\label{eq:range of sr}
\operatorname{R}_{H,T,M}(\mathbb{A}) \in[1,M].
\end{equation}
For example, when $H(\mathbb{A})$ is a subset of an $r$-dimensional subspace (say with $r\ll M$), then \eqref{eq:s rank of h(A)} dictates that $\operatorname{R}_{H,T,M}(\mathbb{A})\le r\ll M$. On the other hand, if the rows of $X_{H,T,M}-Y_{H,T,M}$  have similar lengths and diverse directions (for every $x,y\in\mathbb{A}$), then $\operatorname{R}_{H,T,M}(\mathbb{A})$ might be close to $M$. As we see next, the larger $\operatorname{R}_{H,T,M}(\mathbb{A})$, the better.

\subsection{Main Result}
\label{sec:main results}
We are now in position to state the main result of this work.

\begin{thm}
\label{thm:manifold2}\emph{\textbf{(Stable Takens' embedding theorem)}}
Let $\mathbb{A}\subset\mathbb{R}^N$ be a  smooth, bounded, and boundary-less submanifold of $\mathbb{R}^N$.  For a fixed sampling interval $T>0$, assume that $\phi_{T}(\cdot)$ is a diffeomorphism on $\mathbb{A}$ and that the singular values of the derivative of $\phi_{T}(\cdot)$ belong to the interval $[\sigma_{\min},\sigma_{\max}]\subset(0,\infty)$. For an integer $P$, fix the measurement basis functions $h_p:\mathbb{A}\rightarrow\mathbb{R}$ for $p\in\{1,2,\dots,P\}$ and let $\mathcal{H}$ be the linear span of $\{h_p(\cdot)\}_p$. The random coefficient vector $\alpha\in\mathbb{R}^P$ is assumed to have entries that are i.i.d.\ zero-mean and unit-variance subgaussian random variables with subgaussian norm $\theta$.\footnote{A random variable $Z$ is subgaussian if its subgaussian norm $\|Z\|_{\psi_2}$ is finite, where $\|Z\|_{\psi_2}:=\sup_{p\ge 1} \left( \mathbb{E}|Z|^p \right)^{\frac{1}{p}}$. Qualitatively speaking, the tail of (the distribution of) a subgaussian random variable is similar to that of a Gaussian random variable, hence the name~\cite{Vershynin2012}. In particular, Gaussian random variables are subgaussian.}

Consider the map $H:\mathbb{A} \rightarrow\mathbb{R}^P$ constructed in (\ref{eq:def of h}), and suppose that $H(\cdot)$ satisfies the assumptions A1--A3 listed in Section~\ref{sec:measurement}.  Let $\operatorname{R}_{H,T,M}(\mathbb{A})$ denote the stable-rank of $\mathbb{A}$ as defined in
(\ref{eq:s rank of h(A)}). For arbitrary isometry constant $\delta\in(0,1)$ and failure probability $\rho\in(0,1)$, suppose that
\begin{align}
\label{eq:stable rank thm}
& \operatorname{R}_{H,T,M}\left(\mathbb{A}\right)
  \ge C_\theta \cdot
   \max
   \Bigg[
    \delta^{-2} \cdot  \operatorname{dim}(\mathbb{A})
	\cdot \log
 \left(
 \eta_{\max}\sqrt{ \operatorname{dim}(\mathbb{A})}
   \left( \frac{\sigma_{\min}^{-2M\cdot \operatorname{dim}\left(\mathbb{A}\right)}-1}
   {\sigma_{\min}^{-2\cdot \operatorname{dim}(\mathbb{A})}-1}
   \right)^{\frac{1}{2\cdot \operatorname{dim}(\mathbb{A})}}
    \frac{\operatorname{vol}\left(\mathbb{A}\right) ^{\frac{1}{\operatorname{dim}(\mathbb{A})}}}
    {\operatorname{rch}(\mathbb{A}_{H,T,M})}
  \right)
  \nonumber\\
  &
  \qquad \qquad \qquad \qquad \qquad \qquad ,
  e^{-\min W\left(\frac{-\delta^2}{\operatorname{dim}(\mathbb{A})} \right)}
   ,\delta^{-2} \log\left(\frac{1}{\rho} \right)
	\Bigg]
\end{align}
where $C_\theta$ is a constant that depends only on $\theta$, and make the mild assumption\footnote{This assumption requires the volume of $\mathbb{A}$ not to be too small. Similar assumptions have appeared in earlier works, e.g., \cite{Eftekhari2015}.} that
\begin{equation}
\frac{\operatorname{vol}\left(\mathbb{A}\right)^{\frac{1}{\operatorname{dim}(\mathbb{A})}}}{\operatorname{rch}\left(\mathbb{A}_{H,T,M}\right)}
 \gtrsim
\frac{1}{\eta_{\min}\sqrt{\operatorname{dim}(\mathbb{A})}}
\cdot
 \left(
  \frac{\sigma_{\max}^{-2M\cdot\operatorname{dim}(\mathbb{A})}-1}{\sigma_{\max}^{-2\cdot\operatorname{dim}(\mathbb{A})}-1}
\right)^{-\frac{1}{2\cdot \operatorname{dim}(\mathbb{A})}},
\label{eq:assumptionmain}
\end{equation}
with $\mathbb{A}_{H,T,M}\subset\mathbb{R}^{MP}$ defined in \eqref{eq:traj attr} and $W(\cdot)$ denoting the Lambert W-function.\footnote{ See Appendix \ref{sec:proof_manifold} or \cite[\textsection 4.13]{Olver2010} for the exact definition. Very roughly, the exponential term involving the Lambert W-function in \eqref{eq:stable rank thm} scales like $\operatorname{dim}(\mathbb{A}) \cdot \log(\operatorname{dim}(\mathbb{A}))$.
}

Then, except with a probability of at most $\rho$ (over  the choice of $\alpha$), the delay-coordinate map $F_{h_\alpha,T,M}(\cdot)$ stably embeds $\mathbb{A}$ in the sense that
\begin{equation}\label{eq:final pre thm}
(1-\delta)\cdot l_H^2 \cdot \operatorname{geo}(\mathbb{A})^{-2} \cdot \frac{\sigma_{\max}^{-2M}-1}{M(\sigma_{\max}^{-2}-1)}  \le \frac{\left\| F_{h_{\alpha},T,M}(x)-F_{h_{\alpha},T,M}(y) \right\|_2^2}{M \cdot \|x-y\|_2^2} \le (1+\delta)\cdot u_H^2\cdot  \operatorname{geo}(\mathbb{A})^2\cdot \frac{\sigma_{\min}^{-2M}-1}{M(\sigma_{\min}^{-2}-1)} ,
\end{equation}
for every pair $x,y\in\mathbb{A}$ with $x \neq y$.
\end{thm}

The proof of this result is found in Appendix~\ref{sec:proof_manifold}. In~\eqref{eq:final pre thm}, regarding the behavior of the terms involving $\sigma_{\max}$ and $\sigma_{\min}$ (the largest and smallest singular values of the derivative of $\phi_{T}(\cdot)$), we note that these terms are close to $1$ if the singular values cluster near $1$. In particular,
\[
\lim_{\sigma \rightarrow 1} \frac{\sigma^{-2M}-1}{M(\sigma^{-2}-1)} = 1.
\]

\subsection{Observations and Interpretation}

Several remarks are in order to help shape our understanding of Theorem \ref{thm:manifold2}.

\begin{rem}
\textbf{(Comparison with Takens' theorem)} \label{rem:compare}  Let us fix the measurement basis functions $\{h_p(\cdot)\}$. Note that the distribution of the random coefficient vector $\alpha\in\mathbb{R}^P$ in Theorem \ref{thm:manifold2} induces a distribution on the space of measurement functions, $\mathcal{H}=\mbox{span}[\{h_p(\cdot)\}]$. Qualitatively speaking, Theorem~\ref{thm:manifold2} establishes that, except on a subset with an exponentially small measure, every function in $\mathcal{H}$ forms a delay-coordinate map that stably embeds $\mathbb{A}$, if $\operatorname{R}_{H,T,M}(\mathbb{A})$ is proportional to $\operatorname{dim}(\mathbb{A})$ with a proportionality constant that depends chiefly on the geometry of $\mathbb{A}$ and the flow $\phi_{T}(\cdot)$.

In contrast, Takens' original theorem (Theorem~\ref{thm:Takens}) established that generic choices of the flow $\phi_T$ and measurement function $h$ will yield an embedding so long as that $M>2\cdot \operatorname{dim}(\mathbb{A})$. The refinement of Takens' theorem by Sauer et al.~\cite{Sauer1991} established that, for a fixed flow $\phi_T$ and a random choice of $h$ from a certain probe space, one will obtain an embedding with probability one. This result also required that $M>2\cdot \operatorname{dim}(\mathbb{A})$ but placed certain restrictions on the periodicities of the orbits of $\phi_T$ on $\mathbb{A}$.

Thus, Theorem~\ref{thm:manifold2} provides a stronger embedding guarantee than the topological and probabilistic Takens' theorems, but it does so with a nonzero failure probability and it is contingent on a condition involving the stable rank $\operatorname{R}_{H,T,M}(\mathbb{A})$.  If this condition can be satisfied for a given attractor $\mathbb{A}$, flow $\phi_T$, and space of measurement functions $\mathcal{H}$, it may require choosing the number of delays $M$ larger than $2\cdot \operatorname{dim}(\mathbb{A})$.
\end{rem}

\begin{rem} \textbf{(Stable rank)} \label{rem:stable rank} The requirement on the stable rank of $\mathbb{A}$ in (\ref{eq:stable rank thm}) merits special attention. Let us fix the measurement basis functions (and consequently the map $H(\cdot)$). The condition in~\eqref{eq:stable rank thm} must be satisfied to stably embed $\mathbb{A}$, which may require the user to sufficiently increase $\operatorname{R}_{H,T,M}(\mathbb{A})$ by adjusting the sampling interval $T$ and the number of delays $M$. In fact, (\ref{eq:stable rank thm}) helps justify certain design rules that are commonly employed in constructing delay-coordinate mappings.

Suppose for the moment that an oracle could inform the user of $\operatorname{R}_{H,T,M}(\mathbb{A})$ for a given pair $(T,M)$ and let us examine the behavior of the stable rank under these variables. If $P < M$, then $\operatorname{R}_{H,T,M}(\mathbb{A})$ is upper bounded by $P$. However, if $P\ge M$, recall from (\ref{eq:range of sr}) that $\operatorname{R}_{H,T,M}(\mathbb{A}) \in [1,M]$. If $\operatorname{R}_{H,T,M}(\mathbb{A})\approx M$, the user could eventually enforce (\ref{eq:stable rank thm}) by increasing $M$ (thereby stably embedding $\mathbb{A}$). But how can the user enforce $\operatorname{R}_{H,T,M}(\mathbb{A}) \approx M$ by adjusting $T$? From (\ref{eq:s rank of h(A)}), $\operatorname{R}_{H,T,M}(\mathbb{A})\approx M$ means that the rows of $X_{H,T,M}-Y_{H,T,M}\in\mathbb{R}^{M\times P}$ are nearly  orthonormal for every pair $x,y\in\mathbb{A}$ (see the discussion in Section \ref{sec:stable rank sec}). Roughly speaking, the following considerations are relevant:
\begin{itemize}
\item For the rows of $X_{H,T,M}-Y_{H,T,M}$ (see \eqref{eq:other def of XM}) to have nearly the same length, $T$ must be substantially smaller (in magnitude) than the Lyapunov exponents of the flow $\phi_{T}(\cdot)$ on $\mathbb{A}$ so that
    \begin{equation}\label{eq:irrelavancy}
    \|\phi_{H,T}^{-m}(H(x))-\phi_{H,T}^{-m}(H(y))\|_2 \approx  \|H(x)-H(y)\|_2,\qquad \forall x,y\in\mathbb{A}, \, m\in \{0,1,\dots,M-1\},
    \end{equation}
    by the invariance of Lyapunov exponents under the diffeomorphism $H(\cdot)$~\cite[Section 11.2]{Kantz2004}.
    Note that when $T$ is comparable to the Lyapunov exponents (in magnitude), then the rows of $X_{H,T,M}-Y_{H,T,M}$ might differ considerably in length, and $\operatorname{R}_{H,T,M}(\mathbb{A})$ is likely to be small (leading to a poor embedding of $\mathbb{A}$).

This aspect of our theoretical result mirrors the well-recognized phenomenon of \emph{irrelevancy} \cite{Casdagli1991,Uzal2011,Kugiumtzis1996}. Indeed, when $T$ is comparable to the Lyapunov exponents (in magnitude), entries of the delay vector $F_{h_\alpha,T,M}(x(t))\in\mathbb{R}^M$ are likely to be ``causally independent.'' In this case, the trajectory $F_{h_\alpha,T,M}(x(\cdot))\subset\mathbb{R}^M$ in the reconstruction space will be unnecessarily more complex than the original trajectory $x(\cdot)\subset\mathbb{R}^N$ in the state space.

\item For the rows of $X_{H,T,M}-Y_{H,T,M}$ to be nearly orthogonal for every $x,y\in\mathbb{A}$, the trajectories of the flow $\phi_{H,T}(\cdot)$ on $H(\mathbb{A})$ should be ``diverse'' in that they should ``visit'' different dimensions as time progresses. Adjusting $T$ here might help push the rows of $X_{H,T,M}-Y_{H,T,M}$ to become nearly orthogonal. However, when $T$ is very small, the rows of $X_{H,T,M}-Y_{H,T,M}$ (for a pair $x,y\in\mathbb{A}$) are similar in direction and length, and consequently $\operatorname{R}_{H,T,M}(\mathbb{A})$ is likely to be small (resulting in a poor embedding).

    Similarly, this aspect of our theoretical result mirrors a known phenomenon called \emph{redundancy} \cite{Casdagli1991,Uzal2011,Kugiumtzis1996}. Indeed, when $T$ is very small, the adjacent entries of a delay vector $F_{h_\alpha,T,M}(x(t))\in\mathbb{R}^M$ are likely to be highly similar (or ``correlated'') and the information contained in $F_{h_\alpha,T,M} (x(t))$ is largely redundant. In this case, the trajectory $F_{h_\alpha,T,M}(x(\cdot))\subset\mathbb{R}^M$ in the reconstruction space will be stretched out along the identity line (regardless of the geometry of the attractor $\mathbb{A}\subset\mathbb{R}^N$).
\end{itemize}
To summarize the main points, if $T$ is chosen too large or too small, then $\operatorname{R}_{H,T,M}(\mathbb{A})$ will rapidly plateau when the user increases $M$. Thus, our theoretical findings echo the (mainly heuristically-investigated) trade-off between irrelevancy and redundancy in the literature, suggesting the user may improve the embedding quality if a sampling interval in this ideal intermediate range can be found.

\begin{rem}\textbf{(Choice of $T$ and $M$)} \label{rem:choice of T M}
The discussion in Remark \ref{rem:stable rank} raises the following question:
\begin{itemize}
\item Can the user experimentally find the right range for $T$ and $M$ without prior knowledge of the quantities involved in \eqref{eq:stable rank thm}?
\end{itemize}
To answer this question, we first point out that a similar issue has arisen in the past with choosing the number of delays $M$ for Takens' original theorem.  As a practical method for setting this parameter~\cite{Kantz2004}, the community has observed that $\operatorname{dim}({\mathbb{A}})$ is preserved under delay-coordinate mapping effectively \emph{as long as} $M> 2\cdot \operatorname{dim}(\mathbb{A})$ and there is no noise.  This observation suggests the following procedure for estimating $\operatorname{dim}(\mathbb{A})$ and consequently estimating the required number of delays $M$ in Takens' theorem.
For fixed $T$ and every $M$ within a fixed range $\{M_1,\dots,M_2\}$, the user constructs a sequence of delay-coordinate maps for many example test observations from the system. For each $M$, the user applies the Grassberger-Procaccia algorithm \cite{Grassberger2004}
to estimate  $\mbox{dim}(F_{h,T,M}(\mathbb{A}))$ and searches for a range of values of $M$ where the graph of $\mbox{dim}(F_{h,T,M}(\mathbb{A}))$ (versus $M$) plateaus. This plateau is an estimate for $\operatorname{dim}(\mathbb{A})$, and a reasonable choice of $M$  immediately follows.  When noise is present, this plateau may disappear at large values of $M$ as well, resulting in a ``sweet spot'' in the graph where $M=O(2\cdot \operatorname{dim}(\mathbb{A}))$ and $\mbox{dim}(F_{h,T,M}(\mathbb{A}))=\mbox{dim}(\mathbb{A})$.

Returning to the present problem, a similar approach can be used. If \eqref{eq:stable rank thm} indeed holds, then (with high probability) the delay-coordinate map with parameters $T$ and $M$ stably embeds $\mathbb{A}$ into the reconstruction space $\mathbb{R}^M$ and the volume of $\mathbb{A}$ is preserved.  In general, \eqref{eq:RIP ideal} implies that\footnote{This claim is proved similar to those in Appendix \ref{sec:proof of lemma props of A_h,M,T}.} \begin{equation}
(\epsilon_l\cdot M)^{\frac{\operatorname{dim}(\mathbb{A})}{2}} \cdot \operatorname{vol}(\mathbb{A})
\le
\operatorname{vol}\left( F_{h_\alpha,T,M}(\mathbb{A})\right)
\le
(\epsilon_u\cdot M)^{\frac{\operatorname{dim}(\mathbb{A})}{2}} \cdot \operatorname{vol}(\mathbb{A}).
\end{equation}
This observation implies a variant of the algorithm described above where volume is used in place of dimension to find the correct range of $T$ and $M$, which we detail in Table~\ref{table:prescription}.
\end{rem}

\begin{table}[t]
\caption{\small\sl Prescription to find the proper range of the sampling interval $T$ and the number of delays $M$ in delay-coordinate mapping.}
\label{table:prescription}
\begin{center}
\fbox{\begin{minipage}[t]
{.9\columnwidth}%
\textbf{How to choose $T$ and $M$ in delay-coordinate mapping}
\begin{enumerate}
\item Given a time-series $\{s_i\}_i$ and a scalar measurement function $h(\cdot)$, compute the delay vectors $\{F_{h,T,M}(x(i\cdot T))\}_i\subset \mathbb{R}^M$ for every pair $(T,M)$ in the window $[T_{\min},T_{\max}]\times [M_{\min}:M_{\max}]$.
\item For each pair $(T,M)$, empirically compute the dimension $d_{T,M}$ and volume $V_{T,M}$
of the surface formed by the delay vectors $\{F_{h,T,M}(x(i\cdot T))\}_i$, and plot
$\frac{V_{T,M}}{\sqrt{M^{d_{T,M}}}}$ for various pairs $(T,M)$ in the above window.
\item Find the range of $(T,M)$ for which the graph is nearly constant. This provides the recommended range for $T$ and $M$ in delay-coordinate mapping of the system under study.
\end{enumerate}
\end{minipage}
}
\end{center}
\end{table}

\begin{rem}\textbf{(Quality of embedding)} Let us again fix the basis functions (and thus $H(\cdot)$), and suppose that (\ref{eq:stable rank thm}) holds for a given isometry constant $\delta\in(0,1)$ and failure probability $\rho\in(0,1)$.  Then the quality of embedding in (\ref{eq:final pre thm}) clearly depends on
\begin{itemize}
\item the bi-Lipschitz constants of $H(\cdot)$ (i.e.,  $l_H,u_H$);
\item the spectrum of the derivative of the flow $\phi_{T}(\cdot)$ (through $\sigma_{\min},\sigma_{\max}$); and
\item the geodesic regularity of the attractor $\mathbb{A}$ (i.e., $\operatorname{geo}(\mathbb{A})$).
\end{itemize}
Large values of $\frac{u_H}{l_H}$, $\frac{\sigma_{\max}}{\sigma_{\min}}$, and $\operatorname{geo}(\mathbb{A})$ in (\ref{eq:final pre thm}) all result in a poor embedding guarantee for $\mathbb{A}$ (i.e., a large disparity between the upper and lower bounds in (\ref{eq:final pre thm})). In particular, when the dynamical system is highly unpredictable (e.g., has a large Lyapunov exponent), then $\frac{\sigma_{\max}}{\sigma_{\min}}$ is likely to be very large and the embedding guarantee (and, indeed, the embedding itself) is likely to be poor. In a nutshell, stably embedding unpredictable systems (e.g., chaotic systems) is often difficult.
\label{rem:quality}
\end{rem}

\begin{rem} \textbf{(Orbits and other pathologies)}
The flow $\phi_T(\cdot)$ has an orbit with period $n$ if $\phi^n_T(x)=\phi_{nT}(x)=x$ for some $x\in\mathbb{A}$. As noted in Remark~\ref{rem:compare}, the probabilistic statement of Takens' theorem by Sauer et al.~\cite{Sauer1991} placed certain restrictions on the periodicities of the orbits of $\phi_T$. Indeed, the existence of orbits also typically deteriorates the stable rank of a system. As an extreme example, consider an orbit of period one, otherwise known as a \emph{fixed point}: $\phi_T(x)=x$ for some $x\in\mathbb{A}$. Using (\ref{eq:s rank of h(A)}), we may easily verify that $\operatorname{R}_{H,T,M}(\mathbb{A})=1$ for any choice of basis functions and any number of delays $M$. That is, the stable rank of $\mathbb{A}$ does not increase at all when the  user  increases $M$. In view of (\ref{eq:stable rank thm}), this leads to a very poor embedding of the attractor $\mathbb{A}$. We note that orbits of period one are explicitly forbidden by Sauer et al.~\cite{Sauer1991} and implicitly forbidden in Theorem~\ref{thm:Takens} through the genericity of $\phi_T$.

\end{rem}

\subsection{Extensions to Strange Attractors}
\label{sec:chaoticextension}

While our discussion thus far has focused on attractors that comprise smooth submanifolds of $\mathbb{R}^n$, many dissipative dynamical systems (e.g., chaotic systems) converge onto attractors that are not smooth submanifolds of the Euclidean space.
In this section, we discuss what changes when considering the stable embedding of more general (e.g., strange) attractors. In what follows, we continue to assume that the state lies on the attractor $\mathbb{A}$ so that for every time $t$, $x(t) \in \mathbb{A}$.

\subsubsection{Global enveloping manifolds}

The easiest scenario arises when there exists a \emph{global enveloping manifold} $\mathbb{M}$ that subsumes the attractor $\mathbb{A}$. Roughly speaking, we say that $\mathbb{M} \subset \mathbb{R}^n$ is a global enveloping manifold of an attractor $\mathbb{A} \subset \mathbb{R}^n$ if $\mathbb{A} \subset \mathbb{M}$ and at every point $x \in \mathbb{A}$, $\mathbb{T}_x \mathbb{A} = \mathbb{T}_x \mathbb{M}$ (see~\cite{Ott2003,Brin2004} for a more precise definition). Here, $\mathbb{T}_x \mathbb{M}$ denotes the conventional tangent space of $\mathbb{M}$ at $x$ (recall Section~\ref{sec:Differential Geometry}), and $\mathbb{T}_x \mathbb{A}$ denotes a generalized tangent space of $\mathbb{A}$ at $x$, defined as follows.

\begin{definition}\emph{\textbf{(Generalized tangent space~\cite{Brin2004})}}
\label{def:gts}
Consider an attractor $\mathbb{A} \subset \mathbb{R}^{N}$ and a point $x \in \mathbb{A}$. The generalized tangent space of $\mathbb{A}$ at $x$, denoted $\mathbb{T}_x \mathbb{A}$, is the smallest linear space containing all unit vectors of the form $(z_i - y_i)/\|z_i-y_i\|_2$ generated by sequences $\{y_i\}$ and $\{z_i\}$ in $\mathbb{A}$ with $y_i \rightarrow x$ and $z_i \rightarrow x$.
\end{definition}

In scenarios where there does exist a global enveloping manifold $\mathbb{M}$ for $\mathbb{A}$, Theorem~\ref{thm:manifold2} can be naturally extended to provide conditions for the stable embedding of $\mathbb{M}$ (and thus $\mathbb{A}$). In order to prove this result, one merely replaces $\mathbb{A}$ with $\mathbb{M}$ throughout the statement and the proof of Theorem~\ref{thm:manifold2}; consequently, all of the geometric quantities that appear in the resulting bound---dimension, volume, reach, and so on---will refer to $\mathbb{M}$ instead of $\mathbb{A}$.\footnote{As noted in Remark~\ref{rem:quality}, when the dynamical system is highly unpredictable (e.g., has a large Lyapunov exponent), then $\frac{\sigma_{\max}}{\sigma_{\min}}$ is likely to be very large and the embedding guarantee (and, indeed, the embedding itself) is likely to be poor. In such a case, Remark~\ref{rem:poor geod} may have some value.} However, because it may be unreasonable to assume that the enveloping manifold is invariant under the flow (i.e., that $\phi_T(\mathbb{M}) = \mathbb{M}$), one may relax this assumption in the statement and proof of the theorem; all that is needed is that $\phi_T$ acts as a diffeomorphism between $\mathbb{M}$ and $\phi_T(\mathbb{M})$ (or, more precisely, between $\mathbb{M}$ and each of $\phi_T^{-1}(\mathbb{M}), \dots, \phi_T^{-M+1}(\mathbb{M})$), and that the assumptions on $H$ listed in Section~\ref{sec:measurement} hold not only on $\mathbb{M}$ but also on each of $\phi_{T}^{-1}(\mathbb{M}), \dots, \phi_T^{-M+1}(\mathbb{M})$.

\subsubsection{More general attractors}

Alas, a counter-example in~\cite{Brin2004} shows that not all subsets of Euclidean space, and thus potentially not all attractors of dynamical systems, can have a global enveloping manifold. When the attractor $\mathbb{A}$ does not have a global enveloping manifold, we require a few additional definitions that will endow the attractor with certain geometric properties that make it amenable for our analysis. We first describe these properties in terms of a general subset $\mathbb{B} \subset \mathbb{R}^{N}$.

\begin{definition}\emph{\textbf{(Box-counting dimension~\cite{Sauer1991})}}
\label{def:boxdim}
Consider a set $\mathbb{B} \subset \mathbb{R}^{N}$. Suppose $\mathbb{R}^{N}$ is divided into cubes of size $\zeta$ by a grid based at points whose coordinates are $\zeta$-multiples of the integers. Let $\mathcal{N}(\zeta)$ be the number of boxes or cubes of size $\zeta$ that intersect $\mathbb{B}$. Then the box-counting dimension of $\mathbb{B}$, denoted by $\operatorname{boxdim}(\mathbb{B})$, is defined as
\[
\operatorname{boxdim}(\mathbb{B}):= \lim_{\zeta \rightarrow 0} -\frac{\log \mathcal{N}(\zeta)}{\log(\zeta)}.
\]
\end{definition}

\begin{definition}\emph{\textbf{(Covering regularity)}}
We say that the set $\mathbb{B}$ has covering regularity $\operatorname{cov}(\mathbb{B})$ (depending on some maximum size $\zeta_{0}$) if for every $\zeta \le \zeta_{0}$,
\begin{eqnarray*}
\mathcal{N}(\zeta) \le \operatorname{cov}(\mathbb{B}) \zeta^{-\operatorname{boxdim}(\mathbb{B})},
\end{eqnarray*}
where $\mathcal{N}(\zeta)$ is the number of boxes or cubes of size $\zeta$ that intersect $\mathbb{B}$ (see Definition~\ref{def:boxdim}).
\end{definition}

One can think of the covering regularity $\operatorname{cov}(\mathbb{B})$ as a proxy for the volume of $\mathbb{B}$ because volume is proportional to $\mathcal{N}(\zeta) \zeta^{\operatorname{boxdim}(\mathbb{B})}$ in the limit of small $\zeta$ when $\mathbb{B}$ is a submanifold.

\begin{definition}\emph{\textbf{(Tangent covering regularity)}}
We say that the set $\mathbb{B}$ has tangent covering regularity $\operatorname{tancov}(\mathbb{B})$ (depending on some maximum size $\zeta_{0}$) if for every $a \in \mathbb{B}$, whenever $\| x - a \|_{2}, \|y - a\|_{2} \le \zeta \le \zeta_{0}$ for some $x,y \in \mathbb{B}$, we can find a $v \in \mathbb{T}_a \mathbb{B}$ such that
\begin{eqnarray*}
\left\| v - \frac{x - y}{\|x - y\|_2} \right\|_2 \le \operatorname{tancov}(\mathbb{B}) \zeta.
\end{eqnarray*}
\end{definition}

Here $\operatorname{tancov}(\mathbb{B})$ can be thought of as a measure the curvature of $\mathbb{B}$.

\begin{definition}\emph{\textbf{(Tangent dimension)}}
We define the tangent dimension $\operatorname{tandim}(\mathbb{B})$ of the set $\mathbb{B}$ as
\begin{eqnarray*}
\operatorname{tandim}(\mathbb{B}) := \sup_{x \in \mathbb{B}} \dim \left(\mathbb{T}_x \mathbb{B} \right),
\end{eqnarray*}
where $\mathbb{T}_x \mathbb{B}$ refers to the generalized tangent space of $\mathbb{B}$ at $x$ (see Definition~\ref{def:gts}).
\end{definition}

As noted in~\cite{Brin2004}, the tangent dimension bounds the box-counting dimension from above: for any set $\mathbb{B}$, $\operatorname{tandim}(\mathbb{B}) \ge \operatorname{boxdim}(\mathbb{B})$. In what follows, we shall ignore the dependence of the regularity quantities on their maximal resolution $\zeta_0$.

With these properties thus defined, we present our result for the stable embedding of a general (including strange) attractor. The following theorem makes a series of assumptions not on the attractor $\mathbb{A}$ itself, but rather on the trajectory attractor $\mathbb{A}_{H,T,M} \subset \mathbb{R}^{MP}$ defined in~\eqref{eq:traj attr}. We discuss these conditions further after presenting the main result, which is proved in Appendix~\ref{sec:proof_manifold strange}.

\begin{thm}
\label{thm:manifold2strange}\emph{\textbf{(Stable Takens' embedding theorem for strange attractors)}}
Let $\mathbb{A}\subset\mathbb{R}^N$ be an attractor. For a fixed sampling interval $T>0$, assume that $\phi_{T}(\cdot)$ is a flow on $\mathbb{A}$. For an integer $P$, fix the measurement basis functions $h_p:\mathbb{A}\rightarrow\mathbb{R}$ for $p\in\{1,2,\dots,P\}$ and let $\mathcal{H}$ be the linear span of $\{h_p(\cdot)\}_p$. The random coefficient vector $\alpha\in\mathbb{R}^P$ is assumed to have entries that are i.i.d.\ zero-mean and unit-variance subgaussian random variables with subgaussian norm $\theta$.

Consider the map $H:\mathbb{A} \rightarrow\mathbb{R}^P$ constructed in (\ref{eq:def of h}), and suppose that $H(\cdot)$ satisfies assumption A1 listed in Section~\ref{sec:measurement}. Let $\mathbb{A}_{H,T,M} \subset \mathbb{R}^{MP}$ be the associated trajectory attractor defined in~\eqref{eq:traj attr}. Suppose $\mathbb{A}_{H,T,M}$ has box-counting dimension $\operatorname{boxdim}(\mathbb{A}_{H,T,M})$, tangent dimension $\operatorname{tandim}(\mathbb{A}_{H,T,M})$, covering regularity $\operatorname{cov}(\mathbb{A}_{H,T,M}) > 1$, and tangent covering regularity $\operatorname{tancov}(\mathbb{A}_{H,T,M}) > \frac{3}{\sqrt{MP}}$. Finally, let $\operatorname{R}_{H,T,M}(\mathbb{A})$ denote the stable-rank of $\mathbb{A}$ as defined in
(\ref{eq:s rank of h(A)}).

For arbitrary isometry constant $\delta\in(0,1)$ and failure probability $\rho\in(0,1)$, suppose that
\begin{align}
& \operatorname{R}_{H,T,M}\left(\mathbb{A}\right) \ge \nonumber \\
& \quad\quad C'_\theta \cdot \max \Bigg[ \delta^{-2} \operatorname{tandim}(\mathbb{A}_{H,T,M}) \log\left(\sqrt{MP} \operatorname{tancov}(\mathbb{A}_{H,T,M})\left(\operatorname{cov}(\mathbb{A}_{H,T,M})\right)^{1/\operatorname{boxdim}(\mathbb{A}_{H,T,M})} \right), \nonumber \\
  & \quad\quad\quad\quad\quad\quad\quad\quad\quad e^{-\min W\left(\frac{-\delta^2}{\operatorname{tandim}(\mathbb{A}_{H,T,M})} \right)}, \delta^{-2}\cdot \log\left(\frac{1}{\rho} \right) \Bigg]  \label{eq:stable rank thm strange}
\end{align}
where $C'_\theta$ is a constant that depends only on $\theta$. Then except with a probability of at most $\rho$ (over the choice of $\alpha$),
\begin{align}
\label{eq:poor geo strange main}
1-\delta  \le \frac{\left\| F_{h_{\alpha},T,M}(x)-F_{h_{\alpha},T,M}(y)\right\|_F^2}{
\sum_{m=0}^{M-1}
\left\|
H\left(\phi^{-m}_T(x)\right) - H\left(\phi^{-m}_T(y)\right)
\right\|_2^2}
\le 1+\delta
\end{align}
holds for all $x,y\in\mathbb{A}$ with $x \neq y$.

Moreover, if~\eqref{eq:poor geo strange main} holds for all $x,y\in\mathbb{A}$ with $x \neq y$ and if there exist quantities $\operatorname{geo}(\mathbb{A})$, $\sigma_{\min}$, $\sigma_{\max}$ such that for all $x,y \in \mathbb{A}$ and $m = 1,2,\dots,M-1$,
\begin{equation}
\operatorname{geo}(\mathbb{A})^{-1}\cdot \sigma_{\max}^{-m} \cdot \left\|x-y\right\|_2 \le \left\| \phi_T^{-m}(x)-\phi_T^{-m}(y)\right\|_2 \le  \operatorname{geo}(\mathbb{A})\cdot \sigma_{\min}^{-m} \cdot \left\|x-y\right\|_2,
\label{eq:newdistanceassumption2main}
\end{equation}
it follows that
\begin{equation}\label{eq:final 0 strange main}
 (1-\delta)\cdot l_H^2 \cdot \operatorname{geo}(\mathbb{A})^{-2} \cdot \frac{\sigma_{\max}^{-2M}-1}{\sigma_{\max}^{-2}-1}  \le \frac{\left\| F_{h_{\alpha},T,M}(x)-F_{h_{\alpha},T,M}(y) \right\|_2^2}{ \|x-y\|_2^2}\le (1+\delta)\cdot u_H^2\cdot  \operatorname{geo}(\mathbb{A})^2\cdot \frac{\sigma_{\min}^{-2M}-1}{\sigma_{\min}^{-2}-1}
\end{equation}
holds for all $x,y\in\mathbb{A}$ with $x \neq y$.
\end{thm}

Much like our original Theorem~\ref{thm:manifold2}, Theorem~\ref{thm:manifold2strange} guarantees a stable embedding of an attractor with high probability, under the condition that the stable rank $\operatorname{R}_{H,T,M}\left(\mathbb{A}\right)$ is sufficiently large. However, whereas the right hand side of~\eqref{eq:stable rank thm} involves mostly properties of the attractor $\mathbb{A}$ itself, the right hand side of~\eqref{eq:stable rank thm strange} references properties of the trajectory attractor $\mathbb{A}_{H,T,M}$ instead. Indeed, a key step in the proof is bounding the covering number of the set of all normalized secants of the trajectory attractor. In the proof of Theorem~\ref{thm:manifold2}, we used Lemma~\ref{lemma:props of A_h,M,T} to relate properties of $\mathbb{A}_{H,T,M}$ to those of $\mathbb{A}$. In the case of strange attractors, we leave this connection as an open question.

Perhaps interestingly, Theorem~\ref{thm:manifold2strange} does not require any assumptions regarding $\phi_T$ or $H$ being a diffeomorphism, nor any assumptions on the singular values of their derivatives. Such properties do likely affect the quality of the embedding. In the proof of Theorem~\ref{thm:manifold2}, we used these properties both in Lemma~\ref{lemma:props of A_h,M,T} (to relate properties of $\mathbb{A}_{H,T,M}$ to those of $\mathbb{A}$) and to guarantee that a condition equivalent to~\eqref{eq:newdistanceassumption2main} holds. However, the original proof of the condition equivalent to~\eqref{eq:newdistanceassumption2main} required an argument involving geodesic distance, which is not appropriate for a strange attractor. Here, we pull out~\eqref{eq:newdistanceassumption2main} as its own assumption, which could conceivably hold even for a strange attractor. Thus, in~\eqref{eq:newdistanceassumption2main}, the quantities $\operatorname{geo}(\mathbb{A})$, $\sigma_{\min}$, $\sigma_{\max}$ do not necessarily refer to the geodesic regularity of the attractor $\mathbb{A}$ or the singular values of the derivative of $\phi_T(\cdot)$. However, for~\eqref{eq:newdistanceassumption2main} to hold, these parameters would likely play similar roles to those that they played in Theorem~\ref{thm:manifold2}.

\section{Examples}\label{sec:examples}

In this section, we present two examples that support the theoretical findings in Section \ref{sec:main results}, emphasizing the relationship between the stable rank of a system and the number of delays in delay-coordinate mapping.

\subsection{Moment Curve}
\label{sec:second_example}

We begin with an example where we can analytically calculate (or bound) the quantities of interest.
For an  integer $N$, let $\mathbb{A}$ be the moment curve in $\mathbb{C}^N$.\footnote{Strictly speaking, Theorem \ref{thm:manifold2} applies to subsets of $\mathbb{R}^N$ and  not to $\mathbb{A}\subset\mathbb{C}^N$ as in this  example. However, study  of the ``real'' moment curve (formed from the real part of $\gamma(\cdot)$) is far more tedious, and is therefore not pursued here for the sake of the clarity. In fact, we strongly suspect that Lemma 15 in \cite{Eftekhari2015} and consequently Theorem \ref{thm:manifold2} can be extended (with minor changes) to account for complex attractors. }   That is,
\begin{equation}\label{eq:moment curve}
\mathbb{A} = \left\{ \gamma(t) \,:\, t\ge 0 \right\}\subset\mathbb{C}^N,\quad \gamma(t) =
\left[ \begin{array}{c}
	1 \\
	e^{\mbox{i} 2 \pi t} \\
	\vdots \\
	e^{\mbox{i} 2 \pi (N-1) t}
	\end{array}
	\right].
\end{equation}
Note that $\mathbb{A}$ is a closed curve because $\gamma(n)=\gamma(0)$, for every integer $n$.
For a fixed $T>0$, we endow $\mathbb{A}$ with a linear dynamical system with flow $\phi_T(\cdot)$. This linear flow, which we identify with an $N\times N$ matrix, is specified as
\begin{equation}\label{eq:flow moment}
\phi_T = \mbox{diag}\left[ \gamma(T)\right]\in \mathbb{C}^{N\times N},
\end{equation}
where $\mbox{diag}[a]$ returns the diagonal matrix formed from the entries of vector $a$. For any $t\ge 0$, observe that $\phi_T(\gamma(t)) = \phi_T\cdot \gamma(t) = \gamma(t+T)$; that is $\mathbb{A}=\gamma(\cdot)$ is parametrized by time.

Let $\mathcal{H}$ be the space of all linear functionals on $\mathbb{C}^N$, so that every scalar  measurement function may be characterized as $h_\alpha(\cdot) = \langle \cdot,\alpha\rangle$ for some $\alpha\in\mathbb{C}^N$. In the language of Theorem \ref{thm:manifold2}, we set $P=N$ and take  $H(\cdot)$ to be the identity operator (and, in particular, $H(\mathbb{A})= \mathbb{A}$). Assume also that the entries of $\alpha$ are independent Gaussian random variables with zero mean and unit variance.\footnote{The variance of a complex random variable is the sum of the variances of its real and imaginary parts.}

We next compute the relevant geometric quantities. Since $\mathbb{A}$ is a curve, $\operatorname{dim}(\mathbb{A})=1$, and $\operatorname{vol}(\mathbb{A})$ is simply its length:
\begin{align}\label{eq:length moment}
\operatorname{vol}(\mathbb{A}) = \mbox{length}(\gamma(\cdot)) & = \int_0^1 \left\|\frac{d\gamma(t)}{dt}\right\|_2 \, dt \nonumber\\
& = 2\pi \sqrt{\sum_{n=0}^{N-1} n^2}\cdot  \int_0^1 \, dt
\qquad \mbox{(see \eqref{eq:moment curve})}
\nonumber\\
& = \pi  \sqrt{\frac{2}{3}\cdot (N-1)N(2N-1)}.
\end{align}
Next, we turn to the geodesic regularity of the moment curve which involves comparing geodesic and Euclidean distances between an arbitrary pair of points on $\mathbb{A}$. Using  \eqref{eq:length moment} (and the implicit observation therein that $\gamma(\cdot)$ has constant ``speed''), we deduce that the geodesic distance  between $\gamma(t_1),\gamma(t_2)\in\mathbb{A}$ is given by
\begin{align}\label{eq:geod dist moment}
d_{\mathbb{A}} \left(\gamma(t_1),\gamma(t_2)\right)
& = \left|t_1-t_2\right| \cdot \mbox{length}\left(\gamma(\cdot)\right) \nonumber\\
& = \left|t_1-t_2\right|\cdot \pi  \sqrt{\frac{2}{3}\cdot (N-1)N(2N-1)}, \qquad \forall t_1,t_2\in [0,1).
\end{align}
In Appendix \ref{sec:proof of geod reg}, we calculate the Euclidean distance $\|\gamma(t_1)-\gamma(t_2)\|_2$ and estimate the geodesic regularity of the moment curve by comparing the two metrics.

\begin{lemma}\label{lem:geod reg} \textbf{\emph{(Geodesic regularity)}}
For an integer $N$, let $\mathbb{A}$ be the moment curve in $\mathbb{C}^N$ (see (\ref{eq:moment curve})). Then, the geodesic regularity of $\mathbb{A}$ (see \ref{eq:assump on geod diff thm}) is bounded as
\begin{equation}
\operatorname{geo}(\mathbb{A}) \le \frac{2\pi^2}{3(1-\Cl[Bet]{far})}
\cdot N(N-1)
,\qquad \forall N>N_m.
\end{equation}
Above, $\Cr{far}>0$ is a (small) absolute constant, and $N_m$ is a sufficiently large integer (which is subject to change in every appearance).
\end{lemma}
In other words, the geodesic regularity of $\mathbb{A}$ is poor, $\operatorname{geo}(\mathbb{A})= O(N^2)$ which, in light of \eqref{eq:final pre thm},  suggests that delay-coordinate mapping might poorly embed this system (when the dimension of the state space $N$ is large). The guarantees in Theorem \ref{thm:manifold2} appear to be conservative here as our simulations indicate later in this section.

Next, to compute the reach, we borrow from Lemma 1 in \cite{Eftekhari2015}:
\begin{equation}
\Cl[Bet]{rch}\sqrt{N}\le \operatorname{rch}(\mathbb{A}) \le \sqrt{N},\qquad \forall N>N_m.
\end{equation}
Here, $\Cr{rch}< 1$ is independent of $N$, and $N_m$ is a sufficiently large integer.\footnote{The discrepancy in the definition of moment curve here and in \cite{Eftekhari2015} is inconsequential.} That is, fortuitously,   the reach of the moment curve is relatively large. Next, we turn to the stable rank of this system. The estimate below is obtained in Appendix \ref{sec:proof of s rank of moment}.
\begin{lemma}\label{lem:s rank of moment curve}
For an integer $N$, let $\mathbb{A}$ be the moment curve in $\mathbb{C}^N$ (see (\ref{eq:moment curve})). For $T\in (0,\frac{1}{M}]$, equip $\mathbb{A}$ with the linear flow $\phi_T(\cdot)$  specified in (\ref{eq:flow moment}). Then, the stable rank of $\mathbb{A}$ (as defined in (\ref{eq:s rank of h(A)})) satisfies
\begin{equation}
\frac{M}{ 20 +  \frac{40}{N\sin(\pi T)}\cdot \log\left(e/\tan\left( \frac{\pi T}{2}\right) \right) } \le \operatorname{R}_{H,T,M}(\mathbb{A})
\le M,\qquad \forall N>N_m,
\end{equation}
where $N_m$ is a sufficiently large integer.
\end{lemma}
Roughly speaking, as long as $\frac{1}{N}\lesssim T \le \frac{1}{M}$, the stable rank of our system is large ($\operatorname{R}_{H,T,M}(\mathbb{A})\approx M$). (Note that $M\lesssim N$ is necessary for this claim to hold.) Since the stable rank of any system is bounded by the number of delays $M$ (see \eqref{eq:range of sr}), this result is nearly ideal.

Let us now empirically compute the stable rank of this system (see \eqref{eq:s rank of h(A)}) for variable number of delays $M$ and with $N=250$ and $T = 1/(N + \xi)$, where $\xi$ is chosen randomly from a standard normal distribution. The outcome appears in Figure~\ref{fig:sinemanifold}(a). To see the connection between the stable rank and the quality of embedding, we plot in Figure~\ref{fig:sinemanifold}(b) the isometry constants $\epsilon_l\le \epsilon_u$ (recall~\eqref{eq:RIP ideal}) versus the number of delays $M$. To produce the plot, we generated $100$ independent copies of $h_\alpha(\cdot)$ and computed the isometry constants according to \eqref{eq:RIP ideal}. The curve shows the mean isometry constants (over $100$ repetitions). As $M$ increases, the stable rank increases (improves) and the isometry constants tighten (the quality of embedding improves); this matches Theorem \ref{thm:manifold2}.

\begin{figure}[t]
\begin{center}
(a) \includegraphics[height=50mm]{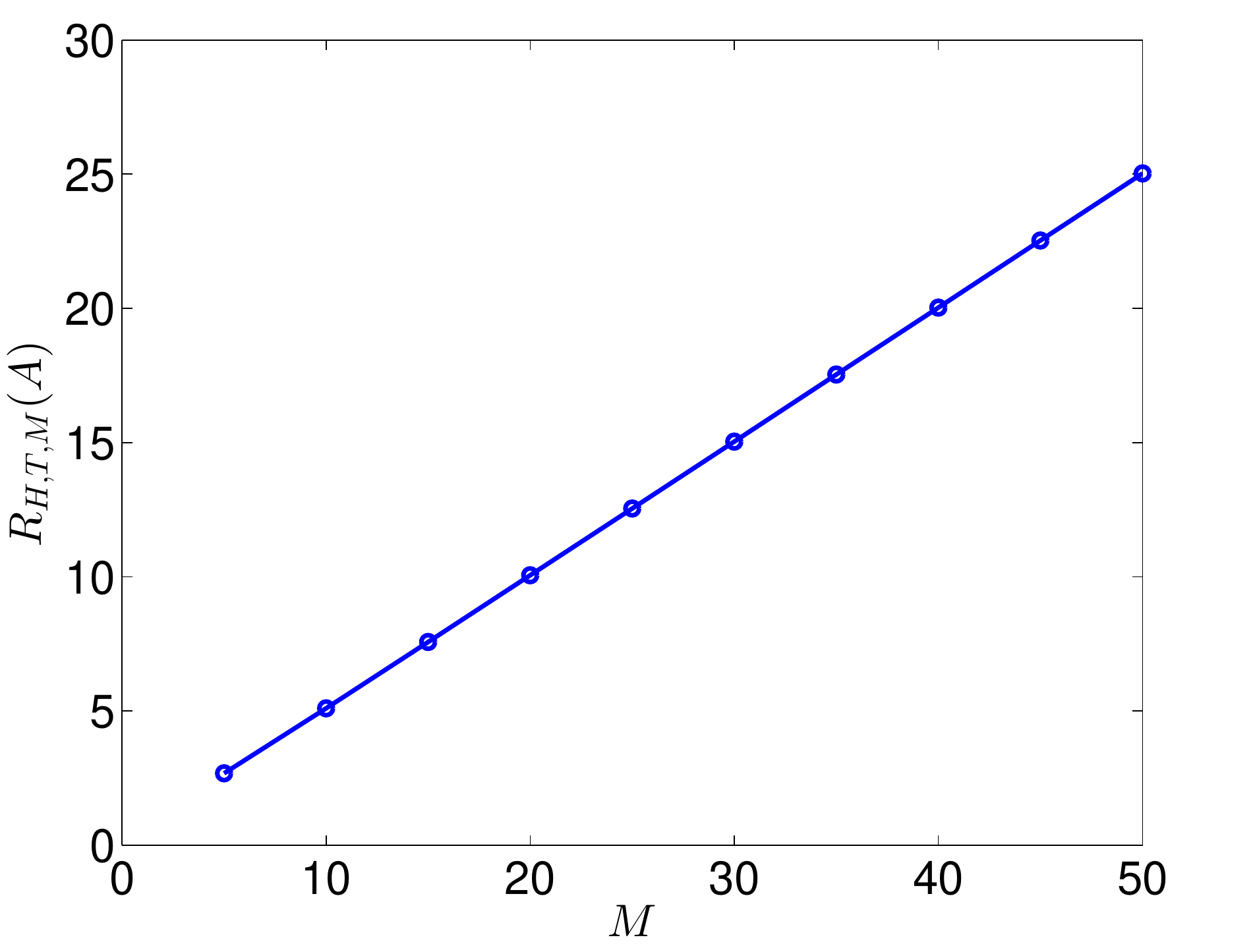} ~~~
(b) \includegraphics[height=50mm]{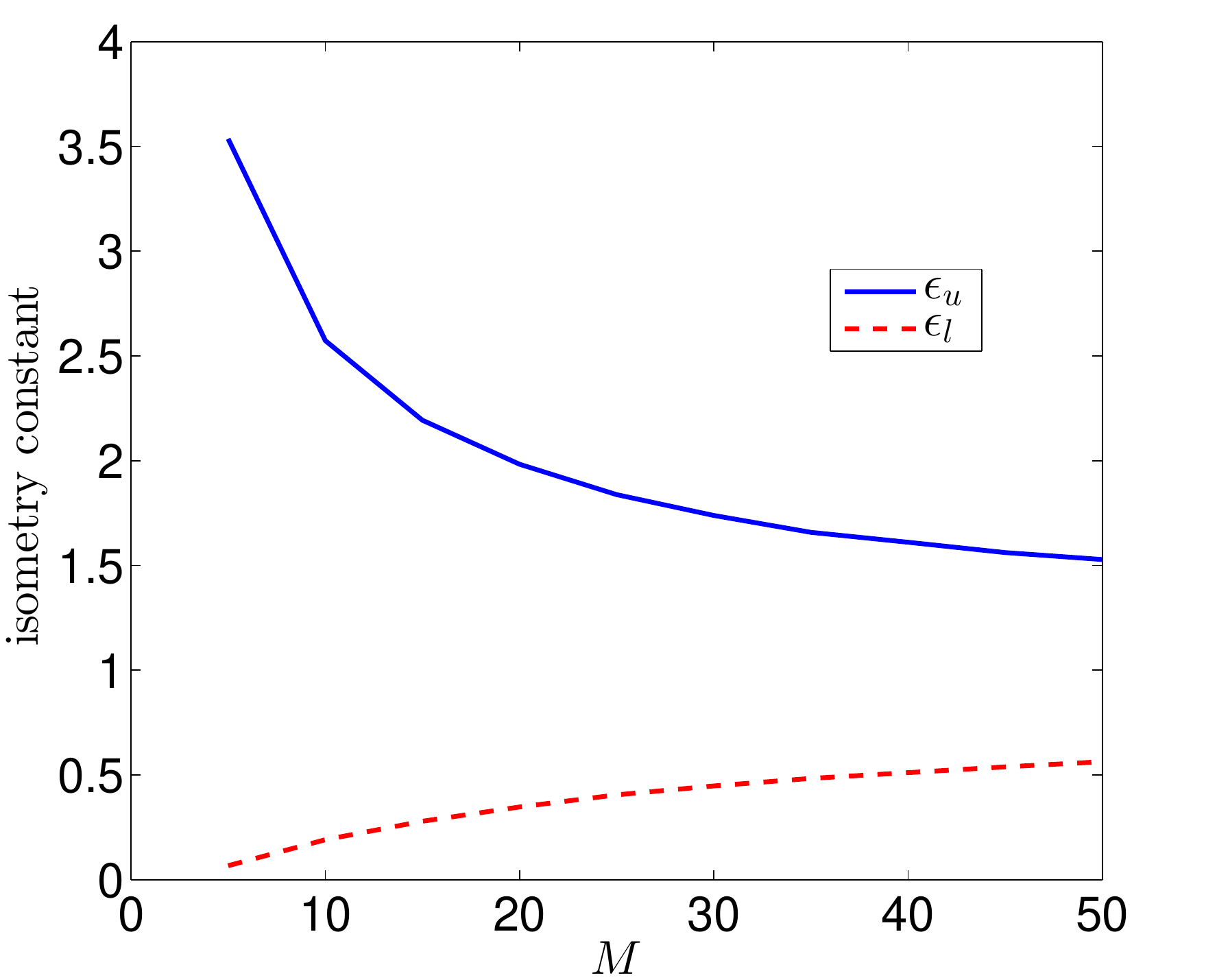}
\end{center}
\vspace{-5mm}
\caption{\small\sl Stable rank and quality of delay-coordinate mapping for the linear system described in Section \ref{sec:second_example}. (a) Stable rank versus $M$ (number of delays) with sampling interval $T \approx 1/250$. Note that the stable rank of the system gradually improves with increasing $M$. (b) Quality of embedding through delay-coordinate mapping as measured by the isometry constants $\epsilon_l\le \epsilon_u$ versus $M$ with $T \approx 1/250$ (see \eqref{eq:RIP ideal}). Note that, like the stable rank, the quality of embedding gradually improves with increasing $M$.}
\label{fig:sinemanifold}
\end{figure}

\subsection{Nonlinear Schr\"odinger System}

As a case study involving a nonlinear system, we consider a sequence of points on a trajectory generated by a certain partial differential equation, the Nonlinear Schr\"odinger (NLS) equation:
\[
\mbox{i} u_t(z,t) + \frac{1}{2} u_{zz}(z,t) + |u(z,t)|^2 u(z,t) = 0.
\]
Here, $t$ denotes the continuous time variable and $z$ denotes the continuous space variable; $u_t$ denotes the partial derivative of $u$ with respect to $t$ and $u_{zz}$ denotes the second order partial derivative of $u$ with respect to $z$; and we adopt the boundary conditions $u \rightarrow 0$ as $z \rightarrow \pm \infty$. Adapting the construction provided in~\cite[Chapter 19]{kutz2013data}, we sample $N = 800$ points between $z = -30$ and $z = 30$ at each time to generate data in $\mathbb{C}^N$. Data is generated with a time step of $0.02$ seconds. The evolution of the trajectory over time is shown in Figure~\ref{fig:nlstraj}, which plots the magnitude of the entries of each data vector. The three different plots correspond to three different integer values of a parameter $S$ which is used in the initial conditions
\[
u(z,0) = S \cdot \text{sech}(z+z_0) \cdot e^{\mbox{i} \Omega t}.
\]
The resulting solutions are known as the $S$-soliton solutions (with $S = 1,2,3$) and have an initial center position $z_0 = 23.5$ and a drift over time due to the group velocity parameter $\Omega = \pi$.

\begin{figure}[t]
	\begin{center}
    \includegraphics[height=48mm]{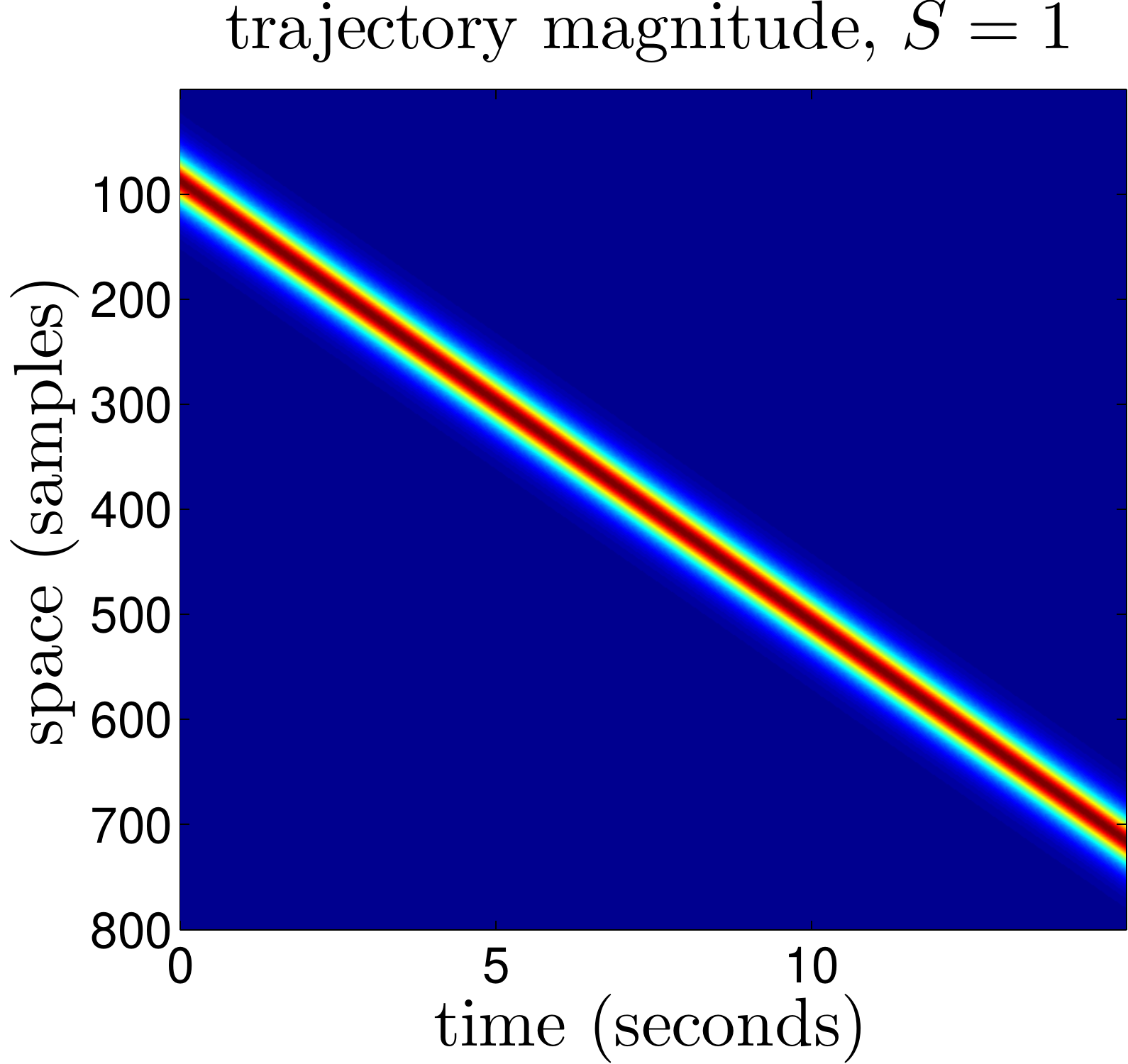} ~~
    \includegraphics[height=48mm]{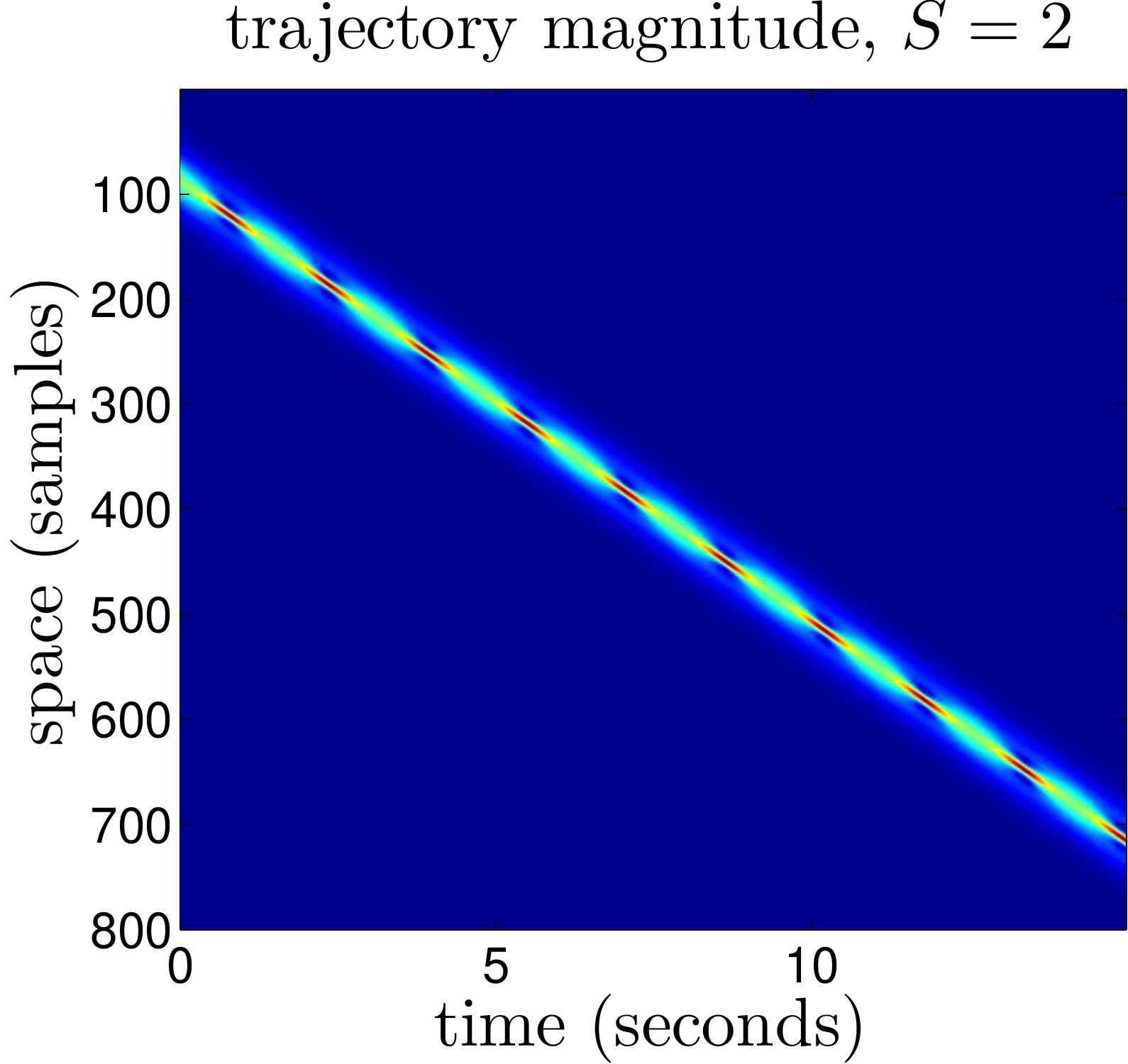} ~~
    \includegraphics[height=48mm]{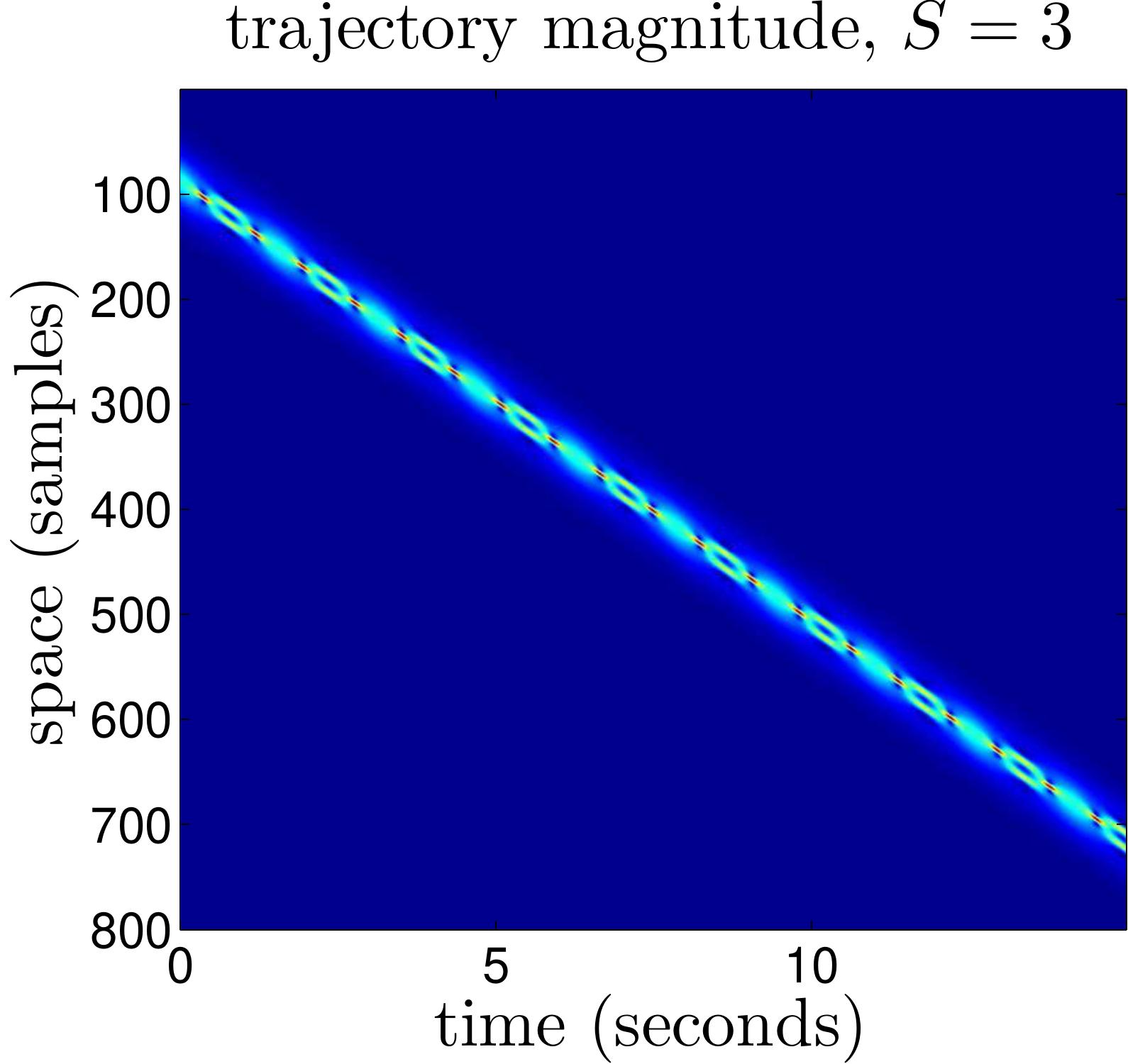}
	\end{center}
    \vspace{-5mm}
	\caption{\small\sl Magnitude of trajectory vectors for the Nonlinear Schr\"odinger system, with $S=1$, $S=2$, and $S=3$-soliton solutions displayed.}
	\label{fig:nlstraj}
\end{figure}

We begin by presenting a specific example involving the $S=2$-soliton solution. Figure~\ref{fig:nlssingledemo1}(a) plots a random projection\footnote{This projection, which is useful for obtaining a generic visualization of the trajectory, is computed by constructing a real-valued $3 \times N$ matrix $\Xi$ populated with independent zero-mean and unit-variance Gaussian random variables. For each data vector $x \in \mathbb{C}^N$, we compute $\Xi x$ and preserve the real part of the resulting vector.} of the data vectors from $\mathbb{C}^N$ to $\mathbb{R}^3$, and Figure~\ref{fig:nlssingledemo1}(b) shows the pairwise distances $\|x-y\|^2_2$ between all pairs $x,y\in\mathbb{C}^N$ on the trajectory. Here, we consider only the final $\sim 5$ seconds of the data; the initial $\sim 10$ seconds are used for populating delay coordinate vectors when needed.

\begin{figure}[t]
	\begin{center}
    (a) \includegraphics[width=75mm]{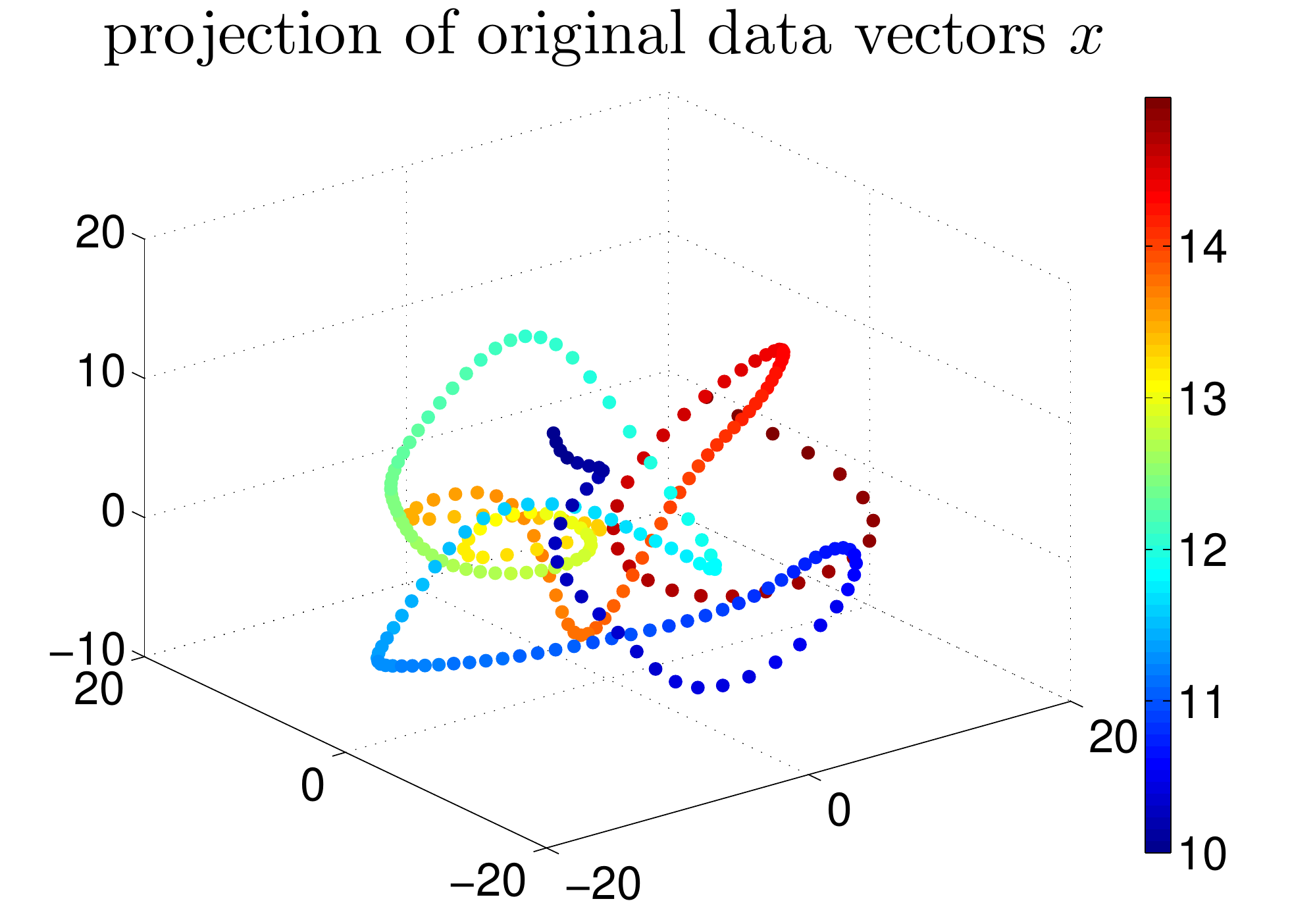} ~~
    (b) \includegraphics[width=70mm]{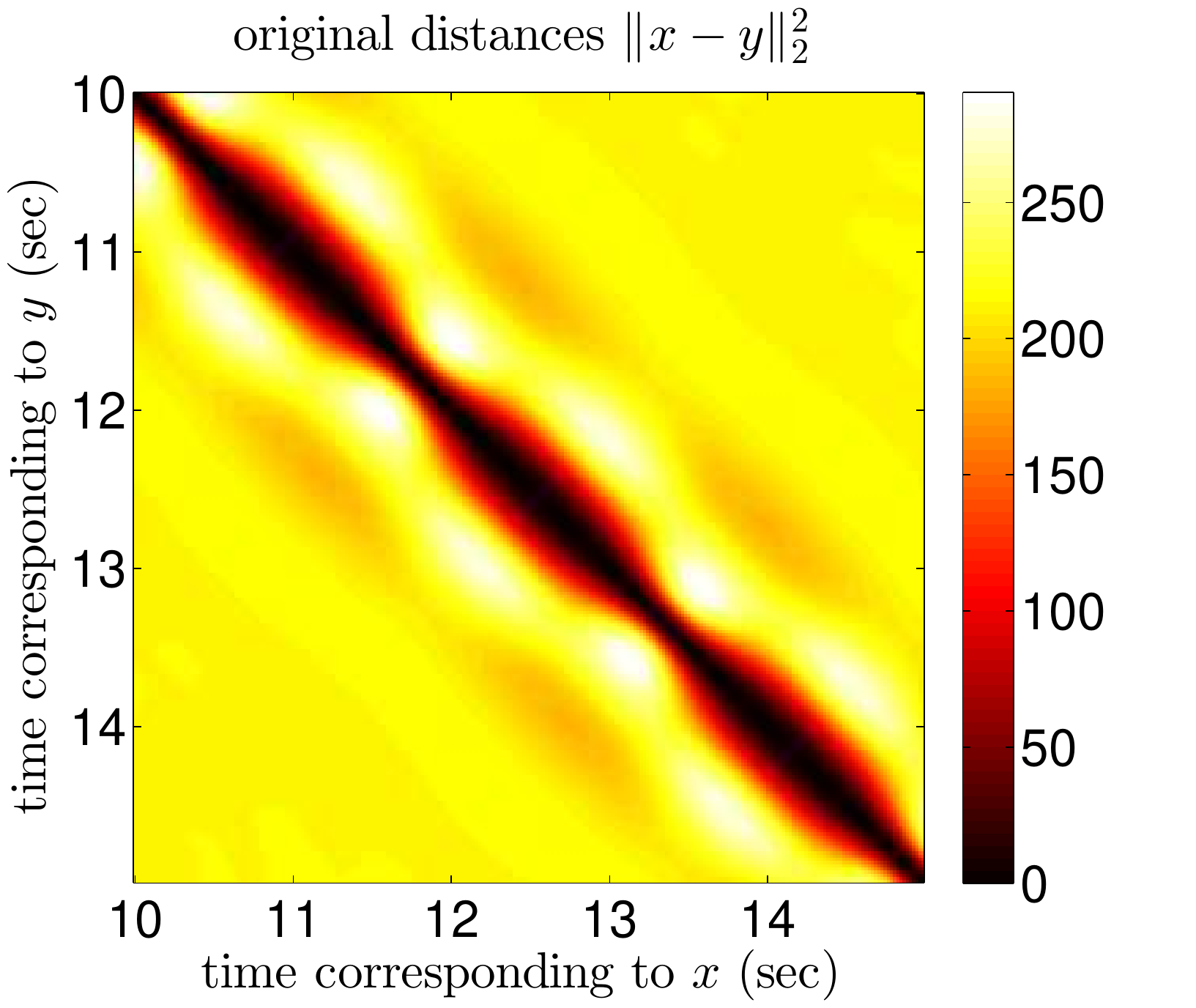} \\
    (c) \includegraphics[width=75mm]{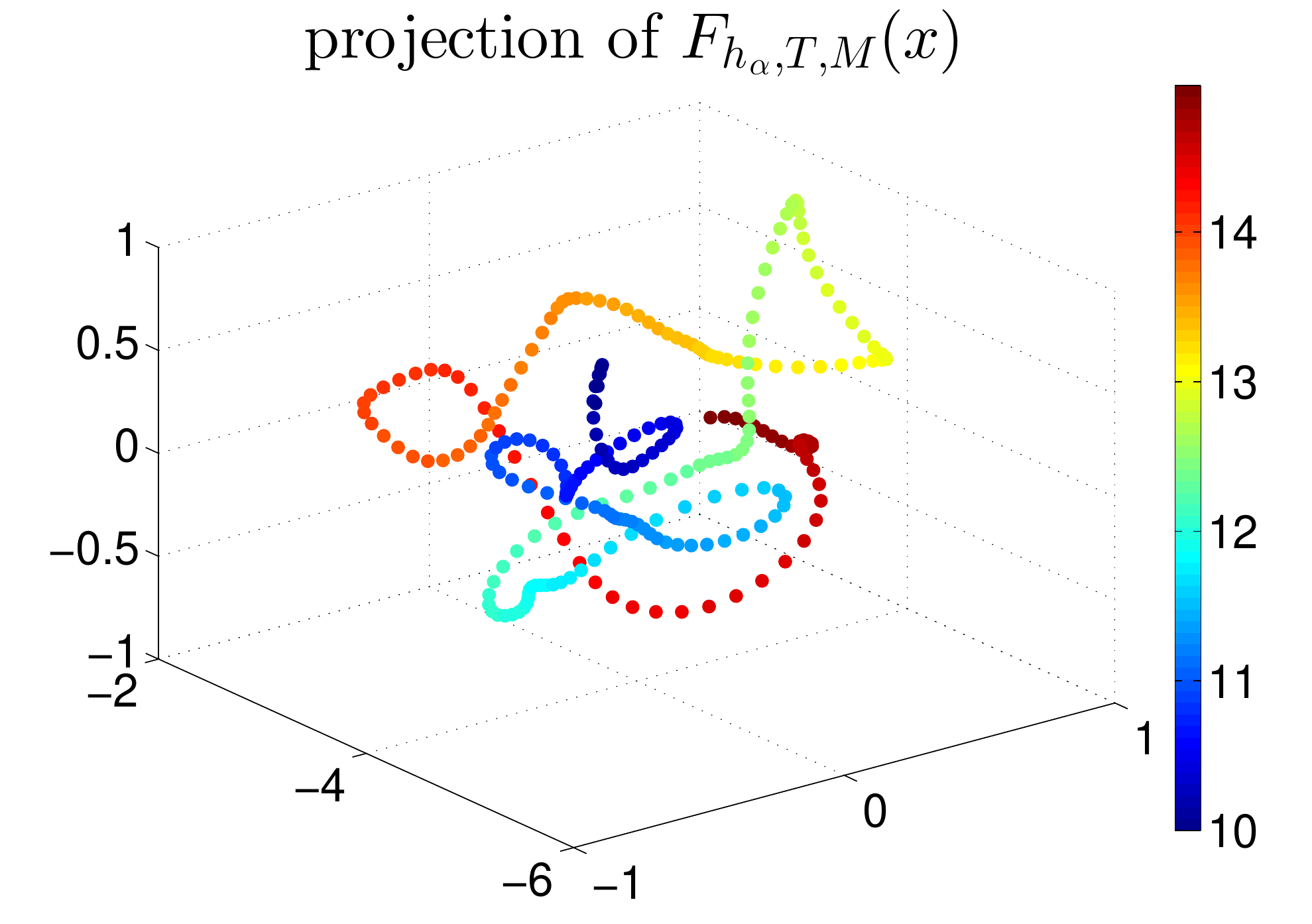} ~~
    (d) \includegraphics[width=70mm]{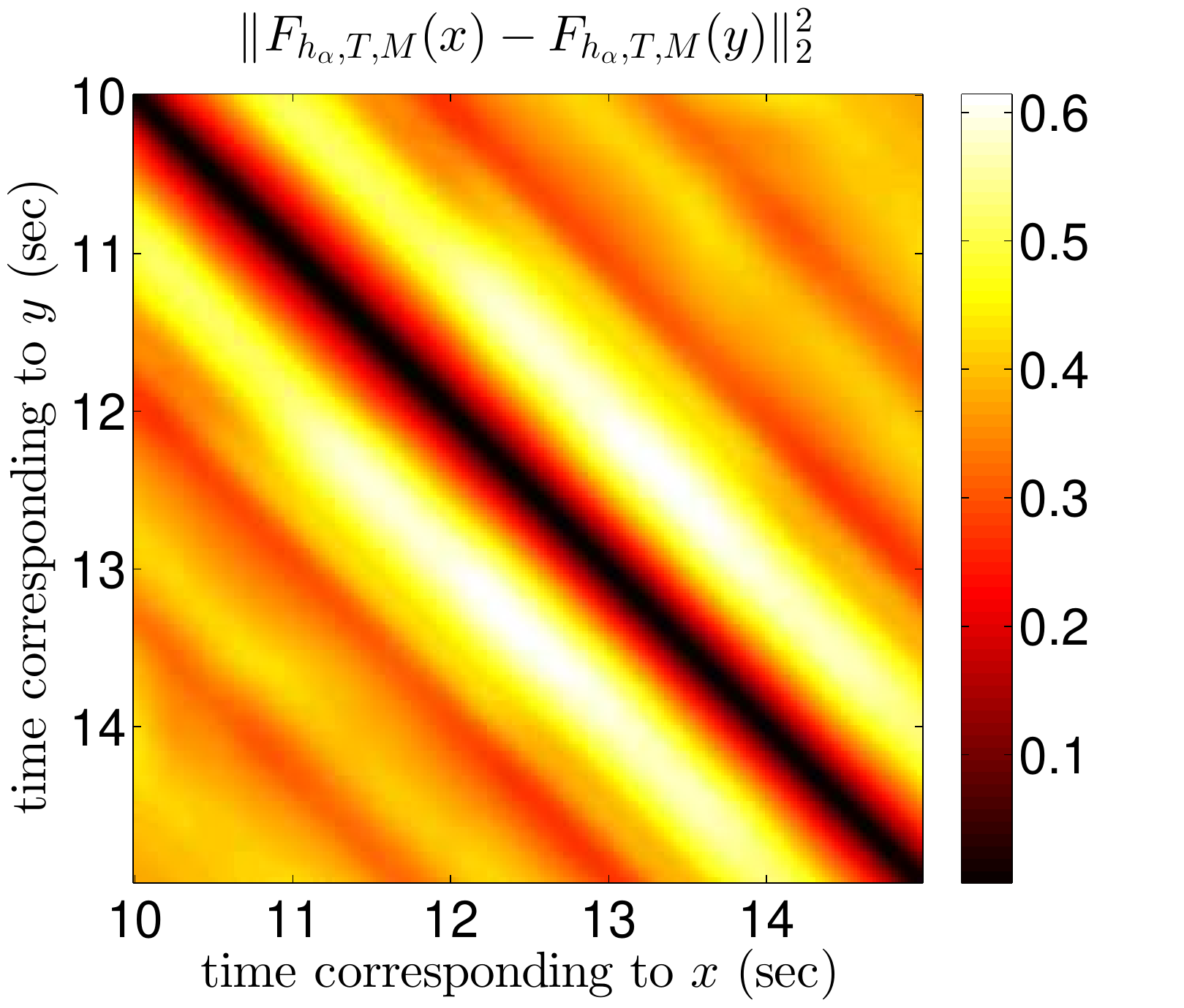}
	\end{center}
    \vspace{-5mm}
	\caption{\small\sl Embedding demonstration for the Nonlinear Schr\"odinger system, $S=2$-soliton solution with nonlinear RBF measurement functions. (a)~Visualization of data vectors $x \in\mathbb{C}^N$ on the trajectory, projected via a random linear map to $\mathbb{R}^3$. Color is used to indicate the time (in seconds) corresponding to each data vector $x$. (b)~Pairwise distances between all pairs $x,y\in\mathbb{C}^N$ on the trajectory. (c)~Visualization of the resulting delay coordinate vectors $F_{h_\alpha,T,M}(x)$, projected via a random linear map to $\mathbb{R}^3$. Again, color is used to indicate the time (in seconds) corresponding to each original data vector $x$. (d)~Pairwise distances $\|F_{h_\alpha,T,M}(x)-F_{h_\alpha,T,M}(y)\|^2_2$ between all points in the reconstruction space.}
	\label{fig:nlssingledemo1}
\end{figure}

In this example, we set $T = 0.06$ seconds and $M = 160$. To construct the class $\mathcal{H}$ of measurement functions, we consider a set of $P = 50$ nonlinear \emph{radial basis functions (RBFs)}, each defined by a center $v_p \in \mathbb{C}^{N}$ randomly chosen from a ball with radius comparable to the data set. The resulting measurement basis functions take the form $h_p(x) = e^{-\|x-v_p\|^2/2\sigma^2}$, where $\sigma$ is a scaling parameter chosen to be comparable to the norm of a typical data vector.

Figure~\ref{fig:nlssingledemo1}(c) shows a random projection of the resulting delay coordinate vectors $F_{h_\alpha,T,M}(x)$, where the entries of $\alpha$ are independent Gaussian random variables with zero mean and unit variance. Figure~\ref{fig:nlssingledemo1}(d) shows the pairwise distances $\|F_{h_\alpha,T,M}(x)-F_{h_\alpha,T,M}(y)\|^2_2$ between all points in the reconstruction space. Figure~\ref{fig:nlssingledemo2}(a) shows a scatter plot comparing the original distances $\|x-y\|_2^2$ between points on the trajectory to the corresponding distances $\|F_{h_\alpha,T,M}(x)-F_{h_\alpha,T,M}(y)\|_2^2$ in the reconstruction space $\mathbb{R}^M$. The dashed lines have slopes equal to the minimum and maximum observed values of the ratio $\|F_{h_\alpha,T,M}(x)-F_{h_\alpha,T,M}(y)\|_2^2/\|x-y\|_2^2$ over all pairs $x,y\in\mathbb{C}^N$ on the trajectory. Under a highly stable embedding (and in particular if the left and right hand sides of~\eqref{eq:final pre thm} were comparable to one another), the two lines in Figure~\ref{fig:nlssingledemo2}(a) would have slopes comparable to one another. In this experiment, the ratio of the larger slope to the smaller slope is approximately $8.80$. Up to some degree of approximation, small pairwise distances remain small, and large pairwise distances remain large.

\begin{figure}[t]
	\begin{center}
    (a) \includegraphics[width=70mm]{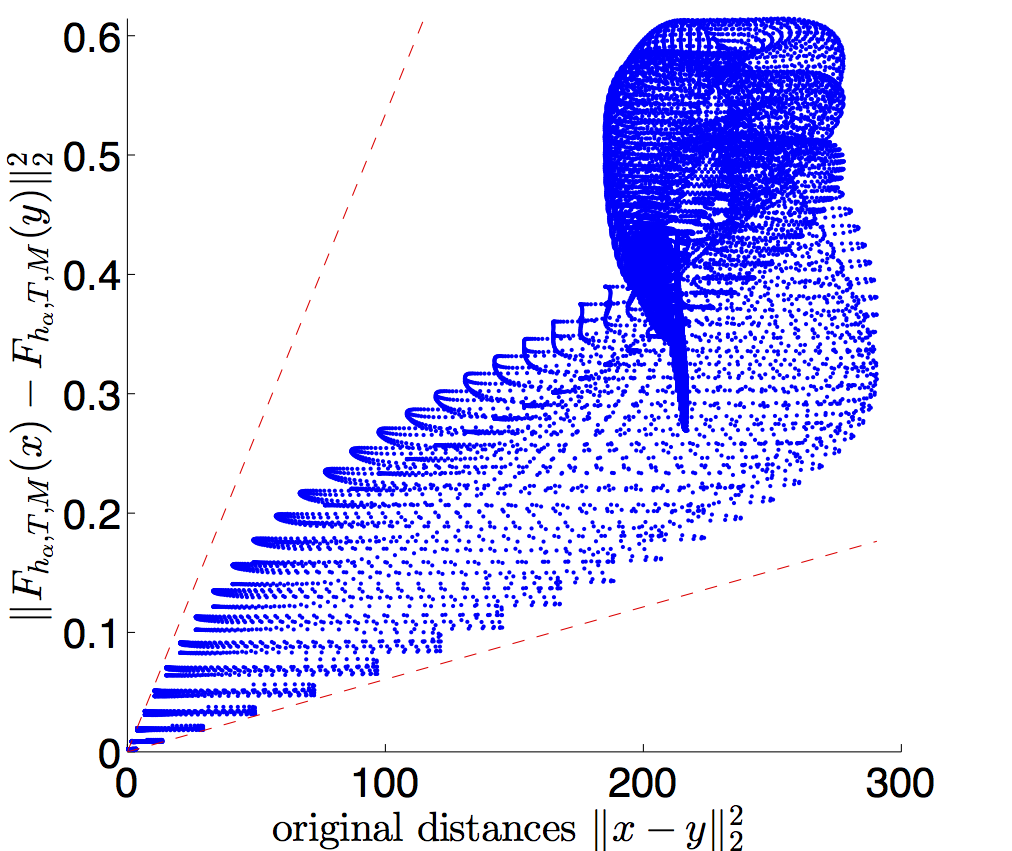} ~~
    (b) \includegraphics[width=70mm]{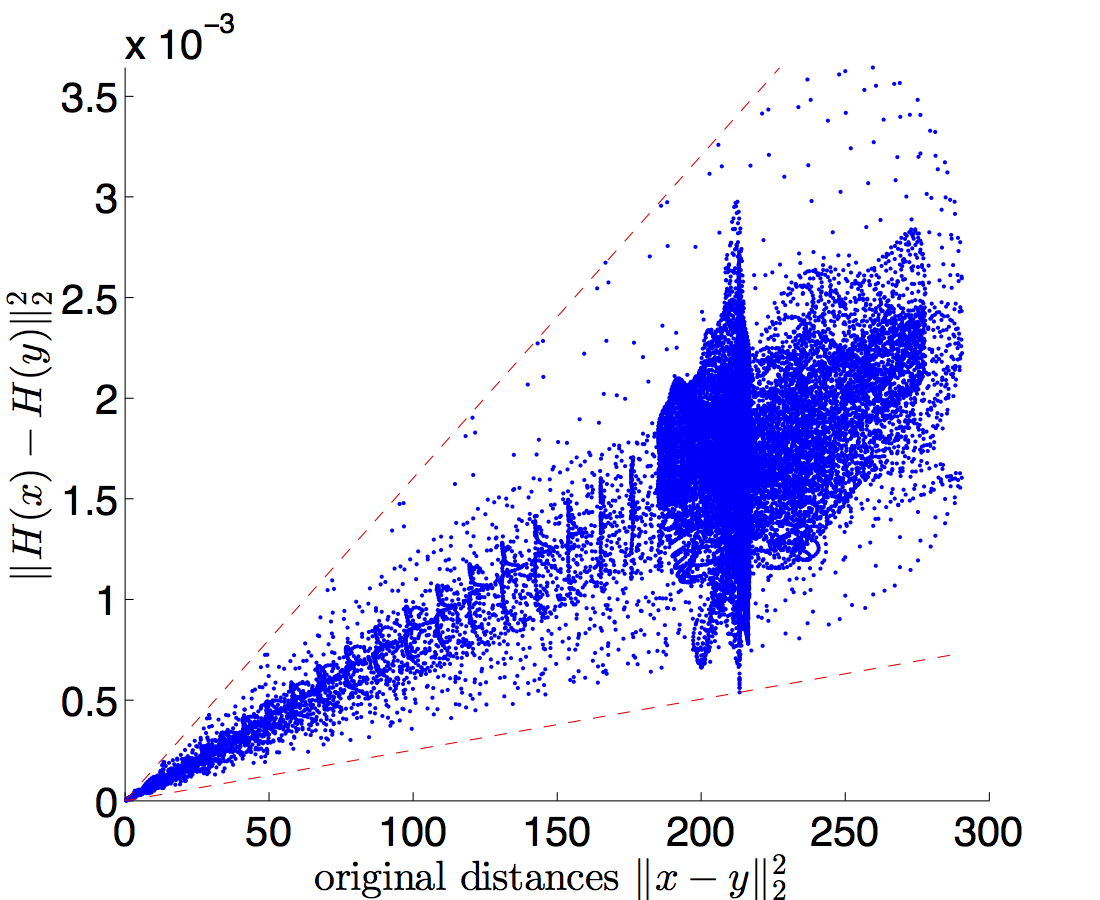} \\
    (c) \includegraphics[width=70mm]{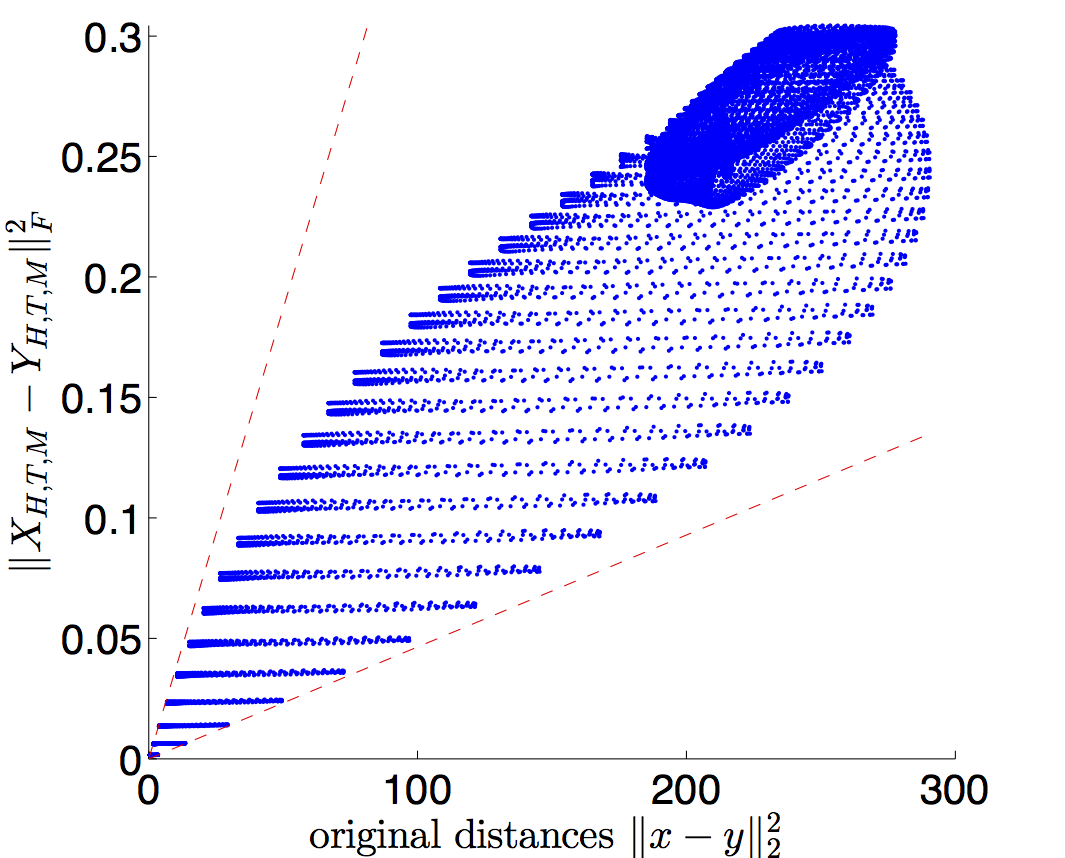} ~~
    (d) \includegraphics[width=70mm]{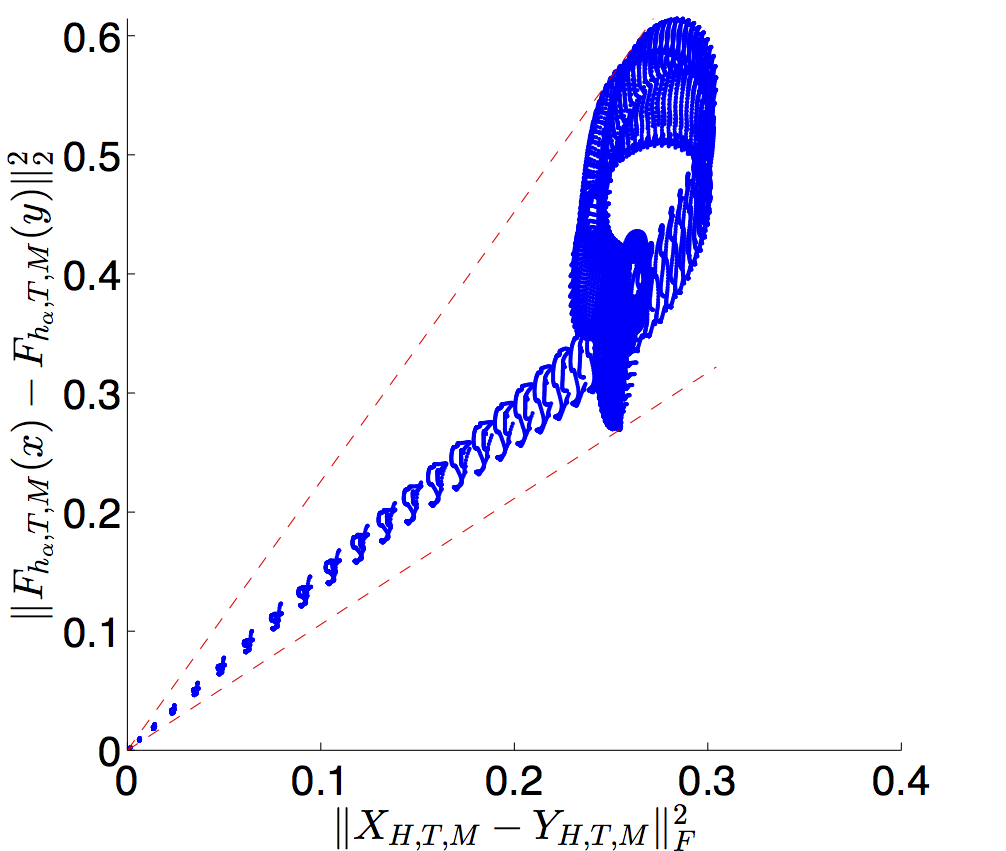}
	\end{center}
    \vspace{-5mm}
	\caption{\small\sl Pairwise distance preservation in the various stages of a delay coordinate embedding, $S=2$-soliton solution with nonlinear RBF measurement functions.}
	\label{fig:nlssingledemo2}
\end{figure}

We can unpack the factors that affect the degree of tightness in this embedding. The careful reader of Appendix~\ref{sec:proof_manifold} and especially~\eqref{eq:final 0} will note that the variability of the ratio $\|F_{h_\alpha,T,M}(x)-F_{h_\alpha,T,M}(y)\|_2^2/\|x-y\|_2^2$ is affected, in turn, by the variability of the ratios $\|H(x)-H(y)\|_2^2/\|x-y\|_2^2$ (see assumption A.1 in Section~\ref{sec:measurement} as well as~\eqref{eq:Hxvsx}), $\|X_{H,T,M}-Y_{H,T,M}\|_F^2/\|x-y\|_2^2$ (see~\eqref{eq:XHTMvsx}), and $\|F_{h_\alpha,T,M}(x)-F_{h_\alpha,T,M}(y)\|_2^2/\|X_{H,T,M}-Y_{H,T,M}\|_F^2$ (see~\eqref{eq:final pre}). Figures~\ref{fig:nlssingledemo2}(b), (c), and (d) show scatter plots corresponding to these three sets of pairwise distances, respectively. Variability in $\|H(x)-H(y)\|_2^2/\|x-y\|_2^2$ can be caused by a large ratio between $u_H$ and $l_H$; the ratio of the slopes in Figure~\ref{fig:nlssingledemo2}(b) is approximately $6.35$. Variability in $\|X_{H,T,M}-Y_{H,T,M}\|_F^2/\|x-y\|_2^2$ is affected not only by $u_H$ and $l_H$, but also by $\sigma_{\min}$, $\sigma_{\max}$, and $\operatorname{geo}(\mathbb{A})$. The ratio of the slopes in Figure~\ref{fig:nlssingledemo2}(c) is approximately $8.05$. Variability in $\|F_{h_\alpha,T,M}(x)-F_{h_\alpha,T,M}(y)\|_2^2/\|X_{H,T,M}-Y_{H,T,M}\|_F^2$ is affected by the stable rank $\operatorname{R}_{H,T,M}(\mathbb{A})$, which in this example is approximately $5.66$. The ratio of the slopes in Figure~\ref{fig:nlssingledemo2}(d) is approximately $2.14$. The tests below further reveal the causes and effects of changing the parameters we have discussed here.

To further our study, we also experiment with the $S=1$-soliton and $S=3$-soliton solutions, and we test additional classes $\mathcal{H}$ of measurement functions. In addition to the RBF kernel already considered, we also take $\mathcal{H}$ to be the space of all linear functionals on $\mathbb{C}^N$, as in Section~\ref{sec:second_example}. We also consider a set of nonlinear monomials of maximum degree $K$ in $N$ variables. Any such monomial can be written as $h_p(x) = x[1]^{\beta_1} \times x[2]^{\beta_2} \times \cdots \times x[N]^{\beta_N}$ for some $\{\beta_n\}_n$ with $\sum_n \beta_n \le K$. We use a set of $P = 200$ randomly-picked monomials with maximum degree $K = 3$.

With a fixed value of $T = 0.06$ seconds, Figure~\ref{fig:nlsmultidemo}(a) plots, as a function of $M$, the ratio of the largest and smallest isometry constants corresponding to $\|X_{H,T,M}-Y_{H,T,M}\|_F^2/\|x-y\|_2^2$. (In the previous example, this corresponded to the ratio of the slopes in Figure~\ref{fig:nlssingledemo2}(c), which was approximately $8.05$.) Figure~\ref{fig:nlsmultidemo}(b) shows the corresponding plot for $\|F_{h_\alpha,T,M}(x)-F_{h_\alpha,T,M}(y)\|_2^2/\|X_{H,T,M}-Y_{H,T,M}\|_F^2$, and Figure~\ref{fig:nlsmultidemo}(c) shows the corresponding plot for $\|F_{h_\alpha,T,M}(x)-F_{h_\alpha,T,M}(y)\|_2^2/\|x-y\|_2^2$, reflecting the tightness of the overall embedding. Figure~\ref{fig:nlsmultidemo}(d) shows the stable rank $\operatorname{R}_{H,T,M}(\mathbb{A})$ as a function of $M$. In these plots, we see several general trends:
 \begin{itemize}
 \item The overall embedding is generally tightest for the $S=1$-soliton solution and weakest for the $S=3$-soliton solution. As illustrated in Figure~\ref{fig:nlstraj}, the complexity of the trajectories generally increases for larger values of $S$. For example, the trajectory for $S=1$ has constant speed, while the instantaneous speed of the trajectory when $S=3$ varies over a dynamic range of approximately $6.95$. This variability affects factors such as $\sigma_{\min}$ and $\sigma_{\max}$, leading to more variability in $\|X_{H,T,M}-Y_{H,T,M}\|_F^2/\|x-y\|_2^2$ as shown in Figure~\ref{fig:nlsmultidemo}(a). There is relatively little effect of $S$ on $\|F_{h_\alpha,T,M}(x)-F_{h_\alpha,T,M}(y)\|_2^2/\|X_{H,T,M}-Y_{H,T,M}\|_F^2$ as shown in Figure~\ref{fig:nlsmultidemo}(b).
 \item The linear measurement functions generally result in the tightest embeddings; partly this is due to the fact that $l_H = u_H = 1$ in the linear case. The nonlinear monomial functions produce the loosest embeddings. However, the nonlinear RBF functions perform nearly as well as the linear functions.
 \item In general, as $M$ increases, the stable rank increases $\operatorname{R}_{H,T,M}(\mathbb{A})$, which reduces the variability of $\|F_{h_\alpha,T,M}(x)-F_{h_\alpha,T,M}(y)\|_2^2/\|X_{H,T,M}-Y_{H,T,M}\|_F^2$ and thus of the overall embedding $\|F_{h_\alpha,T,M}(x)-F_{h_\alpha,T,M}(y)\|_2^2/\|x-y\|_2^2$. This is as expected in light of Theorem \ref{thm:manifold2}.
 \end{itemize}

\begin{figure}[p]
	\begin{center}
    (a) \includegraphics[width=70mm]{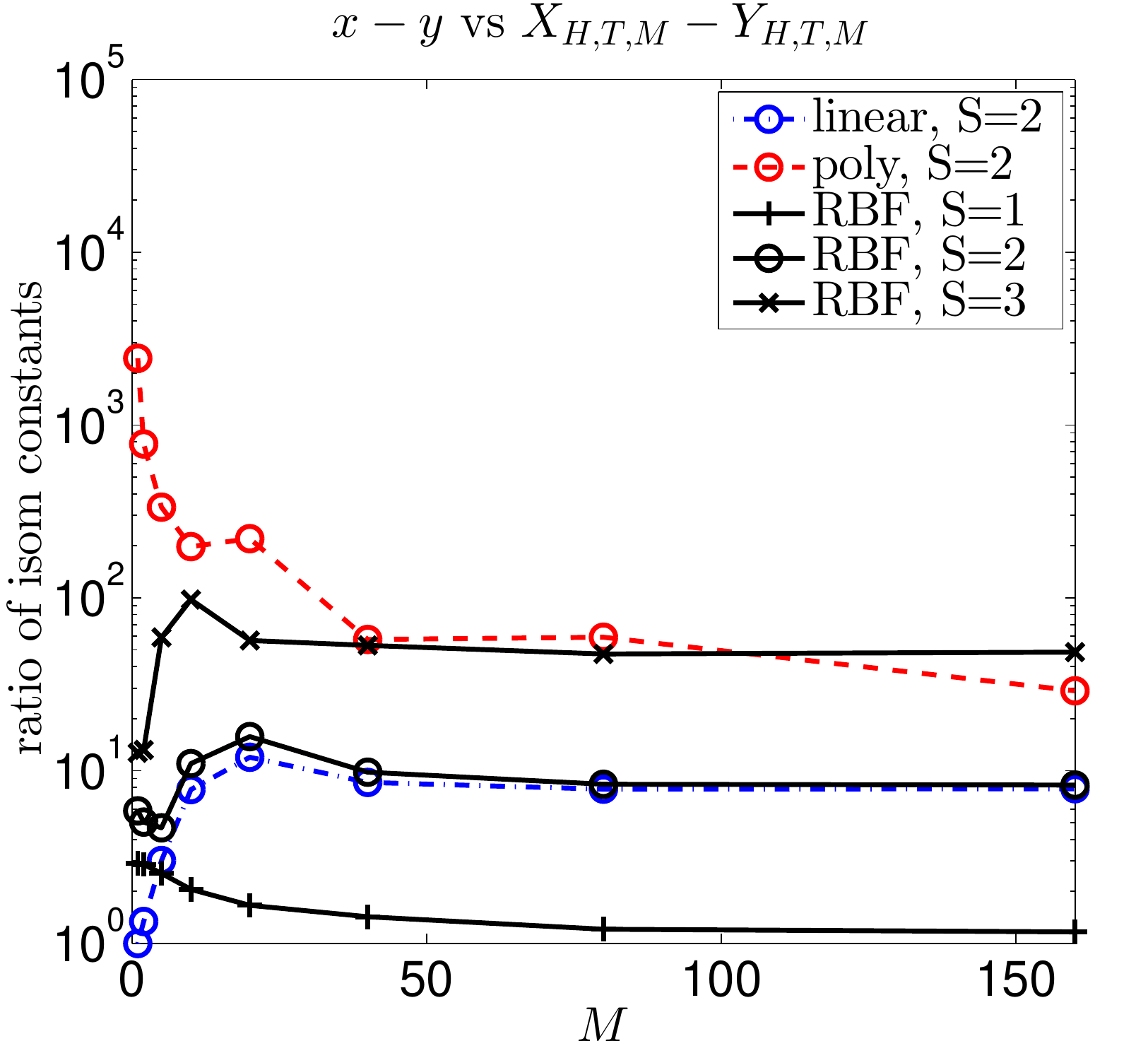} ~~
    (b) \includegraphics[width=70mm]{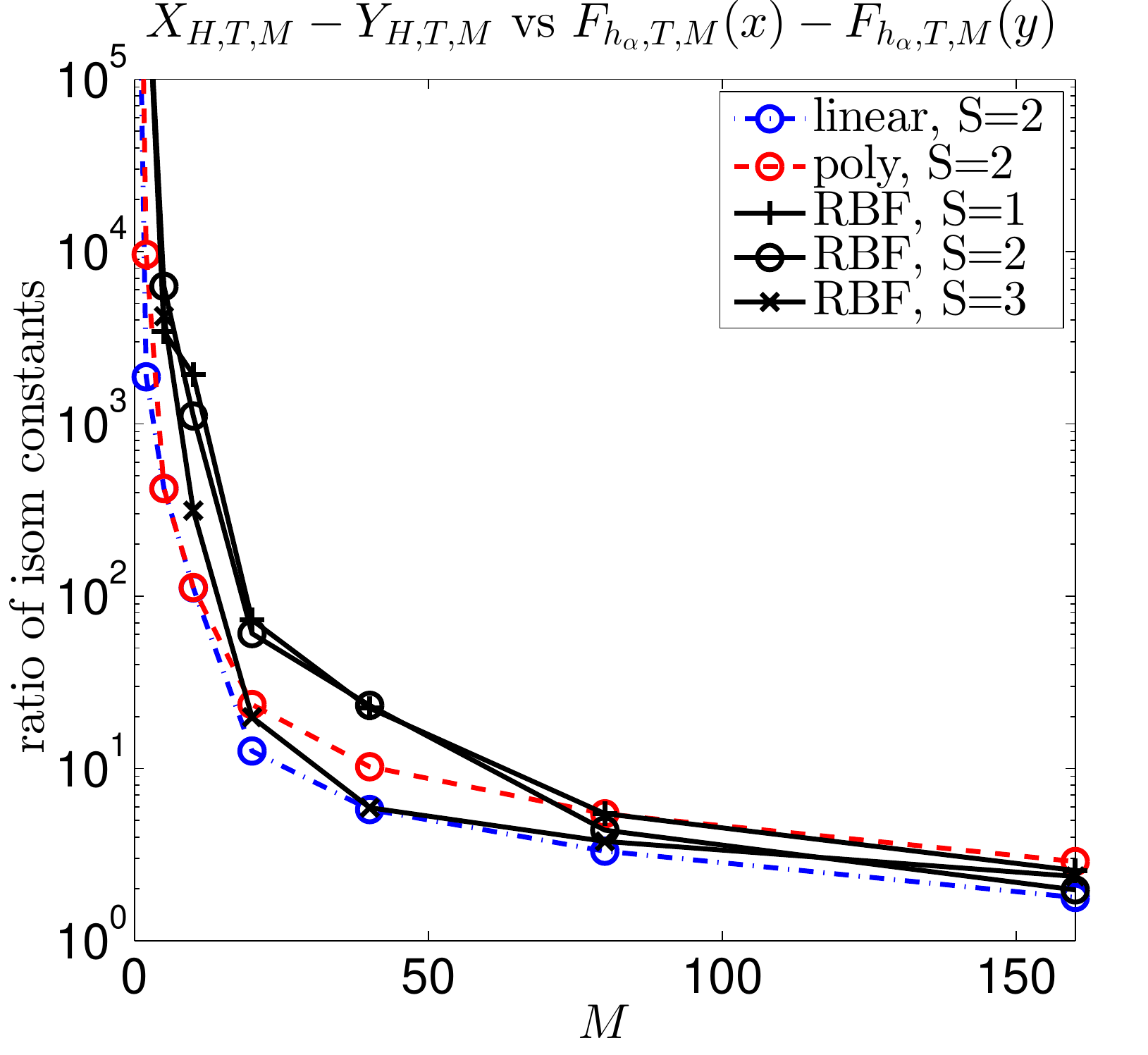} \\
    (c) \includegraphics[width=70mm]{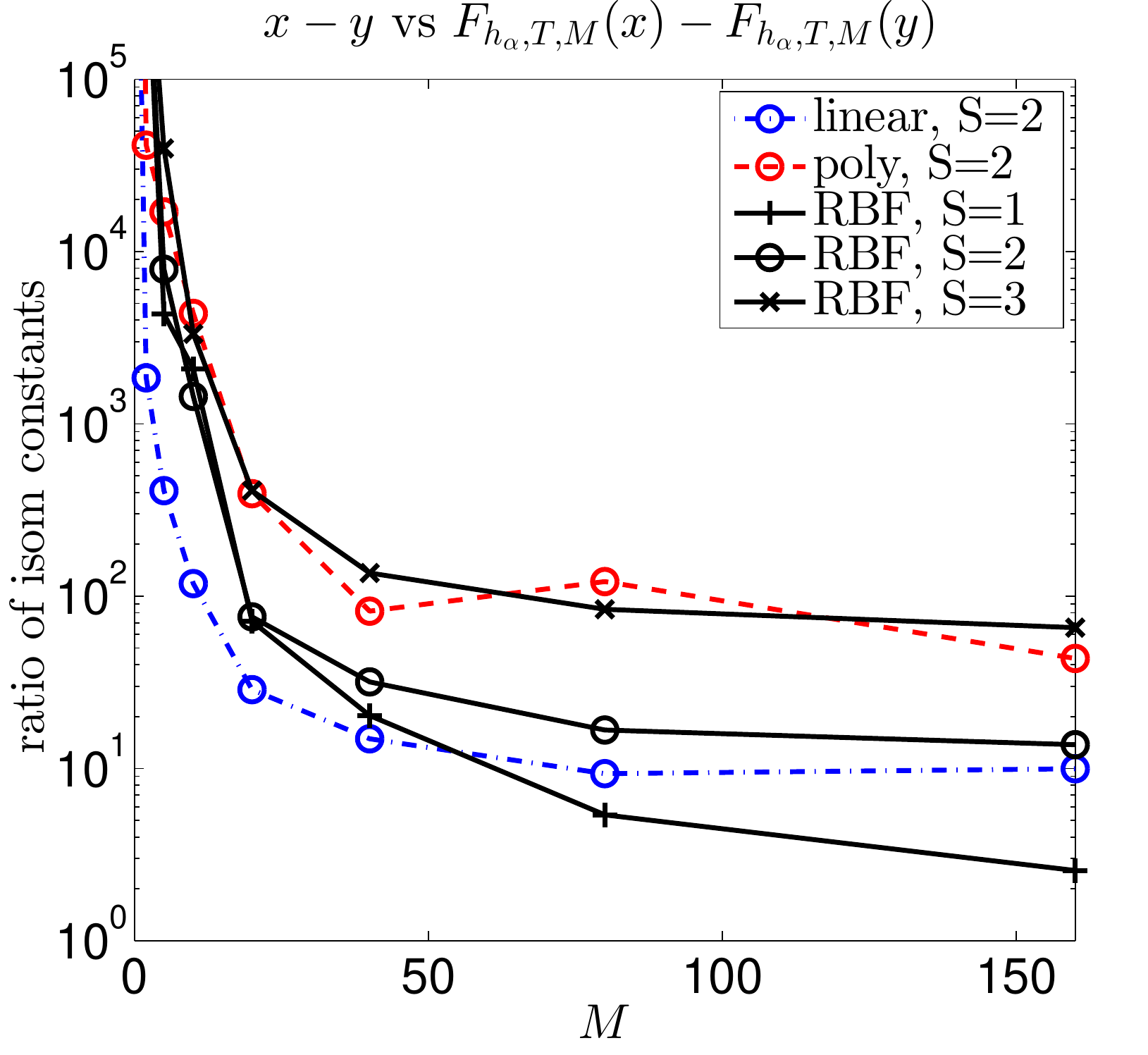} ~~
    (d) \includegraphics[width=70mm]{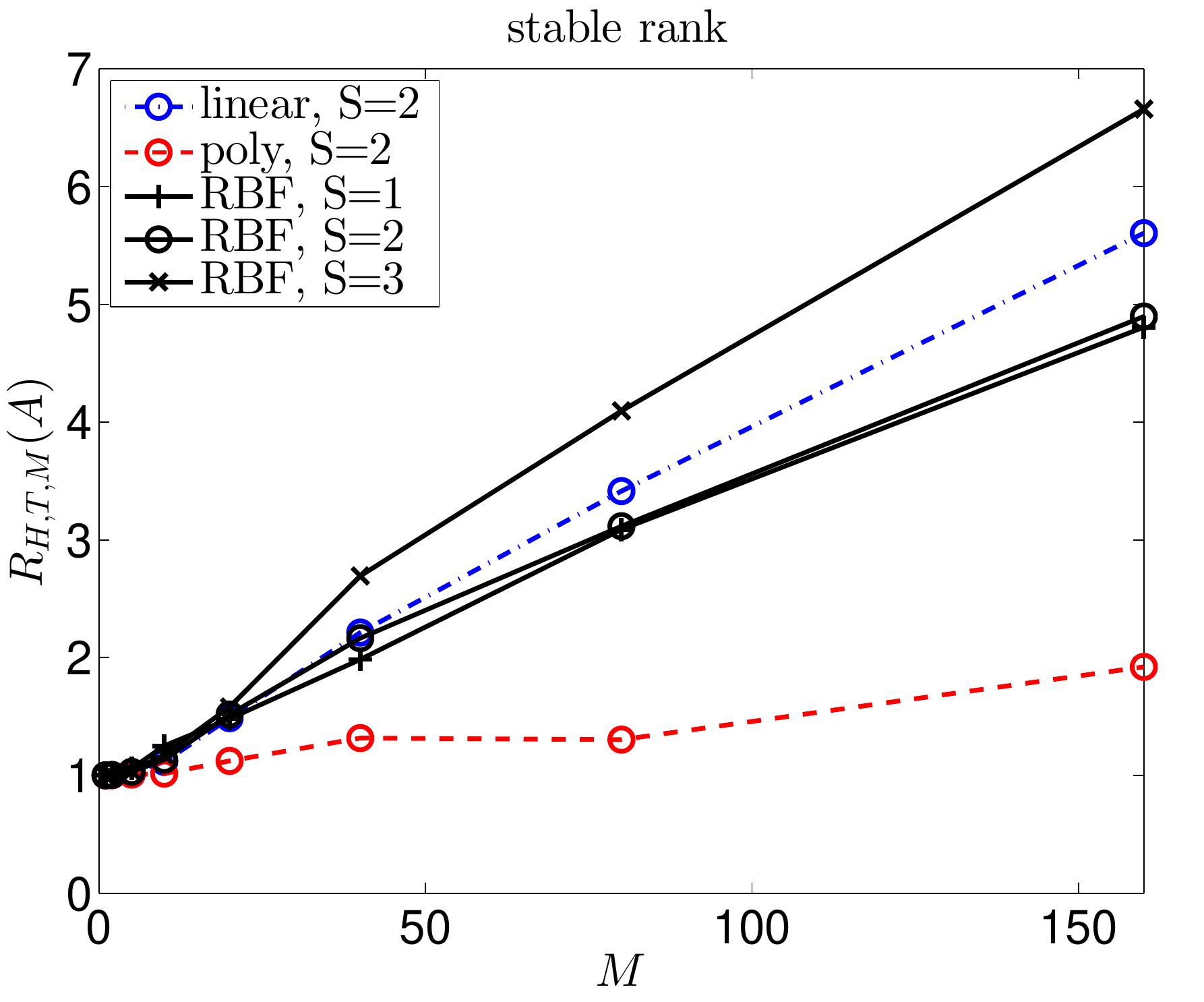} \\
    (e) \includegraphics[width=70mm]{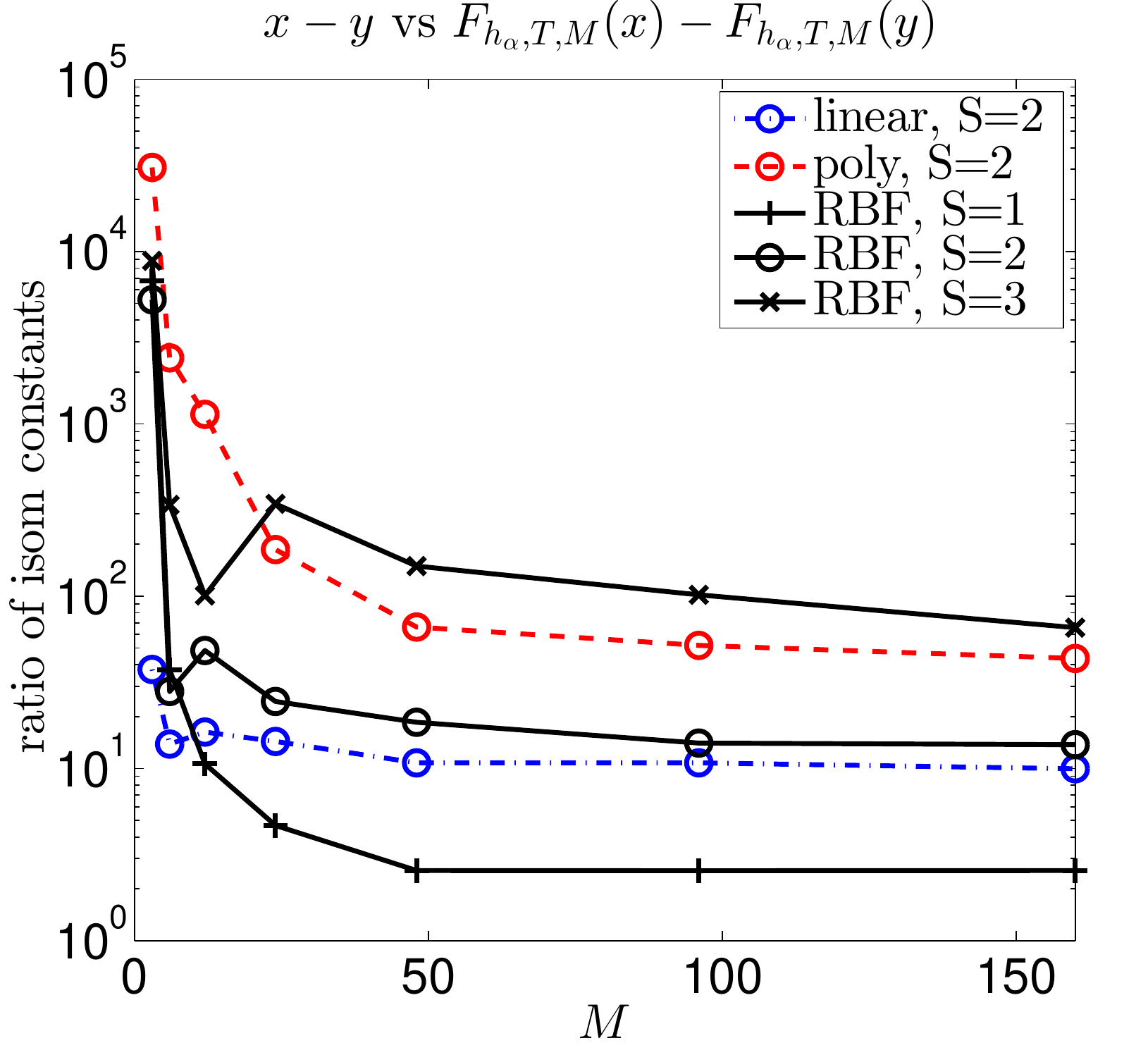} ~~
    (f) \includegraphics[width=70mm]{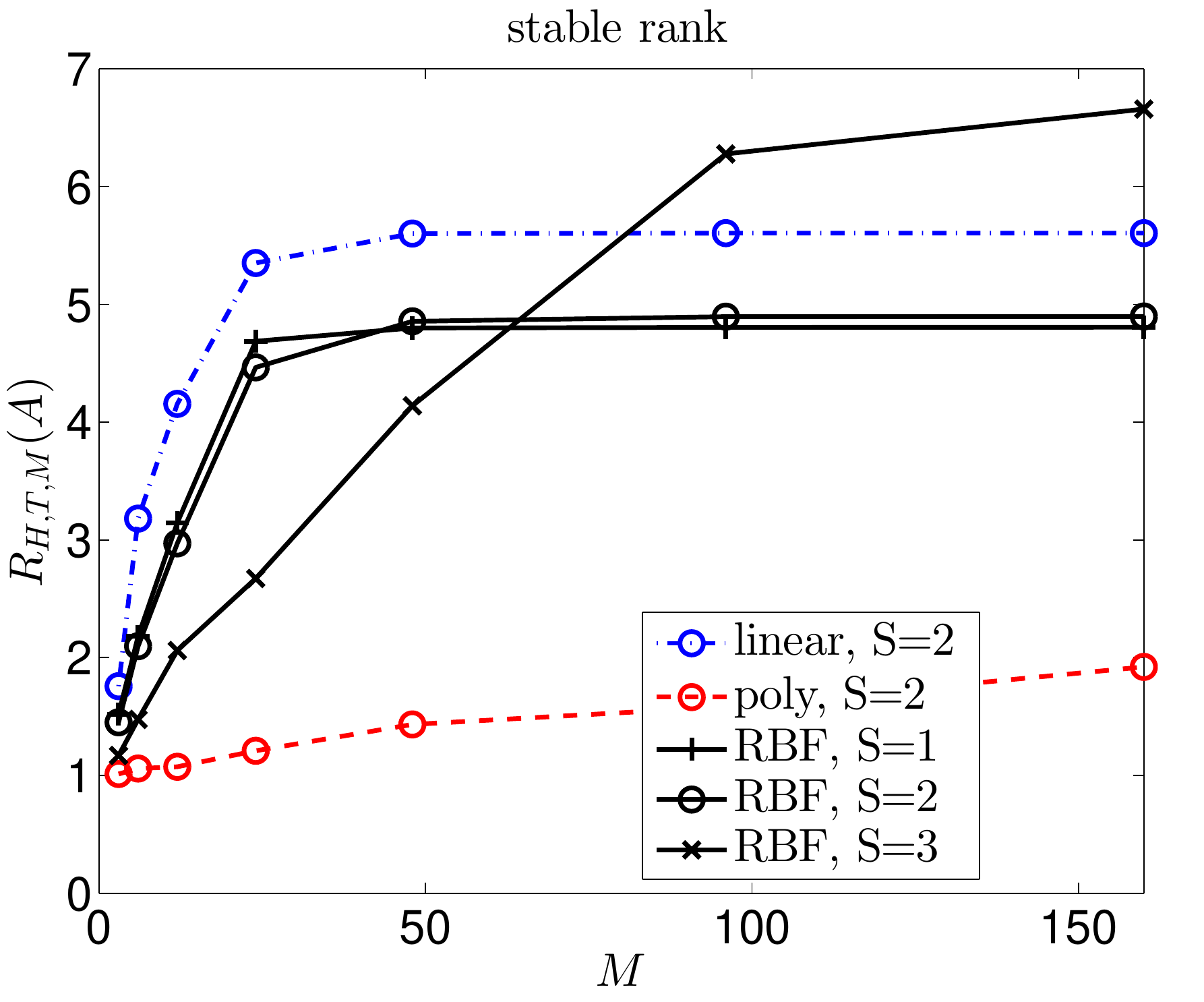}
	\end{center}
    \vspace{-5mm}
	\caption{\small\sl Embedding demonstration for the Nonlinear Schr\"odinger system, including the $S=1$, $S=2$, and $S=3$-soliton solutions and various linear and nonlinear measurement functions. (a)--(d) $T$ fixed to $0.06$ seconds. (e)--(f) $MT$ fixed to $9.6$ seconds.}
	\label{fig:nlsmultidemo}
\end{figure}

Finally, over a fixed total delay of $9.6$ seconds, we experiment with a range of $M$ values. In each case, we choose $T = 9.6/M$. Results are shown in Figure~\ref{fig:nlsmultidemo}(e),(f). These results show that, over this total amount of time it is not necessary to sample densely in time; moderately small values of $M$ (around $40$), corresponding to moderately large values of $T$ (around $0.24$ seconds) lead to delay coordinate embeddings with a reasonable degree of tightness.

\section{Conclusions and Open Problems}
\label{sec:conclusion}

The main result of this paper extends Takens' celebrated embedding theorem to provide conditions when a delay-coordinate map can provide a stable embedding of a dynamical system attractor.  Given the prevalence of these techniques in nonlinear time-series analysis, this result provides a much needed theoretical justification of their numerical performance in the presence of real-world imperfections such as noise and quantization.  While the conditions of this result are restrictive and it may not always be possible to meet them in practice, there is significant value in knowing for which scenarios one can guarantee a given quality level of the embedding.  In fact, researchers have informally conjectured that instability issues may limit the performance of numerical techniques based on delay-coordinate mapping without the theoretical foundations to examine this issue formally (e.g., see the discussion regarding Takens' theorem in the Supporting Online Material of~\cite{jaeger.04}).  The examination of our results has also led to new and insightful interpretations of classical (generally heuristic) techniques for selecting parameters such as the sampling time and number of delays.

Building on these results, there appears to be no shortage of interesting directions for future work. For example:
\begin{itemize}
\item Remark \ref{rem:choice of T M} and Table~\ref{table:prescription} provide a recipe for choosing the sampling interval $T$ and the number of delays $M$ in delay-coordinate mapping. It is of interest to experimentally validate this procedure and perhaps find  alternatives with lower  computational complexity.
\item An open question is whether it is possible to improve (increase) the stable rank of a dynamical system (and hence improve the quality of delay-coordinate mapping) by optimizing over the choice of scalar measurement functions. While we suspect that answer is negative, a rigorous study of this topic does not currently exist.
\item A remaining technical challenge is the role of $\mbox{rch}(\mathbb{A}_{H,T,M})$ (reach of the ``trajectory attractor'') in Theorem \ref{thm:manifold2} (also see \eqref{eq:traj attr}). We  suspect that $\mbox{rch}(\mathbb{A}_{H,T,M})$ can be expressed entirely in terms of $\mbox{rch}(\mathbb{A})$ (and $T$, $M$, and basis functions $\{h_p\}$). Such an expression will substantially simplify and clarify Theorem \ref{thm:manifold2} but has remained elusive despite our efforts.
\item While multivariate time-series have been occasionally discussed in the literature (e.g.,~\cite{cao1998dynamics}), as with our work, most treatments of delay-coordinate maps are restricted to a single scalar measurement function.  An open question is how the presence of multiple measurement functions (producing diverse observations at each sampling time step) would affect the stability of the attractor embedding.
\end{itemize}

\section*{Acknowledgments}
AE and MBW acknowledge Liz Bradley at CU Boulder for helpful discussions. AE also thanks Joshua Garland. Part of this research was conducted when AE was a graduate fellow at the  Statistical and Applied Mathematical Sciences Institute (SAMSI) and later a visitor at  the Institute for Computational and Experimental Research in Mathematics (ICERM). AE is grateful for their hospitality and kindness.

\appendix

\section{Proof of Theorem \protect\ref{thm:manifold2} (Stable Takens' Theorem)}
\label{sec:proof_manifold}

We reserve the letters $C,C_1,C_2,\cdots$ to represent universal positive constants. We adopt the following (semi-)order: $a\lesssim b$ means that there is an absolute constant $\Cl{notation1}$ such that $a\le\Cr{notation1}b$. If, instead of being an absolute constant, $\Cr{notation1}=\Cr{notation1}(\theta)$  depends on some parameter $\theta$, we write $a\lesssim_{\theta}b$. Of course, $a\gtrsim b$ and $a\gtrsim_{\theta}b$ are defined similarly.  Occasionally, we will use the convention that $[a:b]={a,a+1,\cdots,b}$ for integers $a\le b$.

\label{p:proof}
Throughout the proof, the dependence on different quantities might be suppressed if there is no ambiguity. Consider $x\in\mathbb{A}$ and scalar measurement function $h_\alpha(\cdot)=\sum_p \alpha[p]\cdot h_p(\cdot)$ as a linear combination of basis functions. Recall from \eqref{eq:conn btw G and F} and \eqref{eq:other def of XM} that the corresponding delay vector  can be written as
\begin{equation*}
F_{h_\alpha,T,M}(x) = X_{H,T,M}\cdot \alpha,
\end{equation*}
\begin{equation}\label{eq:recalls XM}
X_{H,T,M} =
\left[
\begin{array}{cccc}
H(x) &
H\left(\phi_{T}^{-1}(x)\right) &
\cdots &
H\left(\phi_T^{-M+1}(x)\right)
\end{array}
\right]^*\in\mathbb{R}^{M\times P},
\end{equation}
and where $\phi_T:\mathbb{A}\rightarrow \mathbb{A}$ is the flow on the attractor. For a fixed pair of points $x,y\in\mathbb{A}$, consider the random variable
\begin{equation}\label{eq:main ratio}
\frac{\left\| F_{h_\alpha,T,M}(x)- F_{h_\alpha,T,M}(y) \right\|_2^2}{\left\| X_{H,T,M}-Y_{H,T,M}\right\|_F^2} =
\frac{\left\| X_{H,T,M}\cdot\alpha-Y_{H,T,M}\cdot\alpha \right\|_2^2}{\left\| X_{H,T,M}-Y_{H,T,M}\right\|_F^2},
\end{equation}
and note that
\begin{equation*}
\mathbb{E} \left[ \frac{\left\| F_{h_\alpha,T,M}(x)- F_{h_\alpha,T,M}(y) \right\|_2^2}{\left\| X_{H,T,M}-Y_{H,T,M}\right\|_F^2} \right] =
\frac{\mathbb{E}\left[\left\| \left( X_{H,T,M}-Y_{H,T,M}\right) \cdot\alpha \right\|_2^2\right]}{\left\| X_{H,T,M}-Y_{H,T,M}\right\|_F^2} = 1,
\end{equation*}
where the second identity holds because the entries of $\alpha\in\mathbb{R}^P$ are independent and have unit-variance. This suggests that for \emph{all} pairs of points in $\mathbb{A}$, the  ratio in \eqref{eq:main ratio} might be  close to one. That is, we hope that the following quantity is small with overwhelming probability:
\begin{equation}\label{eq:goal}
\sup_{x,y\in\mathbb{A}} \left| \frac{\left\| F_{h_\alpha,T,M}(x)-F_{h_\alpha,T,M}(y) \right\|_2^2}{\left\| X_{H,T,M}-Y_{H,T,M} \right\|_F^2} -1 \right| =
\sup_{Z\in \mathfrak{Z}} \left| \left\| Z \alpha \right\|_2^2- \mathbb{E}\left[ \left\| Z\alpha \right\|_2^2 \right]\right|.
\end{equation}
Above, we conveniently set
\begin{equation}\label{eq:frak Z}
\mathfrak{Z}:= \left\{ \frac{X_{H,T,M}-Y_{H,T,M}}{\left\| X_{H,T,M}-Y_{H,T,M} \right\|_F}\,:\, x,y\in\mathbb{A} \right\}\subset\mathbb{R}^{M\times P}.
\end{equation}
To control the supremum in \eqref{eq:goal}, we invoke a recent result by Krahmer et al.

\begin{prop}\label{prop:Krahmer}
\emph{\cite[Theorem 3.1]{Krahmer2014}} For integers $M$ and $P$, let $\mathfrak{Z}\subset\mathbb{R}^{M\times P}$ be a collection of matrices. Moreover, let $\alpha\in\mathbb{R}^P$ be a random vector whose entries are independent zero-mean, unit-variance random variables  with subgaussian norm of $\theta$.
Set
\begin{equation*}
d_F(\mathfrak{Z}) = \sup_{Z\in\mathfrak{Z}} \|Z\|_F,\qquad
d_2(\mathfrak{Z}) = \sup_{Z\in\mathfrak{Z}} \|Z\|,
\end{equation*}
where $\|\cdot\|_F$ and $\|\cdot\|$ stand for the Frobenius and spectral  norms, respectively. Also let $\gamma_2(\mathfrak{Z},\|\cdot\|)$ be the Gaussian width of $\mathfrak{Z}$ with respect to the spectral metric,  and define
\begin{equation*}
E_1 := \gamma_2 \left(\mathfrak{Z},\|\cdot\| \right) \cdot \left( \gamma_2\left(\mathfrak{Z},\|\cdot\| \right)+ d_F(\mathfrak{Z}) \right) +  d_F(\mathfrak{Z}) \cdot d_2(\mathfrak{Z}) ,
\end{equation*}
\begin{equation*}
E_2 := d_2^2(\mathfrak{Z}),
\end{equation*}
\begin{equation*}
E_3 := d_2(\mathfrak{Z})\cdot \left(\gamma_2\left(\mathfrak{Z},\|\cdot\| \right) +d_F(\mathfrak{Z})\right).
\end{equation*}
Then, for arbitrary $v>0$, it holds that
\begin{equation*}
\mathbb{P} \left[  \sup_{Z\in\mathfrak{Z}} \left| \left\|Z \alpha \right\|_2^2- \mathbb{E}\left[\left\|Z \alpha\right\|_2^2 \right]\right|> \Cl{Krahmer}(\theta) \cdot E_1+v \right] \le 2 \exp\left( -\Cl{Krahmer2}(\theta)\cdot \min \left[\frac{v}{E_2},\frac{v^2}{E_3^2} \right] \right),
\end{equation*}
where $\Cr{Krahmer}(\theta)$ and $\Cr{Krahmer2}(\theta)$ depend only on $\theta$.
\end{prop}
Without dwelling too much on the concept of Gaussian width above, we recall the following well-known relation \cite{Talagrand2006}:
\begin{equation}\label{eq:Gaussian width ineq}
\gamma_2\left(\mathfrak{Z},\|\cdot\| \right) \lesssim \int_0^\infty \sqrt{\log \left(\# \left(\mathfrak{Z},\|\cdot\|,s  \right)\right)}\, ds.
\end{equation}
Above, $\#(\mathfrak{Z},\|\cdot\|,s)$ is the \emph{covering number} of the set $\mathfrak{Z}$ with respect to the metric $\|\cdot\|$ and at scale $s>0$. That is, $\#(\mathfrak{Z},\|\cdot\|,s)$ is the smallest number of balls of radius $s$ (and with respect to the metric $\|\cdot\|$) needed to cover the set $\mathfrak{Z}$.
In order to apply Proposition \ref{prop:Krahmer} to \eqref{eq:goal}, we must first calculate $d_F(\mathfrak{Z})$, $d_2(\mathfrak{Z})$, and $\#(\mathfrak{Z},\|\cdot\|,s)$ (with $\mathfrak{Z}$ as in \eqref{eq:frak Z}). That, we set out to do next. Observe that
\begin{equation}\label{eq:bnd on dF}
d_F\left( \mathfrak{Z} \right) = \sup_{Z\in\mathfrak{Z}}\|Z\|_F = \sup_{x,y\in\mathbb{A}} \frac{\left\| X_{H,T,M}-Y_{H,T,M}\right\|_F}{\left\| X_{H,T,M}-Y_{H,T,M}\right\|_F} = 1,
\qquad \mbox{(see \eqref{eq:frak Z})}
\end{equation}
\begin{equation}\label{eq:bnd on d2}
d_2\left( \mathfrak{Z} \right) = \sup_{Z\in\mathfrak{Z}}\|Z\|=  \sup_{x,y\in\mathbb{A}} \frac{\left\| X_{H,T,M}-Y_{H,T,M}\right\|}{\left\| X_{H,T,M}-Y_{H,T,M}\right\|_F} = \frac{1}{\sqrt{\operatorname{R}_{H,T,M}\left(\mathbb{A}\right)}}.\qquad \mbox{(see \eqref{eq:s rank of h(A)})}
\end{equation}
Estimating the covering number of $\mathfrak{Z}$ is more involved. From the order between norms $\|\cdot\|\le \|\cdot\|_F$, first deduce that
\begin{equation}\label{eq:spect to frob}
\#\left( \mathfrak{Z} ,\|\cdot\|,s\right) \le \#\left( \mathfrak{Z},\|\cdot\|_F,s\right),\qquad \forall s>0.
\end{equation}
The covering number on the right hand side above is easier to control, as $\mathfrak{Z}\subset{\mathbb{R}^{M\times P}}$  is isometric to another (more malleable) object that we denote here with $U(\mathbb{A}_{H,T,M})$ and define next.
Set
\begin{equation}\label{eq:def of x_h,M,T}
 x_{H,T,M} = x_{H,T,M}(x) := \mbox{vec}\left( X_{H,T,M}\right) = \left[
\begin{array}{c}
H(x) \\
\vdots \\
H\left(\phi_T^{-M+1}(x)\right)
\end{array}
\right]\in\mathbb{R}^{MP}, \qquad \forall x\in \mathbb{A},
\end{equation}
\begin{equation}
\mathbb{A}_{H,T,M}:=\{x_{H,T,M}\,:\,x\in\mathbb{A}\}\subset\mathbb{R}^{MP}.\label{eq:def of AM}
\end{equation}
Then, let $U(\mathbb{A}_{H,T,M})$ denote the set of all directions in $\mathbb{A}_{H,T,M}$, i.e.,
\begin{equation}\label{eq:U of A}
U\left( \mathbb{A}_{H,T,M} \right) := \left\{ \frac{x_{H,T,M}-y_{H,T,M}}{\left\| x_{H,T,M}-y_{H,T,M} \right\|_2}\,: \,x,y\in \mathbb{A}  \right\}\subset\S^{MP-1},
\end{equation}
where $\S^{MP-1}$ is the unit sphere in $\mathbb{R}^{MP}$. Recalling  \eqref{eq:frak Z}, we observe that  the pair $(\mathfrak{Z},\|\cdot\|_F)$  is isometric  to  the pair $(U(\mathbb{A}_{H,T,M}),\|\cdot\|_2)$.
Thanks to this isometry,
 we may continue to simplify \eqref{eq:spect to frob} by writing that
\begin{align}\label{eq:co no of Z}
\#\left(\mathfrak{Z},\|\cdot\|,s \right) & \le \#\left(\mathfrak{Z},\|\cdot\|_F,s \right)\nonumber\\
& =\#\left(U\left(\mathbb{A}_{H,T,M} \right),\|\cdot\|_2,s \right).
\end{align}
Next, we estimate the covering number of $U(\mathbb{A}_{H,T,M})$. Recall that the attractor $\mathbb{A}\subset\mathbb{R}^N$ is a well-behaved manifold and the flow $\phi_T(\cdot)$ is a diffeomorphism on $\mathbb{A}$. Not surprisingly, then,  $\mathbb{A}_{H,T,M}$ (defined in \eqref{eq:def of AM}) too is a well-behaved manifold whose geometrical attributes can be expressed in terms of those of $\mathbb{A}$. This observation is formalized next and proved in Appendix \ref{sec:proof of lemma props of A_h,M,T}.
\begin{lemma}\label{lemma:props of A_h,M,T}
Recall the attractor $\mathbb{A}\subset\mathbb{R}^N$, and the flow $\phi_T:\mathbb{A}\rightarrow\mathbb{A}$, which by assumption is a diffeomorphism on $\mathbb{A}$. Let $D\phi_T(x):\mathbb{T}_x\mathbb{A}\rightarrow\mathbb{T}_{\phi_T(x)}\mathbb{A}$ be the derivative of the flow at $x\in\mathbb{A}$ (see Section \ref{sec:Differential Geometry}). The linear map  $D\phi_T(x)$ may be identified with a $\operatorname{dim}(\mathbb{A})\times\operatorname{dim}(\mathbb{A})$ matrix. Assume that the singular values of this matrix belong to some interval $[\sigma_{\min},\sigma_{\max}]\subset(0,\infty)$. Lastly, recall the properties of the map $H(\cdot)$ listed in Section~\ref{sec:measurement}.

Then, $\mathbb{A}_{H,T,M}\subset\mathbb{R}^{MP}$, as specified in \eqref{eq:def of AM}, is a bounded, boundary-less, and smooth submanifold of $\mathbb{R}^{MP}$ with
$
\operatorname{dim}\left( \mathbb{A}_{H,T,M} \right) = \operatorname{dim} (\mathbb{A})
$
. Moreover,
\begin{equation*}
\eta_{\min}^{\operatorname{dim}(\mathbb{A})}\sqrt{ \frac{\sigma_{\max}^{-2M\cdot\operatorname{dim}(\mathbb{A})}-1}{\sigma_{\max}^{-2\cdot\operatorname{dim}(\mathbb{A})}-1}}\cdot \operatorname{vol}\left( \mathbb{A}\right) \le \operatorname{vol} \left(\mathbb{A}_{H,T,M} \right) \le \eta_{\max}^{\operatorname{dim}(\mathbb{A})}\sqrt{ \frac{\sigma_{\min}^{-2M\cdot\operatorname{dim}(\mathbb{A})}-1}{\sigma_{\min}^{-2\cdot\operatorname{dim}(\mathbb{A})}-1}}\cdot \operatorname{vol}\left( \mathbb{A}\right).
\end{equation*}
\end{lemma}
The above lemma controls the geometric properties of $U(\mathbb{A}_{H,T,M})$---its dimension and volume. By substituting these estimates  into Lemma 15 of \cite{Eftekhari2015}, we can in turn control the  covering number of $U(\mathbb{A}_{H,T,M})$ by writing that
\begin{align*}
& \#\left(U\left(\mathbb{A}_{H,T,M}\right),\|\cdot\|_2,s \right) \nonumber\\
&
\le 2 \left( \frac{6.12\sqrt{\operatorname{dim}\left(\mathbb{A}_{H,T,M}\right)}}{s^2}\right)^{2\cdot\operatorname{dim}\left(\mathbb{A}_{H,T,M}\right)} \\
& \qquad \cdot \left( \frac{\operatorname{vol}\left(\mathbb{A}_{H,T,M}\right)}{\operatorname{rch}\left(\mathbb{A}_{H,T,M}\right)^{\operatorname{dim}(\mathbb{A}_{H,T,M})}}\right)^2\qquad \mbox{(invoke \cite[Lemma 15]{Eftekhari2015})} \nonumber\\
& \le 2 \left( \frac{6.12\sqrt{\operatorname{dim}\left(\mathbb{A}\right)}}{s^2}\right)^{2\cdot\operatorname{dim}\left(\mathbb{A}\right)}
\cdot \eta_{\max}^{2\cdot \operatorname{dim}(\mathbb{A})}\\
& \qquad \cdot \frac{\sigma_{\min}^{-2M\cdot\operatorname{dim}(\mathbb{A})}-1}{\sigma_{\min}^{-2\cdot\operatorname{dim}(\mathbb{A})}-1} \cdot \left( \frac{\operatorname{vol}\left(\mathbb{A}\right)}{\operatorname{rch}\left(\mathbb{A}_{H,T,M}\right)^{\operatorname{dim}(\mathbb{A})}}\right)^2,\, \mbox{(invoke Lemma \ref{lemma:props of A_h,M,T})}
\end{align*}
which holds for any $s\le \frac{1}{2}$, and under the mild assumption that  the volume of $\mathbb{A}$ is sufficiently large:
\begin{align*}
 \frac{\operatorname{vol}\left(\mathbb{A}_{H,T,M}\right)}{\operatorname{rch}\left(\mathbb{A}_{H,T,M}\right)^{\operatorname{dim}\left(\mathbb{A}_{H,T,M}\right)}}
 & \ge \eta_{\min}^{\operatorname{dim}(\mathbb{A})} \sqrt{\frac{\sigma_{\max}^{-2M\cdot\operatorname{dim}(\mathbb{A})}-1}{\sigma_{\max}^{-2\cdot\operatorname{dim}(\mathbb{A})}-1}}\cdot \frac{\operatorname{vol}\left(\mathbb{A}\right)}{\operatorname{rch}\left(\mathbb{A}_{H,T,M}\right)^{\operatorname{dim}\left(\mathbb{A}\right)}} \\
& \ge \left(\frac{21}{2\sqrt{\operatorname{dim}\left(\mathbb{A}_{H,T,M}\right)}} \right)^{\operatorname{dim}\left(\mathbb{A}_{H,T,M}\right)} \\
& = \left(\frac{21}{2\sqrt{\operatorname{dim}\left(\mathbb{A}\right)}} \right)^{\operatorname{dim}\left(\mathbb{A}\right)}.
\end{align*}
In light of \eqref{eq:co no of Z}, we conclude that
\begin{equation*}
\#\left(\mathfrak{Z},\|\cdot\|,s \right)
\le 2 \left( \frac{6.12\sqrt{\operatorname{dim}\left(\mathbb{A}\right)}}{s^2}\right)^{2\cdot\operatorname{dim}\left(\mathbb{A}\right)}
\eta_{\max}^{2\cdot \operatorname{dim}(\mathbb{A})}
 \frac{\sigma_{\min}^{-2M\cdot\operatorname{dim}(\mathbb{A})}-1}{\sigma_{\min}^{-2\cdot\operatorname{dim}(\mathbb{A})}-1}
  \left( \frac{\operatorname{vol}\left(\mathbb{A}\right)}{\operatorname{rch}\left(\mathbb{A}_{H,T,M}\right)^{\operatorname{dim}(\mathbb{A})}}\right)^2,
\end{equation*}
and we denote the right hand side by
\begin{equation}\label{eq:bnd on co no of Z}
\left( \frac{\Delta}{s} \right)^{4\cdot \operatorname{dim}(\mathbb{A})}.
\end{equation}
The above bound holds for every $s\le \frac{1}{2}$, and as long as
\begin{align}
\frac{\operatorname{vol}\left(\mathbb{A}\right)^{\frac{1}{\operatorname{dim}(\mathbb{A})}}}{\operatorname{rch}\left(\mathbb{A}_{H,T,M}\right)}
& \ge
 \eta_{\min}^{-1}
 \left(
\frac{\sigma_{\max}^{-2M\cdot\operatorname{dim}(\mathbb{A})}-1}{\sigma_{\max}^{-2\cdot\operatorname{dim}(\mathbb{A})}-1}
\right)^{-\frac{1}{2\cdot \operatorname{dim}(\mathbb{A})}}
 \cdot \frac{21}{2\sqrt{\operatorname{dim}\left(\mathbb{A}\right)}} .
 \label{eq:assumption1}
\end{align}
With the covering number of $\mathfrak{Z}$ at hand, we now use \eqref{eq:Gaussian width ineq} to control the Gaussian width of $\mathfrak{Z}$:
\begin{align}
& \gamma_2\left(\mathfrak{Z},\|\cdot\| \right) \nonumber\\
&  \lesssim \int_0^\infty \sqrt{\log \left(\# \left(\mathfrak{Z},\|\cdot\|,s  \right)\right)}\, ds\nonumber\\
& = \int_0^{2\cdot d_2\left(\mathfrak{Z}\right)} \sqrt{\log \left(\# \left(\mathfrak{Z},\|\cdot\|,s  \right)\right)}\, ds \qquad \left(\mbox{from \eqref{eq:bnd on d2}: } s\ge 2\cdot d_2\left( \mathfrak{Z}\right)\Longrightarrow \#\left(\mathfrak{Z},\|\cdot\|,s\right) = 1  \right)
\nonumber\\
& \le \int_0^{\frac{2}{ \sqrt{\operatorname{R}_{H,T,M}(\mathbb{A})}}}  \sqrt{\log \left(\# \left(\mathfrak{Z},\|\cdot\|,s  \right)\right)}\, ds \qquad \mbox{(see \eqref{eq:bnd on d2})}\nonumber\\
& \le \sqrt{4\cdot\operatorname{dim}(\mathbb{A})} \int_0^{\frac{2}{ \sqrt{\operatorname{R}_{H,T,M}(\mathbb{A})}}}  \sqrt{\log \left( \frac{\Delta}{s} \right)   }\, ds \qquad \left(\mbox{see \eqref{eq:bnd on co no of Z}}\right)
\nonumber\\
& \le \sqrt{4\cdot\operatorname{dim}(\mathbb{A})} \cdot \left( \frac{4}{\operatorname{R}_{H,T,M}(\mathbb{A})}\right)^{\frac{1}{4}} \nonumber\\
& \quad \sqrt{ \int_0^{\frac{2}{ \sqrt{\operatorname{R}_{H,T,M}(\mathbb{A})}}}   \log \left( \frac{\Delta}{s} \right)   \, ds} \quad \left( \int_0^a\sqrt{f(s)}\,ds \le \sqrt{a \cdot \int_0^a f(s)\,ds} \right)\nonumber\\
& \le \sqrt{4\cdot\operatorname{dim}(\mathbb{A})} \cdot \left( \frac{4}{\operatorname{R}_{H,T,M}(\mathbb{A})}\right)^{\frac{1}{4}} \sqrt{ \int_0^{\frac{2}{ \sqrt{\operatorname{R}_{H,T,M}(\mathbb{A})}}}   \log \left(1+ \frac{\Delta}{s} \right)   \, ds}\nonumber\\
& \le  \sqrt{4\cdot\operatorname{dim}(\mathbb{A})} \cdot \left( \frac{4}{\operatorname{R}_{H,T,M}(\mathbb{A})}\right)^{\frac{1}{4}} \cdot
\sqrt{ \frac{4}{\sqrt{\operatorname{R}_{H,T,M}(\mathbb{A})}}}
 \nonumber\\
&\,  \cdot \sqrt{
\log \left(1+ \frac{\Delta \sqrt{\operatorname{R}_{H,T,M}(\mathbb{A})}}{2} \right)}   \quad  \left(\int_0^a \log\left(1 + \frac{b}{s}\right)
 \, ds \le 2a\cdot  \log\left(1 + \frac{b}{a}\right), \mbox{ if } a \le b
 \right)\nonumber\\
 & \le 8 \sqrt{\frac{\operatorname{dim}(\mathbb{A})}{\operatorname{R}_{H,T,M}(\mathbb{A})} \cdot \log \left(\Delta\sqrt{\operatorname{R}_{H,T,M}(\mathbb{A})}\right)} \qquad  \left(\log(1+a)\le 2 \log(a),\,\forall a\ge 2 \right)
\end{align}
and, by simplifying the last line,
\begin{align}
& \gamma_2\left( \mathcal{Z},\|\cdot\| \right)\nonumber\\
& \lesssim \sqrt{
\frac{\operatorname{dim}(\mathbb{A})}{\operatorname{R}_{H,T,M}(\mathbb{A})}} \nonumber\\
& \cdot \sqrt{
 \log
 \left(
 \sqrt{ \operatorname{dim}(\mathbb{A})}
  \eta_{\max}
   \left(  \frac{\sigma_{\min}^{-2M\cdot \operatorname{dim}(\mathbb{A})}-1}
   {\sigma_{\min}^{-2\cdot \operatorname{dim}(\mathbb{A})}-1}
   \right)^{\frac{1}{2\cdot \operatorname{dim}(\mathbb{A})}}
    \frac{\operatorname{vol}(\mathbb{A}) ^{\frac{1}{\operatorname{dim}(\mathbb{A})}}}{\operatorname{rch}(\mathbb{A}_{H,T,M})}
    \cdot
     \operatorname{R}_{H,T,M}(\mathbb{A})\right)}
 .\, \mbox{(see \eqref{eq:bnd on co no of Z}) } \label{eq:bnd on gamma2}
\end{align}
For the fifth and tenth lines to  hold, we must impose that
\begin{equation}\label{eq:obscure mid step}
\operatorname{R}_{H,T,M}(\mathbb{A})\ge 16\cdot \max\left( 1,\Delta^{-2}\right).
\end{equation}
For \eqref{eq:obscure mid step} to hold, it actually suffices to assume that
\begin{equation}
\operatorname{R}_{H,T,M}(\mathbb{A})\gtrsim 1,
\end{equation}
\begin{equation}
\sqrt{\operatorname{dim}(\mathbb{A})}\cdot
\eta_{\max}\cdot
\left(
 \frac{\sigma_{\min}^{-2M\cdot \operatorname{dim}(\mathbb{A})}-1}{\sigma_{\min}^{-2\cdot \operatorname{dim}(\mathbb{A})}-1}
\right)
^{\frac{1}{2\cdot \operatorname{dim}(\mathbb{A})}}
\cdot
 \frac{ \operatorname{vol}(\mathbb{A})^{\frac{1}{\operatorname{dim}(\mathbb{A})}}}{\operatorname{rch}(\mathbb{A}_{H,T,M})}
 \gtrsim 1.
 \qquad \mbox{(see \eqref{eq:bnd on co no of Z})}
 \label{eq:assumptionagain}
\end{equation}
We note that~\eqref{eq:assumptionagain} is guaranteed to hold if~\eqref{eq:assumption1} (which appears in the theorem statement as~\eqref{eq:assumptionmain}) holds.
Given the estimates of $d_F(\mathfrak{Z})$, $d_2(\mathfrak{Z})$, and $\gamma_2(\mathfrak{Z},\|\cdot\|)$ (see (\ref{eq:bnd on dF}-\ref{eq:bnd on d2}) and \eqref{eq:bnd on gamma2}), we are now in position to apply Proposition  \ref{prop:Krahmer} to $\mathfrak{Z}$ (specified in \eqref{eq:frak Z}). For $\delta,\rho\in(0,1)$, assume that
\begin{align}
\label{eq:condition_on_soft_rank}
	\operatorname{R}_{H,T,M}(\mathbb{A} )
	& \gtrsim \delta^{-2} \cdot  \operatorname{dim}(\mathbb{A})
	\cdot \log
 \Bigg(
 \sqrt{ \operatorname{dim}(\mathbb{A})}
  \cdot
  \eta_{\max} \cdot
   \left( \frac{\sigma_{\min}^{-2M\cdot \operatorname{dim}(\mathbb{A})}-1}
   {\sigma_{\min}^{-2\cdot \operatorname{dim}(\mathbb{A})}-1}
   \right)^{\frac{1}{2\cdot \operatorname{dim}(\mathbb{A})}}
   \cdot
    \frac{\operatorname{vol}(\mathbb{A}) ^{\frac{1}{\operatorname{dim}(\mathbb{A})}}}{\operatorname{rch}(\mathbb{A}_{H,T,M})}
    \nonumber\\
& \qquad      \operatorname{R}_{H,T,M}(\mathbb{A})  \Bigg)
+ \delta^{-2}\cdot \log\left(\frac{1}{\rho} \right).
\end{align}
Under this assumption, we obtain that
\begin{align*}
	d_F(\mathfrak{Z}) & = 1 ,\qquad \mbox{(see \eqref{eq:bnd on dF})} \\
	d_2(\mathfrak{Z}) & = \frac{1}{\sqrt{\operatorname{R}_{H,T,M}(\mathbb{A})}} \lesssim \delta\cdot  \left(\log\left(\frac{1}{\rho} \right)\right)^{-\frac{1}{2}}, \qquad \mbox{(see \eqref{eq:bnd on d2})}\\
	\gamma_2\left(\mathfrak{Z}, \| \cdot \| \right)
& \lesssim \delta. \quad \mbox{(see \eqref{eq:bnd on gamma2})}
	\end{align*}
Subsequently, the quantities $E_1$, $E_2$, and $E_3$ in Proposition \ref{prop:Krahmer} may be bounded as
\begin{align*}
	E_1 &\lesssim \delta\cdot  ( \delta+1) + \delta\cdot \left( \log\left(\frac{1}{\rho} \right)\right)^{-\frac{1}{2}} \lesssim \delta, \qquad \left( \rho<1\right)\\
	E_2 &\lesssim \delta^2 \cdot \left(\log\left(\frac{1}{\rho} \right)\right)^{-1},\\
	E_3 &\lesssim \delta \cdot \left(\log\cdot \left(\frac{1}{\rho} \right)\right)^{-\frac{1}{2}}\cdot ( \delta+1) \lesssim \delta \cdot \left(\log\left(\frac{1}{\rho} \right)\right)^{
	-\frac{1}{2}}.\label{eq:bnds on E123}
\end{align*}
We now recall \eqref{eq:goal}, substitute the above quantities into Proposition \ref{prop:Krahmer} with an arbitrary $v>0$, and finally find that
\begin{align}
& \mathbb{P}\left[\sup_{x,y\in\mathbb{A}} \left| \frac{\left\| F_{h_\alpha,T,M}(x)-F_{h_\alpha,T,M}(y) \right\|_2^2}{\left\| X_{H,T,M}-Y_{H,T,M} \right\|_F^2} -1 \right| > \Cr{Krahmer}(\theta)\cdot \delta+v \right] \nonumber\\
& =
\mathbb{P}\left[ \sup_{Z\in \mathfrak{Z}} \left| \left\| Z \alpha \right\|_2^2- \mathbb{E}\left[ \left\| Z\alpha \right\|_2^2 \right]\right| > \Cr{Krahmer}(\theta)\cdot \delta+v\right]\qquad \mbox{(see \eqref{eq:goal})}\nonumber\\
& \le \mathbb{P} \left[ \sup_{Z\in \mathfrak{Z}} \left| \left\| Z \alpha \right\|_2^2- \mathbb{E}\left[ \left\| Z\alpha \right\|_2^2 \right]\right| > \Cr{Krahmer}(\theta)\cdot E_1+v\right] \qquad \mbox{(see \eqref{eq:bnds on E123})} \nonumber\\
& \le 2 \cdot \exp\left( -\Cr{Krahmer2}(\theta)\cdot \min \left[\frac{v}{E_2},\frac{v^2}{E_3^2} \right] \right)\qquad \mbox{(see Proposition \ref{prop:Krahmer})}\nonumber\\
& \le 2 \cdot \exp \left(- \frac{\Cr{Krahmer2}(\theta)}{\delta^2}  \cdot \log \left(\frac{1}{\rho} \right)\cdot \min\left(v^2, v \right) \right).\qquad \mbox{(see \eqref{eq:bnds on E123})}
\end{align}
By assigning $v = \delta$ above, we conclude that
\begin{align}\label{eq:final pre pre}
& \mathbb{P}\left[\sup_{x,y\in\mathbb{A}} \left| \frac{\left\| F_{h_\alpha,T,M}(x)-F_{h_\alpha,T,M}(y) \right\|_2^2}{\left\| X_{H,T,M}-Y_{H,T,M} \right\|_F^2} -1 \right| > \Cl{Krahmerp}(\theta)\cdot \delta \right]
\nonumber\\
& \le 2 \cdot \exp \left(- \frac{\Cr{Krahmer2}(\theta)}{\delta^2}  \cdot \log \left(\frac{1}{\rho} \right)\cdot \min\left(\delta^2, \delta \right) \right)\nonumber\\
&  \le \Cl{Krahmer2p}(\theta)\cdot   \rho,\qquad \left( \delta<1\right)
\end{align}
for $\Cr{Krahmerp}(\theta)$ and $\Cr{Krahmer2p}(\theta)$ that depend only on $\theta$.  Equivalently, if we replace $\gtrsim$ in \eqref{eq:condition_on_soft_rank} with $\gtrsim_{\theta}$, we can further simplify the above inequality to read
\begin{align}\label{eq:final pre}
\mathbb{P}\left[\sup_{x,y\in\mathbb{A}} \left| \frac{\left\| F_{h_\alpha,T,M}(x)-F_{h_\alpha,T,M}(y) \right\|_2^2}{\left\| X_{H,T,M}-Y_{H,T,M} \right\|_F^2} -1 \right| >  \delta \right]
&  \le \Cl{Krahmer2pp}(\theta) \cdot \rho,
\end{align}
for $\Cr{Krahmer2pp}(\theta)$ that depend only on $\theta$.  Here, $\gtrsim_{\theta}$ hides the explicit dependence on $\theta$ for convenience. This proves the version of Theorem \ref{thm:manifold2} that appears in Remark \ref{rem:poor geod}.

Fix $x,y\in\mathbb{A}$. We can in fact replace $\|X_{H,T,M}-Y_{H,T,M}\|_F$ above with a more approachable quantity as follows. From \eqref{eq:recalls XM}, recall that
\begin{equation}\label{eq:recall XM-YM}
\left\| X_{H,T,M}-Y_{H,T,M}\right\|_F^2 = \sum_{m=0}^{M-1} \left\| H\left(\phi_T^{-m}(x)\right)-H\left(\phi_T^{-m}(y)\right)\right\|_2^2,
\end{equation}
which suggests that we should find a more convenient expression for each summand above. Invoking the assumption in Theorem \ref{thm:manifold2} that the spectrum of $D\phi_T(\cdot)$ (the derivative of the flow) belongs to some interval $[\sigma_{\min},\sigma_{\max}]\in (0,\infty)$, we may easily verify that
\begin{equation}\label{eq:bi-lip pre}
\sigma_{\max}^{-m} \cdot d_{\mathbb{A}}\left(x,y\right) \le d_{\mathbb{A}}\left( \phi_T^{-m}(x),\phi_T^{-m}(y)\right) \le  \sigma_{\min}^{-m} \cdot d_{\mathbb{A}}\left(x,y\right),
\end{equation}
where $d_{\mathbb{A}}(\cdot,\cdot)$ returns the geodesic distance between a pair of points on $\mathbb{A}$ (see Section \ref{sec:Differential Geometry}). To relate the geodesic metric on $\mathbb{A}$ to the Euclidean metric in $\mathbb{R}^N$, we recall the regularity condition \eqref{eq:assump on geod diff thm}, from which it follows that
\begin{equation}
\operatorname{geo}(\mathbb{A})^{-1}\cdot \sigma_{\max}^{-m} \cdot \left\|x-y\right\|_2 \le \left\| \phi_T^{-m}(x)-\phi_T^{-m}(y)\right\|_2 \le  \operatorname{geo}(\mathbb{A})\cdot \sigma_{\min}^{-m} \cdot \left\|x-y\right\|_2.
\end{equation}
Next, recalling the bi-Lipschitz property of $H(\cdot)$ in Section~\ref{sec:measurement} allows us to update the above relation to read:
\begin{equation}
l_H\cdot \operatorname{geo}(\mathbb{A})^{-1}\cdot \sigma_{\max}^{-m} \cdot \left\|x-y\right\|_2 \le \left\| H\left(\phi_T^{-m}(x)\right)-H\left(\phi_T^{-m}(y)\right)\right\|_2 \le u_H\cdot \operatorname{geo}(\mathbb{A})\cdot \sigma_{\min}^{-m} \cdot \left\|x-y\right\|_2.
\label{eq:Hxvsx}
\end{equation}
From \eqref{eq:recall XM-YM}, it then follows that
\begin{equation}
l_H^2\cdot \operatorname{geo}(\mathbb{A})^{-2} \cdot \frac{\sigma_{\max}^{-2M}-1}{\sigma_{\max}^{-2}-1} \cdot \|x-y\|_2^2 \le \left\|X_{H,T,M}-Y_{H,T,M} \right\|_F^2 \le u_H^2\cdot \operatorname{geo}(\mathbb{A})^2 \cdot \frac{\sigma_{\min}^{-2M}-1}{\sigma_{\min}^2-1} \cdot \|x-y\|_2^2.
\label{eq:XHTMvsx}
\end{equation}
In turn,  \eqref{eq:final pre} now implies that
\begin{equation}\label{eq:final 0}
 (1-\delta)\cdot l_H^2 \cdot \operatorname{geo}(\mathbb{A})^{-2} \cdot \frac{\sigma_{\max}^{-2M}-1}{\sigma_{\max}^{-2}-1}  \le \frac{\left\| F_{h_{\alpha},T,M}(x)-F_{h_{\alpha},T,M}(y) \right\|_2^2}{ \|x-y\|_2^2}
\end{equation}
\begin{equation*}
 \qquad \qquad \qquad \qquad \qquad \le (1+\delta)\cdot u_H^2\cdot  \operatorname{geo}(\mathbb{A})^2\cdot \frac{\sigma_{\min}^{-2M}-1}{\sigma_{\min}^{-2}-1},
\end{equation*}
except with a probability of at most $\Cr{Krahmer2pp}(\theta)\cdot \rho$. To reiterate, the above relation holds under \eqref{eq:condition_on_soft_rank} (with $\gtrsim_{\theta}$ rather than $\gtrsim$), and under the mild assumption that
\begin{equation}
\frac{\operatorname{vol}\left(\mathbb{A}\right)^{\frac{1}{\operatorname{dim}(\mathbb{A})}}}{\operatorname{rch}\left(\mathbb{A}_{H,T,M}\right)}
 \gtrsim
\frac{1}{\eta_{\min}\sqrt{\operatorname{dim}(\mathbb{A})}}
\cdot
 \left(
  \frac{\sigma_{\max}^{-2M\cdot\operatorname{dim}(\mathbb{A})}-1}{\sigma_{\max}^{-2\cdot\operatorname{dim}(\mathbb{A})}-1}
\right)^{-\frac{1}{2\cdot \operatorname{dim}(\mathbb{A})}}
 .
 \label{eq:gtrexample}
\end{equation}

As our last step, we now remove the stable rank from the right hand side of \eqref{eq:condition_on_soft_rank}. To accomplish that, we focus on the requirement that
\begin{equation}
\operatorname{R}_{H,T,M} \left(\mathbb{A}\right)
\gtrsim_{\theta}
 \delta^{-2}\cdot \operatorname{dim}(\mathbb{A})\cdot \log\left(
     \operatorname{R}_{H,T,M}\left(\mathbb{A}\right)  \right).
\end{equation}
The \emph{Lambert W-function} $W(\cdot)$ \cite[\textsection 4.13]{Olver2010} is defined through the relation $W(z)\cdot e^{W(z)}= z$. Strictly speaking, the Lambert W-function is not a function, as it is multi-valued when $z<0$. In this case, $W(z)$ denotes the preimage of $W(z)\cdot e^{W(z)}=z$. Then, it is not difficult to verify that the requirement above is equivalent to
\begin{equation}
\operatorname{R}_{H,T,M} \left(\mathbb{A}\right) \gtrsim_{\theta}
e^{-\min W\left(-\frac{\delta^2}{\operatorname{dim}(\mathbb{A})} \right)}.
\end{equation}
This allows us to rewrite \eqref{eq:condition_on_soft_rank} as
\begin{align}
\label{eq:condition_on_soft_rank_3}
	& \operatorname{R}_{H,T,M}\left(\mathbb{A}\right)\nonumber\\
& 	\gtrsim_\theta
   \max
   \Bigg[
    \delta^{-2} \cdot  \operatorname{dim}(\mathbb{A})
	\cdot \log
 \left(
 \eta_{\max}\sqrt{ \operatorname{dim}(\mathbb{A})}
   \left( \frac{\sigma_{\min}^{-2M\cdot \operatorname{dim}\left(\mathbb{A}\right)}-1}
   {\sigma_{\min}^{-2\cdot \operatorname{dim}(\mathbb{A})}-1}
   \right)^{\frac{1}{2\cdot \operatorname{dim}(\mathbb{A})}}
    \frac{\operatorname{vol}\left(\mathbb{A}\right) ^{\frac{1}{\operatorname{dim}(\mathbb{A})}}}
    {\operatorname{rch}(\mathbb{A}_{H,T,M})}
  \right)
  \nonumber\\
  &
  \qquad \qquad \qquad ,
  e^{-\min W\left(\frac{-\delta^2}{\operatorname{dim}(\mathbb{A})} \right)}
   ,\delta^{-2} \log\left(\frac{1}{\rho} \right)
	\Bigg].
\end{align}
The proof of Theorem \ref{thm:manifold2} is now complete.

\begin{rem}\label{rem:poor geod}
\textbf{(Poor geodesic regularity)}
If the geodesic regularity of the attractor is poor (i.e., if $\operatorname{geo}(\mathbb{A})$ in (\ref{eq:assump on geod diff thm}) is large), also, if the singular values have a high ratio (as in a chaotic system), then perhaps the following slightly weaker result is more useful. Theorem \ref{thm:manifold2} holds verbatim but with the following replacing (\ref{eq:final pre thm}):
\begin{align}\label{eq:poor geo}
1-\delta  \le \frac{\left\| F_{h_{\alpha},T,M}(x)-F_{h_{\alpha},T,M}(y)\right\|_F^2}{
\sum_{m=0}^{M-1}
\left\|
H\left(\phi^{-m}_T(x)\right) - H\left(\phi^{-m}_T(y)\right)
\right\|_2^2}
\le 1+\delta ,\qquad \forall x,y\in\mathbb{A}.
\end{align}
\end{rem}

\section{Proof of Lemma \protect\ref{lemma:props of A_h,M,T}}\label{sec:proof of lemma props of A_h,M,T}

Recall that  $\mathbb{A}\subset\mathbb{R}^N$ is a bounded, boundary-less, and smooth manifold. Also, both $\phi_T:\mathbb{A}\rightarrow\mathbb{A}$ and $H:\mathbb{A}\rightarrow H(\mathbb{A})$ are diffeomorphisms. It follows that $\mathbb{A}_{H,T,M}\subset\mathbb{R}^{MP}$ (defined in \eqref{eq:def of AM}) too is a bounded, boundary-less, and smooth manifold, and that $\operatorname{dim}(\mathbb{A}_{H,T,M})  = \operatorname{dim}(\mathbb{A})$.

As for $\operatorname{vol}(\mathbb{A}_{H,T,M})$, we argue as follows. For $x\in\mathbb{A}$, let $DH(x):\mathbb{T}_x\mathbb{A}\rightarrow\mathbb{T}_{H(x)} H(\mathbb{A})$ be the derivative of $H(\cdot)$ at $x\in\mathbb{A}$ (see Section \ref{sec:Differential Geometry}). Each tangent space may be identified with $\mathbb{R}^{\operatorname{dim}(\mathbb{A})}$ and, consequently, $DH(x)$ may be identified with a $\operatorname{dim}(\mathbb{A})\times \operatorname{dim}(\mathbb{A})$ matrix. By assumption, the singular spectrum of $DH(x)$ belongs to the interval $[\eta_{\min},\eta_{\max}]\subset(0,\infty)$ (see Section~\ref{sec:measurement}). Then, the volume element of $\mathbb{A}$ under $H(\cdot)$ deforms as
\begin{equation}\label{eq:vol element under A}
\eta_{\min}^{\operatorname{dim}(\mathbb{A})}\cdot d\operatorname{vol}(x) \le d\operatorname{vol}\left( H(x)\right) \le \eta_{\max}^{\operatorname{dim}(\mathbb{A})}\cdot d\operatorname{vol}(x),\qquad \forall x\in\mathbb{A}.
\end{equation}
Similarly, let $D\phi_T(x):\mathbb{T}_x\mathbb{A} \rightarrow\mathbb{T}_{\phi(x)}\mathbb{A}$ be the derivative of the flow at $x\in\mathbb{A}$. By assumption, the singular spectrum of $D\phi(x)$ belongs to the interval  $[\sigma_{\min},\sigma_{\max}]\subset(0,\infty)$. Then, the volume element of $\mathbb{A}$ under $\phi_T^{-1}(\cdot)$ deforms as
\begin{equation}\label{eq:vol element under phi}
\sigma_{\max}^{-\operatorname{dim}(\mathbb{A})}\cdot d\operatorname{vol}(x)\le d\operatorname{vol}\left(\phi_T^{-1}(x) \right) \le {\sigma_{\min}^{-\operatorname{dim}(\mathbb{A})}} \cdot d\operatorname{vol}(x),\qquad \forall x\in \mathbb{A}.
\end{equation}
Predicated on the above observations, we have that
\begin{align*}
\operatorname{vol}\left( \mathbb{A}_{H,T,M} \right)
& = \int_{x\in\mathbb{A}}\, d\operatorname{vol}\left( x_{H,T,M}\right)\qquad \mbox{(see \eqref{eq:def of AM})}
\nonumber\\
& = \int_{x\in\mathbb{A}} \, \sqrt{\sum_{m=0}^{M-1}d\operatorname{vol}\left(H\left(\phi_T^{-m}\left(x\right)\right)\right)^2}\qquad \mbox{(see \eqref{eq:def of x_h,M,T})}\nonumber\\
& \le \int_{x\in\mathbb{A}} \, \sqrt{\eta_{\max}^{2\cdot \operatorname{dim}(\mathbb{A})}\sum_{m=0}^{M-1}  \sigma_{\min}^{-2m\cdot\operatorname{dim}(\mathbb{A})} }\cdot d\operatorname{vol}\left(x\right)\qquad \mbox{(see \eqref{eq:vol element under A},\eqref{eq:vol element under phi})}
\nonumber\\
& =   \eta_{\max}^{\operatorname{dim}(\mathbb{A})}\int_{x\in\mathbb{A}} \, \sqrt{\frac{\sigma_{\min}^{-2M\cdot\operatorname{dim}(\mathbb{A})}-1}{\sigma_{\min}^{-2\cdot\operatorname{dim}(\mathbb{A})}-1}}\cdot d\operatorname{vol}\left(x\right)
\nonumber\\
& = \eta_{\max}^{\operatorname{dim}(\mathbb{A})} \sqrt{\frac{\sigma_{\min}^{-2M\cdot\operatorname{dim}(\mathbb{A})}-1}{\sigma_{\min}^{-2\cdot\operatorname{dim}(\mathbb{A})}-1}}\cdot \operatorname{vol}\left( \mathbb{A}\right).
\end{align*}
A similar argument establishes that
\begin{equation*}
\operatorname{vol}\left(\mathbb{A}_{H,T,M}\right) \ge \eta_{\min}^{\operatorname{dim}(\mathbb{A})}\sqrt{ \frac{\sigma_{\max}^{-2M\cdot\operatorname{dim}(\mathbb{A})}-1}{\sigma_{\max}^{-2\cdot\operatorname{dim}(\mathbb{A})}-1}}\cdot \operatorname{vol}\left( \mathbb{A}\right).
\end{equation*}

\section{Proof of Theorem~\ref{thm:manifold2strange} (Stable Takens' Theorem for Strange Attractors)}
\label{sec:proof_manifold strange}

The proof follows the same arguments as outlined in Appendix~\ref{sec:proof_manifold}. We define the set $\mathfrak{Z}$ as in \eqref{eq:frak Z}, and we aim to control the supremum in \eqref{eq:goal} by invoking Proposition~\ref{prop:Krahmer}. To invoke this proposition, we must compute $d_F(\mathfrak{Z})$, $d_2(\mathfrak{Z})$, and $\gamma_2\left(\mathfrak{Z},\|\cdot\| \right)$. As in~\eqref{eq:bnd on dF}, we have
\[
d_F\left( \mathfrak{Z} \right) = 1,
\]
and as in \eqref{eq:bnd on d2}, we have
\[
d_2\left( \mathfrak{Z} \right) = \frac{1}{\sqrt{\operatorname{R}_{H,T,M}\left(\mathbb{A}\right)}}.
\]
To bound $\gamma_2\left(\mathfrak{Z},\|\cdot\| \right)$, we have
\begin{align}
& \gamma_2\left(\mathfrak{Z},\|\cdot\| \right) \nonumber\\
&  \lesssim \int_0^\infty \sqrt{\log \left(\# \left(\mathfrak{Z},\|\cdot\|,s  \right)\right)}\, ds \qquad \mbox{(see \eqref{eq:Gaussian width ineq})}\nonumber\\
& = \int_0^{2\cdot d_2\left(\mathfrak{Z}\right)} \sqrt{\log \left(\# \left(\mathfrak{Z},\|\cdot\|,s  \right)\right)}\, ds \qquad \left(\mbox{from \eqref{eq:bnd on d2}: } s\ge 2\cdot d_2\left( \mathfrak{Z}\right)\Longrightarrow \#\left(\mathfrak{Z},\|\cdot\|,s\right) = 1  \right)
\nonumber\\
& \le \int_0^{\frac{2}{ \sqrt{\operatorname{R}_{H,T,M}(\mathbb{A})}}}  \sqrt{\log \left(\# \left(\mathfrak{Z},\|\cdot\|,s  \right)\right)}\, ds \qquad \mbox{(see \eqref{eq:bnd on d2})}\nonumber\\
& \le \int_0^{\frac{2}{ \sqrt{\operatorname{R}_{H,T,M}(\mathbb{A})}}}  \sqrt{\log \left(\#\left(U\left(\mathbb{A}_{H,T,M} \right),\|\cdot\|_2,s \right)\right)}\, ds \qquad \mbox{(see \eqref{eq:co no of Z})} \label{eq:strangecov}
\end{align}
This allows us to focus on estimating the covering number of $U\left(\mathbb{A}_{H,T,M} \right)$. The following lemma is proved in Appendix \ref{sec:strangecovering}.

\begin{lemma}\label{lemma:strangecovering}
Under the assumptions of Theorem~\ref{thm:manifold2strange}, for all $0 < s < 2$,
\begin{equation}
\#\left(U\left(\mathbb{A}_{H,T,M} \right),\|\cdot\|_2,s \right) \le \left( \frac{\Delta}{s} \right)^{7 \operatorname{tandim}(\mathbb{A}_{H,T,M})},
\label{eq:covassume}
\end{equation}
where
\[
\Delta = \sqrt{24 \sqrt{MP} \operatorname{tancov}(\mathbb{A}_{H,T,M})\left(\operatorname{cov}(\mathbb{A}_{H,T,M})\right)^{1/\operatorname{boxdim}(\mathbb{A}_{H,T,M})}}.
\]
\end{lemma}

Now, with~\eqref{eq:covassume}, we may further bound the right hand side of~\eqref{eq:strangecov}. Omitting some intermediate steps, we conclude that
\begin{equation}
\gamma_2\left(\mathfrak{Z},\|\cdot\| \right) \lesssim \sqrt{ \frac{7 \operatorname{tandim}(\mathbb{A}_{H,T,M})}{\operatorname{R}_{H,T,M}(\mathbb{A})} \log\left( \Delta^2 \operatorname{R}_{H,T,M}(\mathbb{A}) \right)}
\label{eq:strangecovfull}
\end{equation}
as long as $\operatorname{R}_{H,T,M}(\mathbb{A}) \ge 16 \Delta^{-2}$. (This is guaranteed since $\Delta \ge 4$.) Then, if we assume that
\begin{equation}
\operatorname{R}_{H,T,M}(\mathbb{A}) \gtrsim_{\theta} \delta^{-2} \cdot 7 \operatorname{tandim}(\mathbb{A}_{H,T,M}) \log\left( \Delta^2 \operatorname{R}_{H,T,M}(\mathbb{A}) \right)  + \delta^{-2}\cdot \log\left(\frac{1}{\rho} \right),
\label{eq:strangeprelambert}
\end{equation}
we can guarantee that the following inequalities hold:
\begin{equation}
d_2(\mathfrak{Z})  = \frac{1}{\sqrt{\operatorname{R}_{H,T,M}(\mathbb{A})}} \lesssim_{\theta} \delta\cdot  \left(\log\left(\frac{1}{\rho} \right)\right)^{-\frac{1}{2}}
\label{eq:strangecond1}
\end{equation}
and
\begin{equation}
\gamma_2\left(\mathfrak{Z}, \| \cdot \| \right) \lesssim_{\theta} \delta.
\label{eq:strangecond2}
\end{equation}

Subsequently, the quantities $E_1$, $E_2$, and $E_3$ may be bounded as in~\eqref{eq:bnds on E123} (with $\lesssim$ replaced by $\lesssim_{\theta}$ throughout), and finally applying Proposition \ref{prop:Krahmer} with $v = \delta$ yields
\begin{align}\label{eq:final pre strange}
\mathbb{P}\left[\sup_{x,y\in\mathbb{A}} \left| \frac{\left\| F_{h_\alpha,T,M}(x)-F_{h_\alpha,T,M}(y) \right\|_2^2}{\left\| X_{H,T,M}-Y_{H,T,M} \right\|_F^2} -1 \right| >  \delta \right]
&  \le \Cl{Krahmer2ppp}(\theta) \cdot \rho,
\end{align}
for $\Cr{Krahmer2ppp}(\theta)$ that depends only on $\theta$. This gives one conclusion (analogous to Remark~\ref{rem:poor geod}), which may be of some value: with probability at least $1-\rho$, \eqref{eq:poor geo strange main} holds for all $x, y \in \mathbb{A}$ with $x \neq y$.

We may further strengthen this conclusion by following the remaining steps in Appendix~\ref{sec:proof_manifold}. If we suppose that~\eqref{eq:newdistanceassumption2main} holds,  then we can use the bi-Lipschitz property of $H(\cdot)$ in Section~\ref{sec:measurement} to conclude~\eqref{eq:Hxvsx}, \eqref{eq:XHTMvsx}, and thus \eqref{eq:final 0 strange main}.

Finally, as in Appendix~\ref{sec:proof_manifold}, we can remove the stable rank from the right hand side of~\eqref{eq:strangeprelambert} using the Lambert W-function to obtain~\eqref{eq:stable rank thm strange}.

\section{Proof of Lemma~\ref{lemma:strangecovering}}
\label{sec:strangecovering}

To bound the covering number of $U\left(\mathbb{A}_{H,T,M} \right)$, we start by defining the sets of long and short chords as
\begin{align*}
U_\gamma^l &= \left\{ \frac{a-b}{\|a-b\|_2}: ~ a, b \in U\left(\mathbb{A}_{H,T,M} \right), ~ \| a - b \|_2 > \gamma \right\}, \\
U_\gamma^s &= \left\{ \frac{a-b}{\|a-b\|_2}: ~ a, b \in U\left(\mathbb{A}_{H,T,M} \right), ~ \| a - b \|_2 \le \gamma \right\},
\end{align*}
where $\gamma > 0$ is a parameter to be set below. Noting that $U\left(\mathbb{A}_{H,T,M} \right) = U_\gamma^l \cup U_\gamma^s$, it suffices to bound the covering numbers of $U_\gamma^l$ and $U_\gamma^s$ separately.

We first bound the covering number of $U_\gamma^l$. Let $K'$ denote a $\left(\frac{\gamma s}{8}, \| \cdot \|_2 \right)$-cover of $\mathbb{A}_{H,T,M}$. To each point in $K'$ (which has distance $\frac{\gamma s}{4}$ or less from $\mathbb{A}_{H,T,M}$), we may associate its closest point that belongs to $\mathbb{A}_{H,T,M}$. Gathering these points, we obtain a new covering we will denote by $K$ such that $\# K \le \# K'$, $K \subset \mathbb{A}_{H,T,M}$, and $K$ is a $\left(\frac{\gamma s}{4}, \| \cdot \|_2 \right)$-cover of $\mathbb{A}_{H,T,M}$.

Now, for an arbitrary $\frac{a-b}{\|a-b\|_2} \in U_\gamma^l$, we have $\| a - b \|_2 > \gamma$ by the definition of $U_\gamma^l$. Also, by the covering construction above, there exist points $a', b' \in K$ such that
\[
\| a - a' \|_2 \le \frac{\gamma s}{4} ~~~ \text{and} ~~~ \| b - b' \|_2 \le \frac{\gamma s}{4}.
\]
Now, consider the Euclidean distance between $\frac{a-b}{\|a-b\|_2}$ and $\frac{a'-b'}{\|a'-b'\|_2}$. Following the proof techniques of Lemma 4.1 in \cite{Clarkson2008}, we have
\begin{eqnarray*}
	\left\|\frac{a - b}{\|a - b\|_2} - \frac{a' - b'}{\|a' - b'\|_2} \right\|_2
 	&\le& \left\| \frac{a - b}{\|a - b\|_{2}} - \frac{a' - b'}{\|a - b\|_{2}}\right\|_{2}
	+ \left\|\frac{a' - b'}{\|a - b\|_{2}} - \frac{a' - b'}{\|a' - b'\|_{2}}\right\|_{2} \\
	&=& \frac{\left\| \left(a - a'\right) - \left(b - b'\right) \right\|_2}{\|a - b\|_{2}} + \frac{\left|\|a'-b'\|_{2}-\|a-b\|_{2}\right|}{\|a-b\|_{2}\|a'-b'\|_{2}}\left\| a'-b'\right\| _{2}\\
	&\le& \frac{\left\| \left(a - a'\right) - \left(b - b'\right) \right\|_2}{\|a - b\|_{2}} + \frac{\left|\|a'-b'\|_{2}-\|a-b\|_{2}\right|}{\|a-b\|_{2}} \\
	&\le& \frac{\left\| \left(a - a'\right) - \left(b - b'\right) \right\|_2}{\|a - b\|_{2}} + \frac{\left\| \left(a-a'\right)-\left(b-b'\right)\right\|_{2}}{\|a-b\|_{2}} \\
	&\le& 2\frac{\left(\left\| a-a'\right\| _{2}+\left\| b-b'\right\| _{2}\right)}{\|a-b\|_{2}} \\
	&<& 2 \gamma^{-1} \cdot \frac{\gamma s}{2} = s,
\end{eqnarray*}
where the triangle and inverse triangle inequality were used several times. Since the choice of $\frac{a-b}{\|a-b\|_2} \in U_\gamma^l$ was arbitrary, it follows that the set
\[
\left\{ \frac{a'-b'}{\|a'-b' \|_2}: ~ a', b' \in K \right\}
\]
is an $(s, \|\cdot\|_2)$-cover of $U_\gamma^l$. Therefore, $\#\left(U_\gamma^l,\|\cdot\|_2,s \right) \le (\# K)^2 \le (\# K')^2$, where it remains to bound $\# K'$. Recalling the definitions of box-counting dimension and covering regularity, we know that $\mathbb{A}_{H,T,M}$ can be covered with cubes such that
\begin{eqnarray*}
\mathcal{N}(\mathbb{A}_{H,T,M},\zeta) \le \operatorname{cov}(\mathbb{A}_{H,T,M}) \zeta^{-\operatorname{boxdim}(\mathbb{A}_{H,T,M})},
\end{eqnarray*}
where we use $\mathcal{N}(\mathbb{A}_{H,T,M},\zeta)$ to denote the number of boxes or cubes of size $\zeta$ that intersect $\mathbb{A}_{H,T,M}\subset\mathbb{R}^{MP}$. To construct a covering with Euclidean balls of radius $r$, one can begin with a covering of cubes with sidelength $\frac{2r}{\sqrt{MP}}$ and inscribe each of these cubes in a ball of radius $r$. Thus, there exists a $\left(\frac{\gamma s}{8}, \| \cdot \|_2 \right)$-cover $K'$ of $\mathbb{A}_{H,T,M}$ with
\begin{equation}
\# K' \le \mathcal{N}(\mathbb{A}_{H,T,M},\frac{\gamma s}{4\sqrt{MP}}) \le \operatorname{cov}(\mathbb{A}_{H,T,M}) \left(\frac{\gamma s}{4\sqrt{MP}}\right)^{-\operatorname{boxdim}(\mathbb{A}_{H,T,M})}.
\label{eq:Kprime}
\end{equation}
Finally,
\[
\#\left(U_\gamma^l,\|\cdot\|_2,s \right) \le (\# K')^2 \le \left(\operatorname{cov}(\mathbb{A}_{H,T,M})\right)^2 \left(\frac{\gamma s}{4\sqrt{MP}}\right)^{-2\operatorname{boxdim}(\mathbb{A}_{H,T,M})}.
\]

We now bound the covering number of $U_\gamma^s$. The idea is to use the generalized tangent vectors of $\mathbb{A}_{H,T,M}$ to form a cover of $U_\gamma^s$. For every $d \in K$, let $\mathcal{S}_d$ denote the unit sphere in the generalized tangent space of $\mathbb{A}_{H,T,M}$ at $d$, $\mathbb{T}_d \mathbb{A}_{H,T,M}$. Let $\mathcal{C}_d$ denote a $\left( \frac{s}{2}, \| \cdot \|_2\right)$-cover for $\mathcal{S}_d$, and consider the finite set
\[
\mathcal{C} := \bigcup_{d \in K} \mathcal{C}_d.
\]
Observe that
\begin{align*}
\# \mathcal{C} & \le (\# K) \cdot \sup_{d \in K} \# \mathcal{C}_d \\
& \le \operatorname{cov}(\mathbb{A}_{H,T,M}) \left(\frac{\gamma s}{4\sqrt{MP}}\right)^{-\operatorname{boxdim}(\mathbb{A}_{H,T,M})} \cdot \sup_{d \in K} \left(1 + \frac{4}{s}\right)^{\text{dim}(\mathbb{T}_d \mathbb{A}_{H,T,M}))} \\
& \le \operatorname{cov}(\mathbb{A}_{H,T,M}) \left(\frac{\gamma s}{4\sqrt{MP}}\right)^{-\operatorname{boxdim}(\mathbb{A}_{H,T,M})} \cdot \left(1 + \frac{4}{s}\right)^{\operatorname{tandim}(\mathbb{A}_{H,T,M})} ,
\end{align*}
where the second line uses $\# K \le \#K'$, \eqref{eq:Kprime}, and a well-known bound on the covering number of the Euclidean ball (see, e.g.,~\cite[Lemma 1]{Vershynin2012}). The third line holds by the definition of the tangent dimension $\operatorname{tandim}(\mathbb{A}_{H,T,M})$.

Now, for an arbitrary $\frac{a-b}{\|a-b\|_2} \in U_\gamma^s$, we have $\| a - b \|_2 \le \gamma$ by the definition of $U_\gamma^s$. Pick $d \in K$ such that $\| d - a \|_2 \le \frac{\gamma s}{4}$. Using the triangle inequality, it follows that $\| d - b \|_2 \le \gamma(1 + s/4)$. Thus, both $a$ and $b$ are within a distance of $\gamma(1+s/4)$ from $d$. By the definition of tangent covering regularity, it follows that there exists $v \in \mathbb{T}_d \mathbb{A}_{H,T,M}$ such that
\[
\| v - \frac{a-b}{\|a-b\|_2} \|_2 \le \operatorname{tancov}(\mathbb{A}_{H,T,M}) \gamma (1+s/4).
\]
To achieve an $(s,\|\cdot\|_2)$-cover for $U_\gamma^s$, we must keep the right-hand side of the above smaller than $s$. Since $s < 2$, this is guaranteed by choosing
\[
\gamma = \gamma(s) = \frac{s}{3 \operatorname{tancov}(\mathbb{A}_{H,T,M})}.
\]
With this choice of $\gamma$, we have that $\mathcal{C}$ is an $(s,\|\cdot\|_2)$-cover for $U_\gamma^s$.

Adding the covering numbers for $U_\gamma^l$ and $U_\gamma^s$ completes the proof: for $0 < s < 2$,
\begin{align*}
& \#\left(U\left(\mathbb{A}_{H,T,M} \right),\|\cdot\|_2,s \right)\\  & \quad \le \left(\operatorname{cov}(\mathbb{A}_{H,T,M})\right)^2 \left(\frac{12 \sqrt{MP} \operatorname{tancov}(\mathbb{A}_{H,T,M})}{s^2}\right)^{2\operatorname{boxdim}(\mathbb{A}_{H,T,M})} \\ & \quad \quad + \operatorname{cov}(\mathbb{A}_{H,T,M}) \left(\frac{12 \sqrt{MP} \operatorname{tancov}(\mathbb{A}_{H,T,M})}{s^2}\right)^{\operatorname{boxdim}(\mathbb{A}_{H,T,M})} \cdot \left( 1 + \frac{4}{s} \right)^{\operatorname{tandim}(\mathbb{A}_{H,T,M})} \\
& \quad \le \left(\operatorname{cov}(\mathbb{A}_{H,T,M})\right)^2 \left(\frac{12 \sqrt{MP} \operatorname{tancov}(\mathbb{A}_{H,T,M})}{s^2}\right)^{2\operatorname{tandim}(\mathbb{A}_{H,T,M})} \\ & \quad \quad + \operatorname{cov}(\mathbb{A}_{H,T,M}) \left(\frac{12 \sqrt{MP} \operatorname{tancov}(\mathbb{A}_{H,T,M})}{s^2}\right)^{\operatorname{tandim}(\mathbb{A}_{H,T,M})} \cdot \left( \frac{6}{s} \right)^{\operatorname{tandim}(\mathbb{A}_{H,T,M})} \\
\\ & \quad \le \left(\frac{\sqrt{12 \sqrt{MP} \operatorname{tancov}(\mathbb{A}_{H,T,M})\left(\operatorname{cov}(\mathbb{A}_{H,T,M})\right)^{1/\operatorname{boxdim}(\mathbb{A}_{H,T,M})}}}{s}\right)^{6\operatorname{tandim}(\mathbb{A}_{H,T,M})} \cdot \left( \frac{6}{s} \right)^{\operatorname{tandim}(\mathbb{A}_{H,T,M})} \\
\\ & \quad \le \left(\frac{\sqrt{24 \sqrt{MP} \operatorname{tancov}(\mathbb{A}_{H,T,M})\left(\operatorname{cov}(\mathbb{A}_{H,T,M})\right)^{1/\operatorname{boxdim}(\mathbb{A}_{H,T,M})}}}{s}\right)^{7\operatorname{tandim}(\mathbb{A}_{H,T,M})}, \end{align*}
where the second inequality follows because $\operatorname{tandim}(\mathbb{A}_{H,T,M}) \ge \operatorname{boxdim}(\mathbb{A}_{H,T,M})$, because $s < 2$, and because we assume $\operatorname{tancov}(\mathbb{A}_{H,T,M}) > \frac{3}{\sqrt{MP}}$. The third inequality follows from multiplying the two summands from the second inequality, both of which are greater than or equal to $2$. The fourth inequality follows because we assume $\operatorname{tancov}(\mathbb{A}_{H,T,M}) > \frac{3}{\sqrt{MP}}$ and $\operatorname{cov}(\mathbb{A}_{H,T,M}) > 1$.

\section{Proof of Lemma \protect\ref{lem:geod reg}}\label{sec:proof of geod reg}

We begin by calculating the Euclidean distances on $\mathbb{A}$. For $t_1,t_2\in[0,1)$, note that
\begin{align}\label{eq:Euc dist moment}
& \left\| \gamma(t_1)-\gamma(t_2) \right\|_2^2 \nonumber\\
& = \sum_{n=0}^{N-1} \left| e^{\mbox{i} 2\pi nt_1}-e^{\mbox{i} 2\pi nt_2}\right|^2_2 \qquad \mbox{(see \eqref{eq:moment curve})}\nonumber\\
& = 4\sum_{n=0}^{N-1} \sin^2\left( \pi n(t_1-t_2)\right) \nonumber\\
& =  (2N-1)\cdot \left( 1-\frac{\mbox{Dirichlet}_{2N-1}(t_1-t_2)}{2N-1}\right).
\qquad \mbox{(trigonometric identity)}\nonumber\\
\end{align}
Above, for integer $N'$, $\mbox{Dirichlet}_{N'}(\cdot)$ is the Dirichlet kernel of width $\sim \frac{2}{N'}$, that is
\begin{equation}\label{eq:def of Dirichlet}
\mbox{Dirichlet}_{N'}(t) := \frac{\sin\left(\pi N't\right)}{\sin\left( \pi t\right)},\qquad \forall t\in\mathbb{R}.
\end{equation}
We recall an elementary property of the Dirichlet kernel.
\begin{lemma}\label{lem:prop of Dirichlet}
\emph{\cite[Lemma 13]{Eftekhari2015}} For an integer $N'$, let $\mbox{Dirichlet}_{N'}(\cdot)$ be the Dirichlet kernel as defined in (\ref{eq:def of Dirichlet}). Then, it holds that
\begin{equation}
\frac{\left| \mbox{Dirichlet}_{N'}(t)\right|}{N'} \le
\begin{cases}
\Cr{far} ,&  \frac{2}{N'} < |t|\le \frac{1}{2 }\\
\left(1-\frac{\left( \pi N't \right)^2}{40} \right)+\Cl[Bet]{near} t^2 & |t|\le \frac{2}{N'},
\end{cases}\qquad \forall N'>N_m.
\end{equation}
for (small) universal constants $\Cr{far},\Cr{near}>0$. Here, $N_m=N_m(\Cr{near})$ is a sufficiently large integer.
\end{lemma}
In light of this lemma, we may compare the geodesic and Euclidean distances between $\gamma(t_1),\gamma(t_2)\in\mathbb{A}$ by writing that
\begin{align*}
1 & \le \frac{d_{\mathbb{A}}\left(\gamma(t_1),\gamma(t_2) \right)^2}{\left\| \gamma(t_1)-\gamma(t_2) \right\|_2^2} \nonumber\\
& \le
\frac{2\pi^2}{3}\cdot N(N-1)\cdot
\frac{\left(t_1-t_2\right)^2}{ 1-\frac{\mbox{Dirichlet}_{2N-1}(t_1-t_2)}{2N-1}} \qquad \mbox{(see (\ref{eq:geod dist moment}) and (\ref{eq:Euc dist moment}))}
\end{align*}
and, consequently,
\begin{align}\label{eq:geod regular moment pre}
1 & \le \frac{d_{\mathbb{A}}\left(\gamma(t_1),\gamma(t_2) \right)^2}{\left\| \gamma(t_1)-\gamma(t_2) \right\|_2^2} \nonumber\\
& \le \frac{2\pi^2}{3}\cdot N(N-1)\cdot
\begin{cases}
\frac{\left( t_1-t_2\right)^2}{1-\Cr{far}},& \left| t_1-t_2\right|>\frac{2}{2N-1}\\
\frac{\left( t_1-t_2\right)^2}{\frac{\left( \pi (2N-1)\left(t_1-t_2 \right)\right)^2}{40}-\Cr{near}\left(t_1-t_2\right)^2},& \left|t_1-t_2\right|\le \frac{2}{2N-1}
\end{cases}\nonumber\\
& = \frac{2\pi^2}{3}\cdot N(N-1)\cdot
\begin{cases}
\frac{1}{1-\Cr{far}},& \left| t_1-t_2\right|>\frac{2}{2N-1}\\
\frac{1}{\frac{\pi^2 \left(2N-1\right)^2}{40}-\Cr{near}},& \left|t_1-t_2\right|\le \frac{2}{2N-1}
\end{cases}\nonumber\\
& \le \frac{2\pi^2}{3} \cdot N(N-1)\cdot \max \left[ \frac{1}{1-\Cr{far}}, \frac{1}{\frac{\pi^2 \left(2N-1\right)^2}{40}-\Cr{near}}\right]\nonumber\\
& = \frac{2\pi^2}{3(1-\Cr{far})} \cdot N(N-1),
\qquad \left( \mbox{when } N \mbox{ is large enough: } N>N_m\right)
\end{align}
Above, $N_m$ is a sufficiently large integer. This completes the proof of Lemma \ref{lem:geod reg}.

\section{Proof of Lemma \protect\ref{lem:s rank of moment curve}}
\label{sec:proof of s rank of moment}

From \eqref{eq:s rank of h(A)}, observe that
\begin{equation}\label{eq:stable rank moment}
\operatorname{R}_{H,T,M}(\mathbb{A}) = \inf_{t,t'\ge 0} \frac{\left\| G_{t,T,M}-G_{t',T,M}\right\|_F^2}{\left\| G_{t,T,M}-G_{t',T,M} \right\|^2},
\end{equation}
\begin{align*}
& G_{t,T,M}-G_{t',T,M} \\
& := \left[
\begin{array}{cccc}
\gamma(t)-\gamma(t') &
\gamma(t-T)-\gamma(t'-T) &
\cdots &
\gamma(t-(M-1)T)-\gamma(t'-(M-1)T)
\end{array}
\right]\nonumber\\
& \in\mathbb{C}^{N\times M}.\nonumber
\end{align*}
where we have dropped $H$ from the notation since $H(\cdot)$ is the identity operator throughout Section \ref{sec:second_example}.  Let us first compute the Frobenius norm in \eqref{eq:stable rank moment}. Note that
\begin{align*}
\left\|  G_{t,T,M}-G_{t',T,M}\right\|_F^2 &
= \sum_{m=0}^{M-1} \left\|  \gamma(t-mT)-\gamma(t'-mT)\right\|_2^2,
\end{align*}
and, consequently,
\begin{align}\label{eq:fro moment}
\left\|  G_{t,T,M}-G_{t',T,M}\right\|_F^2
& = 4M\sum_{n=0}^{N-1} \sin^2\left( \pi n(t-t')\right).\qquad \mbox{(see \eqref{eq:Euc dist moment})}
\end{align}
Computing the spectral norm in \eqref{eq:stable rank moment} requires a more elaborate argument. Using \eqref{eq:moment curve}, we may verify that
\begin{equation}\label{eq:arrange moment}
G_{t,T,M}-G_{t',T,M} =
\mbox{diag}\left[ \gamma(t)-\gamma(t')\right]
\cdot
\underset{\widetilde{H}\in\mathbb{C}^{N\times M}}{\underbrace{
\left[
\begin{array}{cccc}
\gamma(0) & \gamma(-T) & \cdots & \gamma(-MT)
\end{array}
\right]}},
\end{equation}
from which it immediately follows that
\begin{align}\label{eq:bnd on 2 norm moment}
\left\| G_{t,T,M}-G_{t',T,M}  \right\| & \le \left\|  \gamma(t)-\gamma(t')\right\|_{\infty} \cdot \|
\widetilde{H} \|\nonumber\\
& = \max_{n\in[0:N-1]} \left| \sin\left(\pi n (t-t') \right)\right| \cdot \| \widetilde{H} \|.\qquad \mbox{(see \eqref{eq:moment curve})}
\end{align}
Next, we bound the spectral norm of the \emph{Vandermonde matrix} $\widetilde{H}\in\mathbb{C}^{N\times M}$. In particular, if $T=\frac{1}{N}$ and $M\le N$, then $\widetilde{H}$ simply consists of the first $M$ columns of the (unnormalized) $N\times N$ Fourier matrix. Consequently, $\|\widetilde{H}\|= \sqrt{N}$.

In general, we bound the spectral norm of $\widetilde{H}$ as follows. After some algebraic manipulation, one recognizes that the corresponding Grammian matrix $G\in\mathbb{C}^{M\times M}$ is both Hermitian and Toeplitz, and that (the magnitude of) its entries are specified as
\begin{align}\label{eq:entries of moment gram}
\left|G[m,m']\right| & := \left|\left[ \widetilde{H}^* \widetilde{H} \right][m,m']\right|\nonumber\\
& = \left|\mbox{Dirichlet}_N\left( (m-m') T \right)\right|, \qquad  \forall m,m'\in[0: M-1].
\end{align}
Above,  $\mbox{Dirichlet}_N(\cdot)$ stands for the Dirichlet kernel of width $\sim \frac{2}{N}$ (see \eqref{eq:def of Dirichlet}).
Using the Gershgorin  disc theorem, it then follows that
\begin{equation}\label{eq:Gres s1}
\|\widetilde{H}\|^2  \le \sum_{m=0}^{M-1} \left|\mbox{Dirichlet}_{N}(mT)\right| = N+ \sum_{m=1}^{M-1} \left|\mbox{Dirichlet}_{N}(mT)\right|,
\end{equation}
Assuming that $MT\le 1$, we may use the bound  $\sin(\pi Nt)\le 1$ to further simplify  \eqref{eq:Gres s1} as
\begin{align*}
\|\widetilde{H}\|^2  & = N+ \sum_{m=1}^{M-1} \left|\mbox{Dirichlet}_{N}(mT)\right|\nonumber\\
& \le N+ 2\sum_{mT\le \frac{1}{2}} \frac{1}{\sin(\pi mT)},
\qquad \mbox{(see \eqref{eq:def of Dirichlet})}
\end{align*}
and, consequently,
\begin{align*}
\|\widetilde{H}\|^2
&  \le N +  \frac{2}{\sin(\pi T)}+ \frac{2}{T}\int_{ T}^{\frac{1}{2}} \frac{1}{\sin(\pi t)} \,dt\qquad  \left(\sin(\pi t) \mbox{ is increasing on } [0,1/2] \right)\nonumber\\
& = N +  \frac{2}{\sin(\pi T)}- \frac{2}{\pi T} \log\left(\tan\left( \frac{\pi T}{2}\right) \right)\nonumber\\
& \le N +  \frac{2}{\sin(\pi T)}\cdot \log\left(e/\tan\left( \frac{\pi T}{2}\right) \right).\qquad \left( \sin(\pi T)\le \pi T\right) \nonumber
\end{align*}
After substituting the estimate above back into \eqref{eq:bnd on 2 norm moment}, we obtain that
\begin{align}\label{eq:2 norm moment final}
& \left\| G_{t,T,M}-G_{t',T,M} \right\|^2\nonumber\\
 & \le \max_{n\in[0:N-1]}  \sin^2\left( \pi n (t-t')\right) \cdot \| \widetilde{H} \|^2 \nonumber\\
& \le \max_{n\in[0:N-1]}  \sin^2\left( \pi n (t-t')\right) \cdot \left( N +  \frac{2}{\sin(\pi T)}\cdot \log\left(e/\tan\left( \frac{\pi T}{2}\right) \right) \right).
\end{align}
With the estimates in \eqref{eq:fro moment} and \eqref{eq:2 norm moment final} in hand, we finally find that
\begin{align}\label{eq:s-rank pre}
\operatorname{R}_{H,T,M}(\mathbb{A}) & = \inf_{t,t'\ge 0} \frac{\left\|
G_{t,T,M}-G_{t',T,M}\right\|_F^2}{\left\| G_{t,T,M}-G_{t',T,M} \right\|^2}\qquad \mbox{(see \eqref{eq:stable rank moment})}\nonumber\\
& \ge \inf_{t,t'\ge 0} \frac{\frac{1}{N}\sum_{n=0}^{N-1} \sin^2\left( \pi n(t-t')\right)}{\max_{n\in[0:N-1]} \sin^2\left( \pi n (t-t')\right)} \cdot \frac{4MN}{ N +  \frac{2}{\sin(\pi T)}\cdot \log\left(e/\tan\left( \frac{\pi T}{2}\right) \right) }
\nonumber\\
& =  \inf_{|t|\le \frac{1}{2}} \frac{\frac{1}{N}\sum_{n=0}^{N-1} \sin^2\left( \pi nt \right)}{\max_{n\in[0:N-1]} \sin^2\left( \pi nt\right)} \cdot \frac{M}{ \frac{1}{4} +  \frac{1}{2N\sin(\pi T)}\cdot \log\left(e/\tan\left( \frac{\pi T}{2}\right) \right) }.
\end{align}
We are now left with the task of controlling the infimum in the last line above.  For a fixed $t\in[-\frac{1}{2},\frac{1}{2}]$, observe that
\begin{align*}
& \frac{\frac{1}{N}\sum_{n=0}^{N-1} \sin^2\left( \pi nt \right)}{\max_{n\in[0:N-1]} \sin^2\left( \pi nt\right)}  \nonumber\\
 & =
 \frac{2N-1}{N}\cdot \frac{ 1 - \frac{\mbox{Dirichlet}_{2N-1}(t)}{2N-1} }{\max_{n\in[0:N-1]} \sin^2\left( \pi nt\right)}\qquad \mbox{(trigonometric identity)}\\
  & \ge  \frac{ 1 - \frac{\mbox{Dirichlet}_{2N-1}(t)}{2N-1}}{\min\left[1,\left(\pi Nt\right)^2\right]}\nonumber\\
 & \ge  \frac{1}{\min\left[1,\left(\pi Nt\right)^2\right]}\cdot
 \begin{cases}
  1 - \Cr{far},& |t| >	\frac{2}{2N-1}\\
\frac{\left(\pi (2N-1)t \right)^2}{40}-\Cr{near} t^2,& |t| \le 	\frac{2}{2N-1}
 \end{cases}, \qquad \mbox{(see  Lemma \ref{lem:prop of Dirichlet})}
\end{align*}
and, consequently,
\begin{align}
 & \frac{\frac{1}{N}\sum_{n=0}^{N-1} \sin^2\left( \pi nt \right)}{\max_{n\in[0:N-1]} \sin^2\left( \pi nt\right)}  \nonumber\\
 & \ge
 \begin{cases}
 1 - \Cr{far},& |t| >	\frac{2}{2N-1}\\
 \frac{\frac{\left(\pi (2N-1)t \right)^2}{40}-\Cr{near} t^2}{\left(\pi Nt \right)^2},& |t| \le	\frac{2}{2N-1}\\
 \end{cases}\nonumber\\
 & \ge
 \begin{cases}
1-\Cr{far},& |t| >	\frac{2}{2N-1}\\
 \frac{1}{40}-\frac{\Cr{near}}{\pi^2N^2},& |t| \le	\frac{2}{2N-1}\\
 \end{cases}\nonumber\\
 & \ge \min\left[ 1-\Cr{far}, \frac{1}{40}- \frac{\Cr{near}}{\pi^2 N^2}\right]\nonumber\\
 & \ge \min\left[ 1-\Cr{far}, \frac{1}{80}\right]\nonumber\\
 & = \frac{1}{80},\qquad \left(\Cr{far}\approx 0.23 \right)
\end{align}
where $\Cr{far},\Cr{near}>0$ are (small) absolute constants   and, in particular, $\Cr{far}\approx 0.23$. The fourth and last two lines above hold for sufficiently large $N$:  $N>N_m=N_m(\Cr{near})$. The above estimate is independent of $t$ and, by substituting in \eqref{eq:s-rank pre}, leads us to
\begin{align}
\operatorname{R}_{H,T,M}(\mathbb{A})
& \ge \frac{1}{80}
\cdot \frac{M}{ \frac{1}{4} +  \frac{1}{2N\sin(\pi T)}\cdot \log\left(e/\tan\left( \frac{\pi T}{2}\right) \right) }.\nonumber
\end{align}
This completes the proof of Lemma \ref{lem:s rank of moment curve}.

\bibliographystyle{plain}
\bibliography{References,AdditionalReferences}

\end{document}